\documentclass[aps,prb,twocolumn, superscriptaddress, showpacs]{revtex4}
\usepackage{amsmath}
\usepackage{amssymb}
\usepackage{graphicx}
\usepackage{mathtools}
\usepackage{relsize}
\usepackage{lmodern}
\usepackage{slantsc}
\usepackage{scalefnt}
\usepackage{subfigure}
\usepackage{dsfont}
\usepackage{xspace}
\usepackage{ifpdf}
\usepackage{pgffor}
\usepackage{verbatim}
\usepackage{tikz}
\usetikzlibrary{shapes.geometric}
\usepackage{color}
\usetikzlibrary{arrows,matrix,calc,scopes,decorations.markings}
\allowdisplaybreaks[1]

\newtheorem{theorem}{Theorem}
\newtheorem{proposition}[theorem]{Proposition}
\newtheorem{convention}[theorem]{Convention}
\newenvironment{proof}[1][Proof]{\noindent\textbf{#1.} }{\ \rule{0.5em}{0.5em}}
\newcommand{\be}{\begin{equation}}
\newcommand{\ee}{\end{equation}}
\newcommand{\bse}{\begin{subequations}}
\newcommand{\ese}{\end{subequations}}
\newcommand{\ket}[1]{|{#1}\rangle}
\newcommand{\bra}[1]{\langle{#1}|}
\newcommand{\braket}[2]{\langle{#1}|{#2}\rangle}

\newcommand{\Z}{\mathbb{Z}}
\newcommand{\R}{\mathbb{R}}
\newcommand{\ii}{\mathrm{i}}
\newcommand{\e}{\mathrm{e}}
\newcommand{\Hil}{\mathcal{H}}

\newcommand{\str}{\mathcal{S}}
\newcommand{\ttr}{\mathcal{T}}
\newcommand{\bpm}{\begin{pmatrix}}
\newcommand{\epm}{\end{pmatrix}}
\newcommand{\bmm}{\begin{matrix}}
\newcommand{\emm}{\end{matrix}}

\newcommand{\dlt}[3]{\delta_{#1#2\cdot #2#3 \cdot #3#1}}
\newcommand{\Blangle}{\Biggl\langle\bmm} 
\newcommand{\BRvert}{\emm\Biggr\vert} 
\newcommand{\BLvert}{\Biggl\vert\bmm} 
\newcommand{\Bvert}{\emm\Biggr\vert\bmm} 
\newcommand{\Brangle}{\emm\Biggr\rangle}
\newcommand{\defeq}{\stackrel{\mathrm{def}}{=}}

\newcommand{\x}{\times}
\newcommand{\fc}{$4$-cocycle\xspace}
\newcommand{\fcs}{$4$-cocycles\xspace}
\newcommand{\sgn}[1]{\mathrm{sgn}(#1)}
\makeatletter
\newcommand*{\Relbarfill@}{\arrowfill@\Relbar\Relbar\Relbar}
\newcommand*{\xeq}[2][]{\ext@arrow 0055\Relbarfill@{#1}{#2}}
\makeatother

\tikzset{->-/.style={decoration={
  markings,
  mark=at position .5 with {\arrow{>}}},postaction={decorate}}}
\tikzset{-<-/.style={decoration={
  markings,
  mark=at position .5 with {\arrow{<}}},postaction={decorate}}}

\newcommand{\tetrahedron}[8]{
\begin{tikzpicture}[scale=#8,baseline]
\coordinate (a) at(0,0);
\coordinate (b) at (0.8,-0.75);
\coordinate (c) at (2.5,0);
\coordinate (d) at (1.25,2);
\coordinate (e) at ($ (a) ! 0.3 ! (c) $);
\coordinate (f) at ($ (a) ! 0.45 ! (c) $);
\draw (a) -- (b) -- (c) -- (d) -- cycle;
\draw (b) -- (d);
\draw (a) -- (e);
\draw (f) -- (c);
\draw[-<,>=latex,line width=0.01pt] (a) -- ($ (a) ! 0.5 ! (b) $);
\draw[-<,>=latex,line width=0.01pt] (b) -- ($ (b) ! 0.5 ! (c) $);
\draw[-<,>=latex,line width=0.01pt] (c) -- ($ (c) ! 0.5 ! (d) $);
\draw[<-,>=latex,line width=0.01pt] (f) -- (c);
\draw[-<,>=latex,line width=0.01pt] (b) -- ($ (b) ! 0.5 ! (d) $);
\draw[-<,>=latex,line width=0.01pt] (a) -- ($ (a) ! 0.5 ! (d) $);

\node[left] at(a) {\scalebox{0.7}{$#1$}};
\node[below] at(b) {\scalebox{0.7}{$#2$}};
\node[right] at(c) {\scalebox{0.7}{$#3$}};
\node[above] at(d) {\scalebox{0.7}{$#4$}};

\node[below] at ($ (a) ! 0.5 ! (b) $) {\scalebox{0.7}{$#5$}};
\node[below] at ($ (b) ! 0.5 ! (c) $) {\scalebox{0.7}{$#6$}};
\node[right] at ($ (c) ! 0.5 ! (d) $) {\scalebox{0.7}{$#7$}};
\node[below] at (f) {\scalebox{0.7}{$#5#6$}};
\node[right] at ($ (b) ! 0.5 ! (d) $) {\scalebox{0.7}{$#6#7$}};
\node[left] at ($ (a) ! 0.5 ! (d) $) {\scalebox{0.7}{$#5#6#7$}};
\end{tikzpicture}
}

\newcommand{\tetrahedronSplit}[9]{
\begin{tikzpicture}[scale=1,baseline]
\coordinate (a) at(0,0);
\coordinate (b) at (0.8,-0.75);
\coordinate (c) at (2.5,0);
\coordinate (d) at (1.25,2);
\coordinate (e) at ($ (a) ! 0.3 ! (c) $);
\coordinate (f) at ($ (a) ! 0.45 ! (c) $);
\coordinate (g) at ($ (a) ! 0.5 ! (b) $);
\coordinate (h) at ($ (c) ! 0.5 ! (d) $);
\coordinate (n) at ($ (g) ! 0.5 ! (h) $); 
\draw (a) -- (b) -- (c) -- (d) -- cycle;
\draw (b) -- (d);
\draw (a) -- (e);
\draw (f) -- (c);
\draw[-<,>=latex,line width=0.01pt] (a) -- ($ (a) ! 0.5 ! (b) $);
\draw[-<,>=latex,line width=0.01pt] (b) -- ($ (b) ! 0.5 ! (c) $);
\draw[-<,>=latex,line width=0.01pt] (c) -- ($ (c) ! 0.5 ! (d) $);
\draw[<-,>=latex,line width=0.01pt] (f) -- (c);
\draw[-<,>=latex,line width=0.01pt] (b) -- ($ (b) ! 0.5 ! (d) $);
\draw[-<,>=latex,line width=0.01pt] (a) -- ($ (a) ! 0.5 ! (d) $);

\node[left] at(a) {\scalebox{0.7}{$#1$}};
\node[below] at(b) {\scalebox{0.7}{$#2$}};
\node[right] at(c) {\scalebox{0.7}{$#3$}};
\node[above] at(d) {\scalebox{0.7}{$#4$}};
\node[circle,fill=black,outer sep=0pt, inner sep=0.6pt] at(n) {};
\node[below] at ($ (a) ! 0.5 ! (b) $) {\scalebox{0.7}{$#5$}};
\node[below] at ($ (b) ! 0.5 ! (c) $) {\scalebox{0.7}{$#6$}};
\node[right] at ($ (c) ! 0.5 ! (d) $) {\scalebox{0.7}{$#7$}};
\node at (0.6,-0.15) {\scalebox{0.7}{$#5#6$}};
\node[left] at ($ (b) ! 0.5 ! (d) $) {\scalebox{0.7}{$#6#7$}};
\node[left] at ($ (a) ! 0.5 ! (d) $) {\scalebox{0.7}{$#5#6#7$}};

\draw[dashed] (a) -- (n);
\draw[dashed] (b) -- (n);
\draw[dashed] (c) -- (n);
\draw[dashed] (d) -- (n);

\ifnum #8=1 {
  \node[right, outer sep=0pt, inner sep=1pt] at($ (n) ! 0.07 ! (d) $) {\scalebox{0.7}{$#1'$}};
  \draw[->,>=latex,line width=0.01pt] ($ (a) ! 0.4 ! (n) $) -- ($ (a) ! 0.5 ! (n) $) node[above] {\scalebox{0.7}{$#9$}};
  \draw[->,>=latex,line width=0.01pt] ($ (b) ! 0.4 ! (n) $) node[right] {\scalebox{0.7}{$#9#5$}} -- ($ (b) ! 0.5 ! (n) $) ;
  \draw[->,>=latex,line width=0.01pt] ($ (c) ! 0.4 ! (n) $) -- ($ (c) ! 0.5 ! (n) $) node[above] {\scalebox{0.7}{$#9#5#6$}};
  \draw[->,>=latex,line width=0.01pt] ($ (d) ! 0.4 ! (n) $) -- ($ (d) ! 0.5 ! (n) $) node[right,outer sep=0pt, inner sep=1pt] {\scalebox{0.6}{$#9#5#6#7$}};}
\fi
\ifnum #8=2 {
  \node[right, outer sep=0pt, inner sep=1pt] at($ (n) ! 0.07 ! (d) $) {\scalebox{0.7}{$#2'$}};
  \draw[-<,>=latex,line width=0.01pt] ($ (a) ! 0.4 ! (n) $) -- ($ (a) ! 0.5 ! (n) $) node[above, outer sep=0pt, inner sep=1pt] {\scalebox{0.7}{$#5#9^{-1}$}};
  \draw[->,>=latex,line width=0.01pt] ($ (b) ! 0.4 ! (n) $) node[right] {\scalebox{0.7}{$#9$}} -- ($ (b) ! 0.5 ! (n) $) ;
  \draw[->,>=latex,line width=0.01pt] ($ (c) ! 0.4 ! (n) $) -- ($ (c) ! 0.5 ! (n) $) node[above] {\scalebox{0.7}{$#9#6$}};
  \draw[->,>=latex,line width=0.01pt] ($ (d) ! 0.4 ! (n) $) -- ($ (d) ! 0.5 ! (n) $) node[right,outer sep=0pt, inner sep=1pt] {\scalebox{0.6}{$#9#6#7$}};}
\fi
\ifnum #8=3 {
  \node[right, outer sep=0pt, inner sep=1pt] at($ (n) ! 0.07 ! (d) $) {\scalebox{0.7}{$#3'$}};
  \draw[-<,>=latex,line width=0.01pt] ($ (a) ! 0.4 ! (n) $) -- ($ (a) ! 0.5 ! (n) $) node[above, outer sep=0pt, inner sep=1pt] {\scalebox{0.7}{$#5#6#9^{-1}$}};
  \draw[-<,>=latex,line width=0.01pt] ($ (b) ! 0.4 ! (n) $) node[right, outer sep=0pt, inner sep=1pt] {\scalebox{0.7}{$#6#9^{-1}$}} -- ($ (b) ! 0.5 ! (n) $) ;
  \draw[->,>=latex,line width=0.01pt] ($ (c) ! 0.4 ! (n) $) -- ($ (c) ! 0.5 ! (n) $) node[above] {\scalebox{0.7}{$#9$}};
  \draw[->,>=latex,line width=0.01pt] ($ (d) ! 0.4 ! (n) $) -- ($ (d) ! 0.5 ! (n) $) node[right,outer sep=0pt, inner sep=1pt] {\scalebox{0.6}{$#9#7$}};}
\fi
\ifnum #8=4 {
  \node[right, outer sep=0pt, inner sep=1pt] at($ (n) ! 0.07 ! (d) $) {\scalebox{0.7}{$#4'$}};
  \draw[-<,>=latex,line width=0.01pt] ($ (a) ! 0.4 ! (n) $) -- ($ (a) ! 0.5 ! (n) $) node[above, outer sep=0pt, inner sep=1pt] {\scalebox{0.7}{$#5#6#7#9^{-1}$}};
  \draw[-<,>=latex,line width=0.01pt] ($ (b) ! 0.4 ! (n) $) node[right, outer sep=0pt, inner sep=1pt] {\scalebox{0.7}{$#6#7#9^{-1}$}} -- ($ (b) ! 0.5 ! (n) $) ;
  \draw[-<,>=latex,line width=0.01pt] ($ (c) ! 0.4 ! (n) $) -- ($ (c) ! 0.5 ! (n) $) node[above] {\scalebox{0.7}{$#7#9^{-1}$}};
  \draw[->,>=latex,line width=0.01pt] ($ (d) ! 0.4 ! (n) $) -- ($ (d) ! 0.5 ! (n) $) node[right,outer sep=0pt, inner sep=1pt] {\scalebox{0.6}{$#9$}};}
\fi
\end{tikzpicture}
}

\newcommand{\oneTriangle}[4][0]{
\begin{tikzpicture}[scale=1,baseline]
\draw (0,0) node[left] {\scalebox{0.7}{$#2$}} -- (1,0) node[right] {\scalebox{0.7}{$#3$}} -- (60:1) node[above] {\scalebox{0.7}{$#4$}} -- cycle;
\ifnum #1=1 {
   \draw[-<,>=latex,line width=0.01pt] (0,0) -- (0.5,0) node[below] {\scalebox{0.7}{$[#2#3]$}};
   \draw[-<,>=latex,line width=0.01pt] (0,0) -- (60:0.5) node[left] {\scalebox{0.7}{$[#3#4]$}};
   \draw[->,>=latex,line width=0.01pt] (0,0) +(60:1) -- +(30:0.866) node[right] {\scalebox{0.7}{$[#2#4]$}};}
\else {
   \draw[-<,>=latex,line width=0.01pt] (0,0) -- (0.5,0) ;
   \draw[-<,>=latex,line width=0.01pt] (0,0) -- (60:0.5) ;
   \draw[->,>=latex,line width=0.01pt] (0,0) +(60:1) -- +(30:0.866);}  \fi
\end{tikzpicture}
}

\newcommand{\twoTriangles}[5]{
\begin{tikzpicture}[scale=0.8,baseline]
\coordinate (b) at (0,0);
\coordinate (c) at (30:1);
\coordinate (d) at (0,1);
\coordinate (a) at (150:1);

\draw (a) node[left] {\scalebox{0.7}{$#1$}} -- (b) node[below] {\scalebox{0.7}{$#2$}} -- (c) node[right] {\scalebox{0.7}{$#3$}} -- (d) node[above] {\scalebox{0.7}{$#4$}} -- cycle;
\draw[-<,>=latex,line width=0.01pt] (a) -- ($ (a) ! 0.5 ! (b) $);
\draw[-<,>=latex,line width=0.01pt] (b) -- ($ (b) ! 0.5 ! (c) $);
\draw[-<,>=latex,line width=0.01pt] (c) -- ($ (c) ! 0.5 ! (d) $);
\draw[-<,>=latex,line width=0.01pt] (a) -- ($ (a) ! 0.5 ! (d) $);

\ifnum #5=0 {
   \draw (b)  -- (d);
   \draw[-<,>=latex,line width=0.01pt] (b) -- ($ (b) ! 0.5 ! (d) $);}
\else {
   \draw (a)  -- (c);
   \draw[-<,>=latex,line width=0.01pt] (a) -- ($ (a) ! 0.5 ! (c) $);}
\fi
\end{tikzpicture}
}

\newcommand{\threeTriangles}[5]{
\begin{tikzpicture}[scale=0.8,baseline]
\coordinate (c) at (0,0);
\coordinate (d) at (0,1);
\coordinate (a) at (210:1);
\coordinate (b) at (-30:1);

\draw (a) node[left] {\scalebox{0.7}{$#1$}} -- (b) node[right] {\scalebox{0.7}{$#2$}} -- (c) node[below] {\scalebox{0.7}{$#3$}} -- (d) node[above] {\scalebox{0.7}{$#4$}} -- (a) -- (c);
\draw (b) -- (d);
\draw[-<,>=latex,line width=0.01pt] (a) -- ($ (a) ! 0.5 ! (b) $);
\draw[-<,>=latex,line width=0.01pt] (a) -- ($ (a) ! 0.5 ! (d) $);
\draw[-<,>=latex,line width=0.01pt] (b) -- ($ (b) ! 0.5 ! (d) $);

\ifnum #5=1 {
\draw[-<,>=latex,line width=0.01pt] (b) -- ($ (b) ! 0.5 ! (c) $);
\draw[->,>=latex,line width=0.01pt] (c) -- ($ (c) ! 0.5 ! (d) $);
\draw[-<,>=latex,line width=0.01pt] (a) -- ($ (a) ! 0.5 ! (c) $);}
\fi
\ifnum #5=2 {
\draw[-<,>=latex,line width=0.01pt] (b) -- ($ (b) ! 0.5 ! (c) $);
\draw[-<,>=latex,line width=0.01pt] (c) -- ($ (c) ! 0.5 ! (d) $);
\draw[-<,>=latex,line width=0.01pt] (a) -- ($ (a) ! 0.5 ! (c) $);}
\fi
\ifnum #5=3 {
\draw[->,>=latex,line width=0.01pt] (b) -- ($ (b) ! 0.5 ! (c) $);
\draw[-<,>=latex,line width=0.01pt] (c) -- ($ (c) ! 0.5 ! (d) $);
\draw[-<,>=latex,line width=0.01pt] (a) -- ($ (a) ! 0.5 ! (c) $);}
\fi
\ifnum #5=4 {
\draw[->,>=latex,line width=0.01pt] (b) -- ($ (b) ! 0.5 ! (c) $);
\draw[-<,>=latex,line width=0.01pt] (c) -- ($ (c) ! 0.5 ! (d) $);
\draw[->,>=latex,line width=0.01pt] (a) -- ($ (a) ! 0.5 ! (c) $);}
\fi
\end{tikzpicture}
}
\newcommand{\fourTet}[8]{
\begin{tikzpicture}[scale=#7,baseline]
\coordinate (a) at(0,0);    
\coordinate (b) at (0.8,-0.75);  
\coordinate (c) at (2.5,0);      
\coordinate (d) at (1.25,2);     
\coordinate (e) at ($ (a) ! 0.3 ! (c) $);
\coordinate (f) at ($ (a) ! 0.45 ! (c) $);
\coordinate (g) at ($ (a) ! 0.5 ! (b) $);
\coordinate (h) at ($ (c) ! 0.5 ! (d) $);
\coordinate (n) at ($ (g) ! 0.5 ! (h) $); 
\draw (a) -- (b) -- (c) -- (d) -- cycle;
\draw (b) -- (d);
\draw (a) -- (e);
\draw (f) -- (c);
\ifnum #8=1 {
\draw[-<,>=latex,line width=0.01pt] (a) -- ($ (a) ! 0.5 ! (b) $);
\draw[-<,>=latex,line width=0.01pt] (b) -- ($ (b) ! 0.5 ! (c) $);
\draw[-<,>=latex,line width=0.01pt] (c) -- ($ (c) ! 0.5 ! (d) $);
\draw[<-,>=latex,line width=0.01pt] (f) -- (c);
\draw[-<,>=latex,line width=0.01pt] (b) -- ($ (b) ! 0.5 ! (d) $);
\draw[-<,>=latex,line width=0.01pt] (a) -- ($ (a) ! 0.5 ! (d) $);}
\fi

\node[left] at(a) {\scalebox{0.7}{$#1$}};
\node[below] at(b) {\scalebox{0.7}{$#2$}};
\node[right] at(c) {\scalebox{0.7}{$#4$}};
\node[above] at(d) {\scalebox{0.7}{$#3$}};
\ifnum #5>0 {
\node[circle,fill=black,outer sep=0pt, inner sep=0.6pt] at(n) {};
\draw[dashed] (a) -- (n);
\draw[dashed] (b) -- (n);
\draw[dashed] (c) -- (n);
\draw[dashed] (d) -- (n);
}\fi

\ifnum #5=1 {
  \node[right, outer sep=0pt, inner sep=1pt] at($ (n) ! 0.07 ! (d) $) {\scalebox{0.7}{$#6$}};
  \ifnum #8=1 {
  \draw[->,>=latex,line width=0.01pt] ($ (a) ! 0.4 ! (n) $) -- ($ (a) ! 0.5 ! (n) $) ;
  \draw[->,>=latex,line width=0.01pt] ($ (b) ! 0.4 ! (n) $) -- ($ (b) ! 0.5 ! (n) $) ;
  \draw[->,>=latex,line width=0.01pt] ($ (c) ! 0.4 ! (n) $) -- ($ (c) ! 0.5 ! (n) $) ;
  \draw[->,>=latex,line width=0.01pt] ($ (d) ! 0.4 ! (n) $) -- ($ (d) ! 0.5 ! (n) $) ;}\fi}
\fi
\ifnum #5=2 {
  \node[right, outer sep=0pt, inner sep=1pt] at($ (n) ! 0.07 ! (d) $) {\scalebox{0.7}{$#6$}};
  \ifnum #8=1 {
  \draw[-<,>=latex,line width=0.01pt] ($ (a) ! 0.4 ! (n) $) -- ($ (a) ! 0.5 ! (n) $) ;
  \draw[->,>=latex,line width=0.01pt] ($ (b) ! 0.4 ! (n) $) -- ($ (b) ! 0.5 ! (n) $) ;
  \draw[->,>=latex,line width=0.01pt] ($ (c) ! 0.4 ! (n) $) -- ($ (c) ! 0.5 ! (n) $) ;
  \draw[->,>=latex,line width=0.01pt] ($ (d) ! 0.4 ! (n) $) -- ($ (d) ! 0.5 ! (n) $) ;}\fi}
\fi
\ifnum #5=3 {
  \node[right, outer sep=0pt, inner sep=1pt] at($ (n) ! 0.07 ! (d) $) {\scalebox{0.7}{$#6$}};
  \ifnum #8=1 {
  \draw[-<,>=latex,line width=0.01pt] ($ (a) ! 0.4 ! (n) $) -- ($ (a) ! 0.5 ! (n) $) ;
  \draw[-<,>=latex,line width=0.01pt] ($ (b) ! 0.4 ! (n) $) -- ($ (b) ! 0.5 ! (n) $) ;
  \draw[->,>=latex,line width=0.01pt] ($ (c) ! 0.4 ! (n) $) -- ($ (c) ! 0.5 ! (n) $) ;
  \draw[->,>=latex,line width=0.01pt] ($ (d) ! 0.4 ! (n) $) -- ($ (d) ! 0.5 ! (n) $) ;}\fi}
\fi
\ifnum #5=4 {
  \node[right, outer sep=0pt, inner sep=1pt] at($ (n) ! 0.07 ! (d) $) {\scalebox{0.7}{$#6$}};
  \ifnum #8=1 {
  \draw[-<,>=latex,line width=0.01pt] ($ (a) ! 0.4 ! (n) $) -- ($ (a) ! 0.5 ! (n) $) ;
  \draw[-<,>=latex,line width=0.01pt] ($ (b) ! 0.4 ! (n) $) -- ($ (b) ! 0.5 ! (n) $) ;
  \draw[-<,>=latex,line width=0.01pt] ($ (c) ! 0.4 ! (n) $) -- ($ (c) ! 0.5 ! (n) $) ;
  \draw[->,>=latex,line width=0.01pt] ($ (d) ! 0.4 ! (n) $) -- ($ (d) ! 0.5 ! (n) $) ;}\fi}
\fi
\end{tikzpicture}
}
\newcommand{\AvQuard}[6]{
\begin{tikzpicture}[y=0.80pt, x=0.8pt,yscale=-1, inner sep=0pt, outer sep=0pt,xscale=0.8,scale=0.7]
  \path[draw=black,line join=round,line cap=round]
    (138.5714,58.0765) -- (44.2857,176.6479) -- (142.1429,216.6479) --
    (233.5714,161.6479) -- cycle;
  \path[draw=black,line join=round,line cap=round]
    (142.1354,216.4314) -- (138.5977,58.1338);
  \path[draw=black,line join=round,line cap=round]
    (44.3836,176.6388) -- (119.7109,170.5965)(128.7844,169.8687) --
    (133.8230,169.4645)(147.9552,168.3309) --
    (161.9662,167.2071)(170.1564,166.5501) -- (233.2821,161.4865);
  \path[draw=black,dash pattern=on 2.40pt off 2.40pt,line join=round,line
    cap=round,miter limit=4.00] (138.3929,58.4336) --
    (110.7143,126.8265) -- (44.6429,176.6479);
  \path[draw=black,dash pattern=on 2.40pt off 2.40pt,line join=round,line
    cap=round,miter limit=4.00] (232.5876,161.4076) --
    (179.9004,146.4322)(173.8277,144.7061) --
    (162.7496,141.5574)(155.0092,139.3573) -- (110.8642,126.8099) --
    (141.6739,216.2084);
  \path[draw=black,dash pattern=on 3.20pt off 1.60pt on 0.80pt off 1.60pt,line
    join=round,line cap=round,miter limit=4.00]
    (138.3909,58.8772) -- (183.3427,132.3657) -- (232.8402,161.4076);
  \path[draw=black,dash pattern=on 3.20pt off 1.60pt on 0.80pt off 1.60pt,line
    join=round,line cap=round,miter limit=4.00]
    (141.9264,216.7135) -- (182.8376,132.6183) --
    (142.9555,145.3047)(136.1070,147.4832) --
    (124.9377,151.0362)(116.0930,153.8497) -- (113.7684,154.5891) --
    (44.6992,176.5599);
  \path[draw=black,line join=round,line cap=round]
    (110.6117,126.5574) -- (135.8843,128.5969)(144.4556,129.2886) --
    (182.5851,132.3658);
  \path[fill=black] (151,61) node[above right]  {$#1$};
  \path[fill=black] (236.8,164.9) node[above right]  {$#2$};
  \path[fill=black] (147,226) node[above right]  {$#3$};
  \path[fill=black] (21.6,175) node[above right]  {$#4$};
  \path[fill=black] (117.6,123) node[above right]  {$#5$}; 
  \path[fill=black] (150.6,126) node[above right]  {$#6$}; 

\end{tikzpicture}
}
\newcommand{\BvTri}[5]{
\begin{tikzpicture}[scale=1,baseline]
\coordinate (a) at(0,0);    
\coordinate (b) at (0.8,-0.75);  
\coordinate (c) at (2.5,0);      
\coordinate (d) at (1.25,2);     
\coordinate (e) at ($ (a) ! 0.3 ! (c) $);
\coordinate (f) at ($ (a) ! 0.45 ! (c) $);
\coordinate (g) at ($ (a) ! 0.5 ! (b) $);
\coordinate (h) at ($ (c) ! 0.5 ! (d) $);
\coordinate (n) at ($ (g) ! 0.5 ! (h) $); 
\draw (a) -- (b) -- (c) -- (d) -- cycle;
\draw (b) -- (d);
\draw (a) -- (e);
\draw (f) -- (c);
\draw[-<,>=latex,line width=0.01pt] (a) -- ($ (a) ! 0.5 ! (b) $);
\draw[-<,>=latex,line width=0.01pt] (b) -- ($ (b) ! 0.5 ! (c) $);
\draw[-<,>=latex,line width=0.01pt] (c) -- ($ (c) ! 0.5 ! (d) $);
\draw[<-,>=latex,line width=0.01pt] (f) -- (c);
\draw[-<,>=latex,line width=0.01pt] (b) -- ($ (b) ! 0.5 ! (d) $);
\draw[-<,>=latex,line width=0.01pt] (a) -- ($ (a) ! 0.5 ! (d) $);

\node[left] at(a) {\scalebox{0.7}{$#1$}};
\node[below] at(b) {\scalebox{0.7}{$#2$}};
\node[right] at(c) {\scalebox{0.7}{$#4$}};
\node[above] at(d) {\scalebox{0.7}{$#3$}};
\node[circle,fill=black,outer sep=0pt, inner sep=0.6pt] at(n) {};
\draw[dashed] (a) -- (n);
\draw[dashed] (b) -- (n);
\draw[dashed] (c) -- (n);
\draw[dashed] (d) -- (n);

\ifnum #5=1 {
  \node[right, outer sep=0pt, inner sep=1pt] at($ (n) ! 0.07 ! (d) $) {\scalebox{0.7}{$#1'$}};
  \draw[->,>=latex,line width=0.01pt] ($ (a) ! 0.4 ! (n) $) -- ($ (a) ! 0.5 ! (n) $) ;
  \draw[->,>=latex,line width=0.01pt] ($ (b) ! 0.4 ! (n) $) -- ($ (b) ! 0.5 ! (n) $) ;
  \draw[->,>=latex,line width=0.01pt] ($ (c) ! 0.4 ! (n) $) -- ($ (c) ! 0.5 ! (n) $) ;
  \draw[->,>=latex,line width=0.01pt] ($ (d) ! 0.4 ! (n) $) -- ($ (d) ! 0.5 ! (n) $) ;}
\fi
\ifnum #5=2 {
  \node[right, outer sep=0pt, inner sep=1pt] at($ (n) ! 0.07 ! (d) $) {\scalebox{0.7}{$#2'$}};
  \draw[-<,>=latex,line width=0.01pt] ($ (a) ! 0.4 ! (n) $) -- ($ (a) ! 0.5 ! (n) $) ;
  \draw[->,>=latex,line width=0.01pt] ($ (b) ! 0.4 ! (n) $) -- ($ (b) ! 0.5 ! (n) $) ;
  \draw[->,>=latex,line width=0.01pt] ($ (c) ! 0.4 ! (n) $) -- ($ (c) ! 0.5 ! (n) $) ;
  \draw[->,>=latex,line width=0.01pt] ($ (d) ! 0.4 ! (n) $) -- ($ (d) ! 0.5 ! (n) $) ;}
\fi
\ifnum #5=3 {
  \node[right, outer sep=0pt, inner sep=1pt] at($ (n) ! 0.07 ! (d) $) {\scalebox{0.7}{$#3'$}};
  \draw[-<,>=latex,line width=0.01pt] ($ (a) ! 0.4 ! (n) $) -- ($ (a) ! 0.5 ! (n) $) ;
  \draw[-<,>=latex,line width=0.01pt] ($ (b) ! 0.4 ! (n) $) -- ($ (b) ! 0.5 ! (n) $) ;
  \draw[->,>=latex,line width=0.01pt] ($ (c) ! 0.4 ! (n) $) -- ($ (c) ! 0.5 ! (n) $) ;
  \draw[->,>=latex,line width=0.01pt] ($ (d) ! 0.4 ! (n) $) -- ($ (d) ! 0.5 ! (n) $) ;}
\fi
\ifnum #5=4 {
  \node[right, outer sep=0pt, inner sep=1pt] at($ (n) ! 0.07 ! (d) $) {\scalebox{0.7}{$#4'$}};
  \draw[-<,>=latex,line width=0.01pt] ($ (a) ! 0.4 ! (n) $) -- ($ (a) ! 0.5 ! (n) $) ;
  \draw[-<,>=latex,line width=0.01pt] ($ (b) ! 0.4 ! (n) $) -- ($ (b) ! 0.5 ! (n) $) ;
  \draw[->,>=latex,line width=0.01pt] ($ (c) ! 0.4 ! (n) $) -- ($ (c) ! 0.5 ! (n) $) ;
  \draw[-<,>=latex,line width=0.01pt] ($ (d) ! 0.4 ! (n) $) -- ($ (d) ! 0.5 ! (n) $) ;}
\fi
\end{tikzpicture}
}

\newcommand{\threeTet}[6]{
\begin{tikzpicture}[y=0.80pt, x=0.8pt,yscale=-1, inner sep=0pt, outer sep=0pt,scale=#6]
  \path[draw=black,line join=round,line cap=round]
    (300.0000,972.3622) -- (190.0000,1002.3622) -- (140.0000,872.3622) --
    (250.0000,772.3622) -- (340.0000,862.3622) -- cycle;
  \path[draw=black,line join=round,line cap=round]
    (250.0000,772.3622) -- (300.0000,972.3622) -- (140.0000,872.3622);
  \path[draw=black,line join=round,line cap=round]
    (140.0000,872.3622) -- (219.2839,868.3980)(230.3456,867.8449) --
    (267.4849,865.9879)(279.6331,865.3805) -- (340.0000,862.3622) --
    (289.6211,909.3825)(280.4793,917.9149) --
    (257.2290,939.6151)(248.7351,947.5428) -- (190.0000,1002.3622) --
    (210.7952,922.6473)(213.2870,913.0954) --
    (218.0316,894.9077)(220.1100,886.9406) -- (250.0000,772.3622);
  \path[fill=black] (307,978) node[above right] {\scalebox{0.65}{$#1$}};
  \path[fill=black] (335,852.36218) node[above right]  {\scalebox{0.65}{$#2$}};
  \path[fill=black] (265,782.36218) node[above right]  {\scalebox{0.65}{$#3$}};
  \path[fill=black] (125,867.36218) node[above right]  {\scalebox{0.65}{$#4$}};
  \path[fill=black] (156,1002.3622) node[above right] {\scalebox{0.65}{$#5$}};
\end{tikzpicture}}

\newcommand{\twoTet}[6]{
\begin{tikzpicture}[y=0.80pt, x=0.8pt,yscale=-1, inner sep=0pt, outer sep=0pt,scale=#6]
  \path[draw=black,line join=round,line cap=round]
    (300.0000,972.3622) -- (190.0000,1002.3622) -- (140.0000,872.3622) --
    (250.0000,772.3622) -- (340.0000,862.3622) -- cycle;
  \path[draw=black,line join=round,line cap=round]
    (250.0000,772.3622) -- (300.0000,972.3622) -- (140.0000,872.3622);
  \path[draw=black,line join=round,line cap=round]
    (279.6331,865.3805) -- (340.0000,862.3622) -- (289.6211,909.3825);
  \path[draw=black,line join=round,line cap=round]
    (280.4793,917.9149) -- (257.2290,939.6151);
  \path[draw=black,line join=round,line cap=round]
    (248.7351,947.5428) -- (190.0000,1002.3622);
  \path[draw=black,line join=miter,line cap=round]
    (219.2839,868.3980) -- (230.3456,867.8449)(230.3456,867.8449) --
    (267.4849,865.9879)(140.0000,872.3622) -- (219.2839,868.3980);
  \path[fill=black] (307,978) node[above right] {\scalebox{0.65}{$#1$}};
  \path[fill=black] (335,852.36218) node[above right]  {\scalebox{0.65}{$#2$}};
  \path[fill=black] (265,782.36218) node[above right]  {\scalebox{0.65}{$#3$}};
  \path[fill=black] (125,867.36218) node[above right]  {\scalebox{0.65}{$#4$}};
  \path[fill=black] (156,1002.3622) node[above right] {\scalebox{0.65}{$#5$}};

\end{tikzpicture}}

\newcommand{\torusCube}[9]{
\begin{tikzpicture}[y=0.80pt, x=0.8pt,yscale=-1, inner sep=0pt, outer sep=0pt,scale=#9]
  \begin{scope}[cm={{0.08284,0.0,0.0,0.08284,(38.62164,986.2694)}}]
    \begin{scope}[cm={{1.06299,0.0,0.0,-1.06299,(-186.02362,789.27165)}},draw=black,line join=miter,line cap=butt,miter limit=10.43]
    \end{scope}
  \end{scope}
  \path[draw=black,line join=round,line cap=round,miter limit=4.00,line
    width=0.400pt] (50.2500,1032.1122) -- (50.2500,1002.1122);
  \path[draw=black,line join=round,line cap=round,miter limit=4.00,line
    width=0.400pt] (20.2500,1032.1122) -- (20.2500,1002.1122);
  \path[draw=black,line join=round,line cap=round,miter limit=4.00,line
    width=0.400pt] (50.2500,1002.1122) -- (20.2500,1002.1122);
  \path[draw=black,line join=round,line cap=round,miter limit=4.00,line
    width=0.400pt] (30.2500,992.1122) -- (20.2500,1002.1122);
  \path[draw=black,line join=round,line cap=round,miter limit=4.00,line
    width=0.400pt] (60.2500,992.1122) -- (50.2500,1002.1122);
  \path[draw=black,line join=round,line cap=round,miter limit=4.00,line
    width=0.400pt] (30.2500,992.1122) -- (60.2500,992.1122);
  \path[draw=black,line join=round,line cap=round,miter limit=4.00,line
    width=0.400pt] (60.2500,1022.1122) -- (60.2500,992.1122);
  \path[draw=black,line join=round,line cap=round,miter limit=4.00,line
    width=0.400pt] (50.2500,1032.1122) -- (60.2500,1022.1122);
  \path[draw=black,line join=round,line cap=round,miter limit=4.00,line
    width=0.200pt] (30.2500,992.1122) -- (30.2500,1001.0747)(30.2500,1003.3476) --
    (30.2500,1010.4186)(30.2500,1013.9542) -- (30.2500,1022.1122);
  \path[draw=black,line join=round,line cap=round,miter limit=4.00,line
    width=0.200pt] (60.2500,1022.1122) -- (51.5400,1022.1122)(49.0146,1022.1122)
    -- (41.4385,1022.1122)(38.6605,1022.1122) -- (30.2500,1022.1122);
  \path[draw=black,line join=round,line cap=round,miter limit=4.00,line
    width=0.200pt] (20.2500,1032.1122) -- (30.2500,1022.1122);
  \path[draw=black,line join=round,line cap=round,miter limit=4.00,line
    width=0.400pt] (20.2500,1002.1122) -- (50.2500,1032.1122);
  \path[draw=black,line join=round,line cap=round,miter limit=4.00,line
    width=0.400pt] (50.2500,1002.1122) -- (60.2500,1022.1122);
  \path[draw=black,line join=round,line cap=round,miter limit=4.00,line
    width=0.400pt] (30.2500,992.1122) -- (50.2500,1002.1122);
  \path[draw=black,line join=round,line cap=round,miter limit=4.00,line
    width=0.200pt] (20.2500,1002.1122) -- (30.2500,1022.1122);
  \path[draw=black,line join=round,line cap=round,miter limit=4.00,line
    width=0.200pt] (30.2500,1022.1122) -- (50.2500,1032.1122);
  \path[draw=black,line join=round,line cap=round,miter limit=4.00,line
    width=0.200pt] (30.2500,992.1122) -- (39.4416,1001.3038)(41.2094,1003.0716) --
    (44.4292,1006.2915)(46.1970,1008.0592) --
    (49.2275,1011.0897)(51.3109,1013.1731) -- (60.2500,1022.1122);
  \path[draw=black,line join=round,line cap=round,miter limit=4.00,line
    width=0.200pt] (30.2500,1022.1122) -- (34.4305,1017.9317)(36.0089,1016.3533)
    -- (50.2500,1002.1122);
  \path[fill=black] (17.399113,1032.236) node[above right]  {\scalebox{0.65}{$#1$}};
  \path[fill=black] (53.447334,1032.1368) node[above right] {\scalebox{0.65}{$#2$}};
  \path[fill=black] (25.25,1022.1121) node[above right] {\scalebox{0.65}{$#3$}};
  \path[fill=black] (61.492386,1022.1121) node[above right]  {\scalebox{0.65}{$#4$}};
  \path[fill=black] (17.328045,1003.2214) node[above right]  {\scalebox{0.65}{$#5$}};
  \path[fill=black] (52.366116,1002.9758) node[above right] {\scalebox{0.65}{$#6$}};
  \path[fill=black] (24.891497,994.84943) node[above right]{\scalebox{0.65}{$#7$}};
  \path[fill=black] (61.156353,994.97632) node[above right] {\scalebox{0.65}{$#8$}};
  \path[draw=black,line join=round,line cap=round,miter limit=4.00,line
    width=0.400pt] (20.2500,1032.1122) -- (50.2500,1032.1122);
  \path[fill=black] (57.4,1010) node[above right] {\scalebox{0.65}{$k$}};
  \path[fill=black] (44,1021) node[above right] {\scalebox{0.65}{$g$}};
  \path[fill=black] (57,1029.0608) node[above right] {\scalebox{0.65}{$h$}};
\end{tikzpicture}
}

\newcommand{\torusCubeNB}[9]{
\begin{tikzpicture}[y=0.80pt, x=0.8pt,yscale=-1, inner sep=0pt, outer sep=0pt,scale=#9]
  \begin{scope}[cm={{0.08284,0.0,0.0,0.08284,(38.62164,986.2694)}}]
    \begin{scope}[cm={{1.06299,0.0,0.0,-1.06299,(-186.02362,789.27165)}},draw=black,line join=miter,line cap=butt,miter limit=10.43]
    \end{scope}
  \end{scope}
  \path[draw=black,line join=round,line cap=round,miter limit=4.00,line
    width=0.400pt] (50.2500,1032.1122) -- (50.2500,1002.1122);
  \path[draw=black,line join=round,line cap=round,miter limit=4.00,line
    width=0.400pt] (20.2500,1032.1122) -- (20.2500,1002.1122);
  \path[draw=black,line join=round,line cap=round,miter limit=4.00,line
    width=0.400pt] (50.2500,1002.1122) -- (20.2500,1002.1122);
  \path[draw=black,line join=round,line cap=round,miter limit=4.00,line
    width=0.400pt] (30.2500,992.1122) -- (20.2500,1002.1122);
  \path[draw=black,line join=round,line cap=round,miter limit=4.00,line
    width=0.400pt] (60.2500,992.1122) -- (50.2500,1002.1122);
  \path[draw=black,line join=round,line cap=round,miter limit=4.00,line
    width=0.400pt] (30.2500,992.1122) -- (60.2500,992.1122);
  \path[draw=black,line join=round,line cap=round,miter limit=4.00,line
    width=0.400pt] (60.2500,1022.1122) -- (60.2500,992.1122);
  \path[draw=black,line join=round,line cap=round,miter limit=4.00,line
    width=0.400pt] (50.2500,1032.1122) -- (60.2500,1022.1122);
  \path[draw=black,line join=round,line cap=round,miter limit=4.00,line
    width=0.200pt] (30.2500,992.1122) -- (30.2500,1001.0747)(30.2500,1003.3476) --
    (30.2500,1010.4186)(30.2500,1013.9542) -- (30.2500,1022.1122);
  \path[draw=black,line join=round,line cap=round,miter limit=4.00,line
    width=0.200pt] (60.2500,1022.1122) -- (51.5400,1022.1122)(49.0146,1022.1122)
    -- (41.4385,1022.1122)(38.6605,1022.1122) -- (30.2500,1022.1122);
  \path[draw=black,line join=round,line cap=round,miter limit=4.00,line
    width=0.200pt] (20.2500,1032.1122) -- (30.2500,1022.1122);
  \path[draw=black,line join=round,line cap=round,miter limit=4.00,line
    width=0.400pt] (20.2500,1002.1122) -- (50.2500,1032.1122);
  \path[draw=black,line join=round,line cap=round,miter limit=4.00,line
    width=0.400pt] (50.2500,1002.1122) -- (60.2500,1022.1122);
  \path[draw=black,line join=round,line cap=round,miter limit=4.00,line
    width=0.400pt] (30.2500,992.1122) -- (50.2500,1002.1122);
  \path[draw=black,line join=round,line cap=round,miter limit=4.00,line
    width=0.200pt] (20.2500,1002.1122) -- (30.2500,1022.1122);
  \path[draw=black,line join=round,line cap=round,miter limit=4.00,line
    width=0.200pt] (30.2500,1022.1122) -- (50.2500,1032.1122);
  \path[draw=black,line join=round,line cap=round,miter limit=4.00,line
    width=0.200pt] (30.2500,992.1122) -- (39.4416,1001.3038)(41.2094,1003.0716) --
    (44.4292,1006.2915)(46.1970,1008.0592) --
    (49.2275,1011.0897)(51.3109,1013.1731) -- (60.2500,1022.1122);
  \path[draw=black,line join=round,line cap=round,miter limit=4.00,line
    width=0.200pt] (30.2500,1022.1122) -- (34.4305,1017.9317)(36.0089,1016.3533)
    -- (50.2500,1002.1122);
  \path[fill=black] (17.399113,1032.236) node[above right]  {\scalebox{0.65}{$#1$}};
  \path[fill=black] (53.447334,1032.1368) node[above right] {\scalebox{0.65}{$#2$}};
  \path[fill=black] (25.25,1023) node[above right] {\scalebox{0.65}{$#3$}};
  \path[fill=black] (61.492386,1022.1121) node[above right]  {\scalebox{0.65}{$#4$}};
  \path[fill=black] (17.328045,1003.2214) node[above right]  {\scalebox{0.65}{$#5$}};
  \path[fill=black] (52.366116,1002.9758) node[above right] {\scalebox{0.65}{$#6$}};
  \path[fill=black] (24.891497,994.84943) node[above right]{\scalebox{0.65}{$#7$}};
  \path[fill=black] (61.156353,994.97632) node[above right] {\scalebox{0.65}{$#8$}};
  \path[draw=black,line join=round,line cap=round,miter limit=4.00,line
    width=0.400pt] (20.2500,1032.1122) -- (50.2500,1032.1122);
  \path[fill=black] (44,1021) node[above right]  {\scalebox{0.65}{$g$}};
  \path[fill=black]  (57.4,1010) node[above right]  {\scalebox{0.65}{$k$}};
  \path[fill=black] (33,1029.0608) node[above right] {\scalebox{0.65}{$h'$}};
\end{tikzpicture}
}

\newcommand{\torusCubeNlabel}[9]{
\begin{tikzpicture}[y=0.80pt, x=0.8pt,yscale=-1, inner sep=0pt, outer sep=0pt,scale=#9]
  \begin{scope}[cm={{0.08284,0.0,0.0,0.08284,(38.62164,986.2694)}}]
    \begin{scope}[cm={{1.06299,0.0,0.0,-1.06299,(-186.02362,789.27165)}},draw=black,line join=miter,line cap=butt,miter limit=10.43]
    \end{scope}
  \end{scope}
  \path[draw=black,line join=round,line cap=round,miter limit=4.00,line
    width=0.400pt] (50.2500,1032.1122) -- (50.2500,1002.1122);
  \path[draw=black,line join=round,line cap=round,miter limit=4.00,line
    width=0.400pt] (20.2500,1032.1122) -- (20.2500,1002.1122);
  \path[draw=black,line join=round,line cap=round,miter limit=4.00,line
    width=0.400pt] (50.2500,1002.1122) -- (20.2500,1002.1122);
  \path[draw=black,line join=round,line cap=round,miter limit=4.00,line
    width=0.400pt] (30.2500,992.1122) -- (20.2500,1002.1122);
  \path[draw=black,line join=round,line cap=round,miter limit=4.00,line
    width=0.400pt] (60.2500,992.1122) -- (50.2500,1002.1122);
  \path[draw=black,line join=round,line cap=round,miter limit=4.00,line
    width=0.400pt] (30.2500,992.1122) -- (60.2500,992.1122);
  \path[draw=black,line join=round,line cap=round,miter limit=4.00,line
    width=0.400pt] (60.2500,1022.1122) -- (60.2500,992.1122);
  \path[draw=black,line join=round,line cap=round,miter limit=4.00,line
    width=0.400pt] (50.2500,1032.1122) -- (60.2500,1022.1122);
  \path[draw=black,line join=round,line cap=round,miter limit=4.00,line
    width=0.200pt] (30.2500,992.1122) -- (30.2500,1001.0747)(30.2500,1003.3476) --
    (30.2500,1010.4186)(30.2500,1013.9542) -- (30.2500,1022.1122);
  \path[draw=black,line join=round,line cap=round,miter limit=4.00,line
    width=0.200pt] (60.2500,1022.1122) -- (51.5400,1022.1122)(49.0146,1022.1122)
    -- (41.4385,1022.1122)(38.6605,1022.1122) -- (30.2500,1022.1122);
  \path[draw=black,line join=round,line cap=round,miter limit=4.00,line
    width=0.200pt] (20.2500,1032.1122) -- (30.2500,1022.1122);
  \path[draw=black,line join=round,line cap=round,miter limit=4.00,line
    width=0.400pt] (20.2500,1002.1122) -- (50.2500,1032.1122);
  \path[draw=black,line join=round,line cap=round,miter limit=4.00,line
    width=0.400pt] (50.2500,1002.1122) -- (60.2500,1022.1122);
  \path[draw=black,line join=round,line cap=round,miter limit=4.00,line
    width=0.400pt] (30.2500,992.1122) -- (50.2500,1002.1122);
  \path[draw=black,line join=round,line cap=round,miter limit=4.00,line
    width=0.200pt] (20.2500,1002.1122) -- (30.2500,1022.1122);
  \path[draw=black,line join=round,line cap=round,miter limit=4.00,line
    width=0.200pt] (30.2500,1022.1122) -- (50.2500,1032.1122);
  \path[draw=black,line join=round,line cap=round,miter limit=4.00,line
    width=0.200pt] (30.2500,992.1122) -- (39.4416,1001.3038)(41.2094,1003.0716) --
    (44.4292,1006.2915)(46.1970,1008.0592) --
    (49.2275,1011.0897)(51.3109,1013.1731) -- (60.2500,1022.1122);
  \path[draw=black,line join=round,line cap=round,miter limit=4.00,line
    width=0.200pt] (30.2500,1022.1122) -- (34.4305,1017.9317)(36.0089,1016.3533)
    -- (50.2500,1002.1122);
  \path[fill=black] (15.8,1032.236) node[above right]  {\scalebox{0.65}{$#1$}};
  \path[fill=black] (53.447334,1032.1368) node[above right] {\scalebox{0.65}{$#2$}};
  \path[fill=black] (25.25,1023) node[above right] {\scalebox{0.65}{$#3$}};
  \path[fill=black] (61.492386,1022.1121) node[above right]  {\scalebox{0.65}{$#4$}};
  \path[fill=black] (17,1003.2214) node[above right]  {\scalebox{0.65}{$#5$}};
  \path[fill=black] (52.366116,1002.9758) node[above right] {\scalebox{0.65}{$#6$}};
  \path[fill=black] (24.891497,994.84943) node[above right]{\scalebox{0.65}{$#7$}};
  \path[fill=black] (61.156353,994.97632) node[above right] {\scalebox{0.65}{$#8$}};
  \path[draw=black,line join=round,line cap=round,miter limit=4.00,line
    width=0.400pt] (20.2500,1032.1122) -- (50.2500,1032.1122);
\end{tikzpicture}
}

\newcommand{\twoAvCommute}[8]{
\begin{tikzpicture}[y=0.80pt, x=0.8pt,yscale=-1, inner sep=0pt, outer sep=0pt,scale=#8]
  \path[draw=black,line join=round,line cap=round]
    (300.0000,972.3622) -- (190.0000,1002.3622) -- (140.0000,872.3622) --
    (250.0000,772.3622) -- (340.0000,862.3622) -- cycle;
  \path[draw=black,line join=round,line cap=round]
    (250.0000,772.3622) -- (300.0000,972.3622) -- (140.0000,872.3622);
  \path[draw=black,line join=round,line cap=round]
    (140.0000,872.3622) -- (205.5237,869.0860)(210.5112,868.8366) --
    (211.9629,868.7640)(216.2469,868.5498) --
    (219.2839,868.3980)(235.7072,867.5768) --
    (267.4849,865.9879)(279.6331,865.3805) -- (340.0000,862.3622) --
    (289.6211,909.3825)(280.4793,917.9149) --
    (273.4047,924.5178)(267.3096,930.2066) --
    (257.2290,939.6151)(248.7351,947.5428) --
    (244.6231,951.3807)(238.5496,957.0493) -- (190.0000,1002.3622) --
    (206.0601,940.7987)(208.1271,932.8751) --
    (210.7952,922.6473)(213.2870,913.0954) --
    (218.0316,894.9077)(220.1100,886.9406) -- (250.0000,772.3622);
  \path[fill=black] (305,977.36218) node[above right] {\scalebox{0.65}{$#1$}};
  \path[fill=black] (335,852.36218) node[above right] {\scalebox{0.65}{$#2$}};
  \path[fill=black] (265,782.36218) node[above right] {\scalebox{0.65}{$#3$}};
  \path[fill=black] (130,867.36218) node[above right] {\scalebox{0.65}{$#4$}};
  \path[fill=black] (170,1002.3622) node[above right] {\scalebox{0.65}{$#5$}};
  \path[draw=red,dash pattern=on 2.40pt off 2.40pt,line join=round,line
    cap=round,miter limit=4.00] (250.0000,772.3622) --
    (212.5984,839.7638) -- (228.4193,842.5701)(241.0797,844.8157) --
    (266.6042,849.3433)(272.5936,850.4057) -- (340.0000,862.3622);
  \path[draw=red,dash pattern=on 2.40pt off 2.40pt,line join=round,line
    cap=round,miter limit=4.00] (140.0000,872.3622) --
    (212.5984,839.7638) -- (202.9114,909.4629)(202.4109,913.0641) --
    (200.4502,927.1717)(199.3829,934.8509) -- (190.0000,1002.3622);
  \path[draw=red,dash pattern=on 2.40pt off 2.40pt,line join=round,line
    cap=round,miter limit=4.00] (212.5984,839.7638) --
    (231.0392,867.7406)(234.0380,872.2901) --
    (247.1316,892.1546)(249.2125,895.3116) -- (300.0000,972.3622);
  \path[draw=blue,dash pattern=on 3.20pt off 1.60pt on 0.80pt off 1.60pt,line
    join=round,line cap=round,miter limit=4.00]
    (212.5984,839.7638) -- (215.7044,899.1893)(216.1672,908.0435) --
    (216.5797,915.9351)(216.9711,923.4239) -- (218.1930,946.8012) --
    (190.0000,1002.3622);
  \path[draw=blue,dash pattern=on 3.20pt off 1.60pt on 0.80pt off 1.60pt,line
    join=round,line cap=round,miter limit=4.00]
    (300.0000,972.3622) -- (218.1930,946.8012) --
    (236.0258,934.4392)(239.4879,932.0391) --
    (260.3719,917.5620)(265.9872,913.6693) --
    (280.1977,903.8183)(285.6088,900.0672) -- (340.0000,862.3622);
  \path[draw=blue,dash pattern=on 3.20pt off 1.60pt on 0.80pt off 1.60pt,line
    join=round,line cap=round,miter limit=4.00]
    (140.0000,872.3622) -- (218.1930,946.8012) --
    (221.8078,926.9765)(223.0397,920.2201) -- (250.0000,772.3622);
  \path[fill=black] (193.41379,838.58203) node[above right] {\scalebox{0.65}{$#6$}};
  \path[fill=black] (214.26122,964.43262) node[above right] {\scalebox{0.65}{$#7$}};

\end{tikzpicture}}

\newcommand{\AgAhISAgh}[8]{
\begin{tikzpicture}[y=0.80pt, x=0.8pt,yscale=-1, inner sep=0pt, outer sep=0pt,scale=#8]
  \path[draw=black,line join=round,line cap=round]
    (354.3307,981.4960) -- (318.8976,698.0315) -- (425.1968,910.6299) -- cycle;
  \path[draw=black,line join=round,line cap=round]
    (286.1673,959.8742) -- (283.4646,981.4960) -- (177.1654,910.6299) --
    (265.6701,910.6299)(274.2564,910.6299) --
    (277.7920,910.6299)(284.6105,910.6299) --
    (287.1429,910.6299)(296.0714,910.6299) --
    (301.7831,910.6299)(309.1067,910.6299) --
    (320.9760,910.6299)(328.8046,910.6299) --
    (340.3571,910.6299)(349.6429,910.6299) -- (425.1968,910.6299) --
    (353.7750,946.3408)(346.7750,949.8408) --
    (344.7262,950.8652)(339.8775,953.2896) --
    (338.2613,954.0977)(333.0085,956.7241) --
    (318.2036,964.1265)(310.6322,967.9122) -- (283.4646,981.4960) --
    (354.3307,981.4960) -- (177.1653,910.6299) -- (318.8976,698.0315) --
    (287.4640,949.5005);
  \path[draw=red,dash pattern=on 2.40pt off 2.40pt,line join=round,line
    cap=round,miter limit=4.00] (318.8976,698.0315) --
    (277.8571,795.2193) -- (333.9221,839.1348)(340.5659,844.3388) --
    (425.1968,910.6299);
  \path[draw=red,dash pattern=on 2.40pt off 2.40pt,line join=round,line
    cap=round,miter limit=4.00] (354.3307,981.4960) --
    (317.3899,891.5145)(314.5423,884.5783) --
    (301.6986,853.2931)(298.5998,845.7450) -- (277.8571,795.2193) --
    (282.4813,948.8324)(282.7092,956.4017) -- (283.4646,981.4960);
  \path[draw=red,dash pattern=on 2.40pt off 2.40pt,line join=round,line
    cap=round,miter limit=4.00] (277.8571,795.2193) --
    (177.1653,910.6299);
  \path[draw=blue,dash pattern=on 3.20pt off 1.60pt on 0.80pt off 1.60pt,line
    join=round,line cap=round,miter limit=4.00]
    (277.8571,795.2193) -- (258.5714,842.3622) --
    (276.5668,849.7350)(282.6425,852.2243) --
    (296.7010,857.9842)(308.5476,862.8378) --
    (337.4516,874.6800)(345.5230,877.9869) -- (425.1969,910.6299);
  \path[draw=blue,dash pattern=on 3.20pt off 1.60pt on 0.80pt off 1.60pt,line
    join=round,line cap=round,miter limit=4.00]
    (354.3307,981.4960) -- (317.4918,927.9708)(313.4715,922.1295) --
    (297.0497,898.2693)(292.2418,891.2836) --
    (282.7515,877.4947)(277.5898,869.9950) -- (258.5714,842.3622) --
    (277.4410,947.8290)(278.6028,954.3227) -- (283.4646,981.4960);
  \path[draw=blue,dash pattern=on 3.20pt off 1.60pt on 0.80pt off 1.60pt,line
    join=round,line cap=round,miter limit=4.00]
    (258.5714,842.3622) -- (177.1653,910.6299);
  \path[fill=black] (369.92761,980.34296) node[above right] {\scalebox{0.65}{$#1$}};
  \path[fill=black] (417.69296,930.09631) node[above right] {\scalebox{0.65}{$#2$}};
  \path[fill=black] (330.34152,715.23157) node[above right] {\scalebox{0.65}{$#3$}};
  \path[fill=black] (166.05582,899.51733) node[above right] {\scalebox{0.65}{$#4$}};
  \path[fill=black] (252,979.51733) node[above right]  {\scalebox{0.65}{$#5$}};
  \path[fill=black] (283.85715,797.21936) node[above right] {\scalebox{0.65}{$#6$}};
  \path[fill=black] (239.62723,872.37445) node[above right] {\scalebox{0.65}{$#7$}};
\end{tikzpicture}
}

\newcommand{\torusBeforeS}[9]{
\begin{tikzpicture}[y=0.80pt, x=0.8pt,yscale=-1, inner sep=0pt, outer sep=0pt,scale=#9]
  \path[draw=black,line join=round,line cap=round,miter limit=4.00,line
    width=0.400pt] (65.0000,1022.3622) -- (65.0000,992.3621);
  \path[draw=black,line join=round,line cap=round,miter limit=4.00,line
    width=0.400pt] (35.0000,1022.3622) -- (35.0000,992.3621);
  \path[draw=black,line join=round,line cap=round,miter limit=4.00,line
    width=0.400pt] (65.0000,992.3621) -- (35.0000,992.3621);
  \path[draw=black,line join=round,line cap=round,miter limit=4.00,line
    width=0.400pt] (35.0000,992.3621) -- (65.0000,1022.3622);
  \path[draw=black,line join=round,line cap=round,miter limit=4.00,line
    width=0.400pt] (35.0000,1022.3622) -- (65.0000,1022.3622);
  \path[draw=black,line join=round,line cap=round,miter limit=4.00,line
    width=0.200pt] (60.0000,1012.3623) -- (60.0000,996.4693)(60.0000,994.3265) --
    (60.0000,993.4336)(60.0000,989.9515) -- (60.0000,982.3622);
  \path[draw=black,line join=round,line cap=round,miter limit=4.00,line
    width=0.400pt] (30.0000,1012.3623) -- (30.0000,982.3622);
  \path[draw=black,line join=round,line cap=round,miter limit=4.00,line
    width=0.400pt] (60.0000,982.3622) -- (30.0000,982.3622);
  \path[draw=black,line join=round,line cap=round,miter limit=4.00,line
    width=0.200pt] (30.0000,982.3622) -- (38.8393,991.2015)(41.2500,993.6122) --
    (48.3928,1000.7551)(49.8214,1002.1837) -- (60.0000,1012.3623);
  \path[draw=black,line join=round,line cap=round,miter limit=4.00,line
    width=0.200pt] (30.0000,1012.3623) -- (33.7500,1012.3623)(36.4286,1012.3623)
    -- (53.4821,1012.3623)(56.4286,1012.3623) -- (60.0000,1012.3623);
  \path[draw=black,line join=round,line cap=round,miter limit=4.00,line
    width=0.400pt] (30.0000,1012.3622) -- (35.0000,1022.3622);
  \path[draw=black,line join=round,line cap=round,miter limit=4.00,line
    width=0.400pt] (60.0000,982.3622) -- (65.0000,992.3622);
  \path[draw=black,line join=round,line cap=round,miter limit=4.00,line
    width=0.400pt] (30.0000,982.3622) -- (35.0000,992.3622);
  \path[draw=black,line join=round,line cap=round,miter limit=4.00,line
    width=0.200pt] (60.0000,1012.3622) -- (65.0000,1022.3622);
  \path[draw=black,line join=round,line cap=round,miter limit=4.00,line
    width=0.400pt] (30.0000,982.3622) -- (65.0000,992.3622);
  \path[draw=black,line join=round,line cap=round,miter limit=4.00,line
    width=0.200pt] (30.0000,1012.3622) -- (33.6674,1013.4100)(36.3797,1014.1850)
    -- (65.0000,1022.3622);
  \path[draw=black,line join=round,line cap=round,miter limit=4.00,line
    width=0.400pt] (35.0000,992.3622) -- (30.0000,1012.3622);
  \path[draw=black,line join=miter,line cap=butt,miter limit=4.00,line
    width=0.200pt] (65.0000,992.3622) -- (60.0000,1012.3622);
  \path[draw=black,line join=round,line cap=round,miter limit=4.00,line
    width=0.200pt] (30.0000,1012.3622) -- (33.5385,1010.3402)(36.3846,1008.7138)
    -- (44.3750,1004.1479)(47.2596,1002.4995) -- (65.0000,992.3622);
  \path[fill=black] (60.641499,1013.0139) node[above right] {\scalebox{0.6}{$#1$}};
  \path[fill=black] (62.641499,985.01392) node[above right] {\scalebox{0.6}{$#2$}};
  \path[fill=black] (26.641499,1013.0139) node[above right] {\scalebox{0.6}{$#3$}};
  \path[fill=black] (26.641499,985.01385) node[above right] {\scalebox{0.6}{$#4$}}; 
  \path[fill=black] (65.722595,1022.3622) node[above right] {\scalebox{0.6}{$#5$}};
  \path[fill=black] (65.742088,994.82477) node[above right] {\scalebox{0.6}{$#6$}};
  \path[fill=black] (29,1022.3622) node[above right] {\scalebox{0.6}{$#7$}};
  \path[fill=black] (31.116116,994.6991) node[above right] {\scalebox{0.6}{$#8$}};
\end{tikzpicture}}

\newcommand{\torusAfterS}[9]{
\begin{tikzpicture}[y=0.80pt, x=0.8pt,yscale=-1, inner sep=0pt, outer sep=0pt,scale=#9]
  \path[draw=black,line join=round,line cap=round,miter limit=4.00,line
    width=0.400pt] (65.0000,1022.3622) -- (65.0000,992.3621);
  \path[draw=black,line join=round,line cap=round,miter limit=4.00,line
    width=0.400pt] (35.0000,1022.3622) -- (35.0000,992.3621);
  \path[draw=black,line join=round,line cap=round,miter limit=4.00,line
    width=0.400pt] (65.0000,992.3621) -- (35.0000,992.3621);
  \path[draw=black,line join=round,line cap=round,miter limit=4.00,line
    width=0.400pt] (35.0000,1022.3622) -- (65.0000,1022.3622);
  \path[draw=black,line join=round,line cap=round,miter limit=4.00,line
    width=0.200pt] (60.0000,1012.3623) -- (60.0000,998.7908)(60.0000,996.0229) --
    (60.0000,994.3265) -- (60.0000,994.3265) --
    (60.0000,993.4336)(60.0000,990.9336) -- (60.0000,982.3622);
  \path[draw=black,line join=round,line cap=round,miter limit=4.00,line
    width=0.400pt] (30.0000,1012.3623) -- (30.0000,982.3622);
  \path[draw=black,line join=round,line cap=round,miter limit=4.00,line
    width=0.400pt] (60.0000,982.3622) -- (30.0000,982.3622);
  \path[draw=black,line join=round,line cap=round,miter limit=4.00,line
    width=0.200pt] (30.0000,1012.3623) -- (33.7500,1012.3623)(36.4286,1012.3623)
    -- (40.1786,1012.3623)(42.1429,1012.3623) --
    (43.4821,1012.3623)(46.2500,1012.3623) -- (53.4821,1012.3623) --
    (56.4286,1012.3623) -- (60.0000,1012.3623);
  \path[draw=black,line join=round,line cap=round,miter limit=4.00,line
    width=0.400pt] (30.0000,1012.3622) -- (35.0000,1022.3622);
  \path[draw=black,line join=round,line cap=round,miter limit=4.00,line
    width=0.400pt] (60.0000,982.3622) -- (65.0000,992.3622);
  \path[draw=black,line join=round,line cap=round,miter limit=4.00,line
    width=0.400pt] (30.0000,982.3622) -- (35.0000,992.3622);
  \path[draw=black,line join=round,line cap=round,miter limit=4.00,line
    width=0.200pt] (60.0000,1012.3622) -- (65.0000,1022.3622);
  \path[draw=black,line join=round,line cap=round,miter limit=4.00,line
    width=0.400pt] (35.0000,992.3622) -- (30.0000,1012.3622);
  \path[draw=black,line join=miter,line cap=butt,miter limit=4.00,line
    width=0.200pt] (65.0000,992.3622) -- (60.0000,1012.3622);
  \path[draw=black,line join=round,line cap=round,miter limit=4.00,line
    width=0.400pt] (35.0000,1022.3622) -- (65.0000,992.3622);
  \path[draw=black,line join=round,line cap=round,miter limit=4.00,line
    width=0.200pt] (30.0000,1012.3623) -- (34.0179,1008.3444)(35.8929,1006.4694)
    -- (48.8839,993.4783)(50.9375,991.4247) -- (60.0000,982.3622);
  \path[draw=black,line join=round,line cap=round,miter limit=4.00,line
    width=0.400pt] (35.0000,992.3622) -- (60.0000,982.3622);
  \path[draw=black,line join=round,line cap=round,miter limit=4.00,line
    width=0.200pt] (35.0000,1022.3622) -- (60.0000,1012.3623);
  \path[draw=black,line join=round,line cap=round,miter limit=4.00,line
    width=0.200pt] (60.0000,982.3622) -- (54.3670,991.3750)(53.0929,993.4135) --
    (35.0000,1022.3622);
  \path[fill=black] (60.641499,1013.0139) node[above right] {\scalebox{0.6}{$#1$}};
  \path[fill=black] (62.641499,985.01392) node[above right] {\scalebox{0.6}{$#2$}};
  \path[fill=black] (26.641499,1013.0139) node[above right] {\scalebox{0.6}{$#3$}};
  \path[fill=black] (26.641499,985.01385) node[above right] {\scalebox{0.6}{$#4$}}; 
  \path[fill=black] (65.722595,1022.3622) node[above right] {\scalebox{0.6}{$#5$}};
  \path[fill=black] (65.742088,994.82477) node[above right] {\scalebox{0.6}{$#6$}};
  \path[fill=black] (29,1022.3622) node[above right] {\scalebox{0.6}{$#7$}};
  \path[fill=black] (31.116116,994.6991) node[above right] {\scalebox{0.6}{$#8$}};
\end{tikzpicture}}

\newcommand{\torusAfterT}[9]{
\begin{tikzpicture}[y=0.80pt, x=0.8pt,yscale=-1, inner sep=0pt, outer sep=0pt,scale=#9]
  \path[draw=black,line join=round,line cap=round,miter limit=4.00,line
    width=0.400pt] (64.1296,1022.1122) -- (64.1296,992.1121);
  \path[draw=black,line join=round,line cap=round,miter limit=4.00,line
    width=0.400pt] (34.1296,1022.1122) -- (34.1296,992.1121);
  \path[draw=black,line join=round,line cap=round,miter limit=4.00,line
    width=0.400pt] (64.1296,992.1121) -- (34.1296,992.1121);
  \path[draw=black,line join=round,line cap=round,miter limit=4.00,line
    width=0.400pt] (44.1296,982.1122) -- (34.1296,992.1121);
  \path[draw=black,line join=round,line cap=round,miter limit=4.00,line
    width=0.400pt] (74.1296,982.1122) -- (64.1296,992.1121);
  \path[draw=black,line join=round,line cap=round,miter limit=4.00,line
    width=0.400pt] (44.1296,982.1122) -- (74.1296,982.1122);
  \path[draw=black,line join=round,line cap=round,miter limit=4.00,line
    width=0.400pt] (74.1296,1012.1122) -- (74.1296,982.1122);
  \path[draw=black,line join=round,line cap=round,miter limit=4.00,line
    width=0.400pt] (64.1296,1022.1122) -- (74.1296,1012.1122);
  \path[draw=black,line join=round,line cap=round,miter limit=4.00,line
    width=0.200pt] (44.1296,982.1122) -- (44.1296,991.0746)(44.1296,993.3475) --
    (44.1296,996.1122)(44.1296,998.2550) -- (44.1296,1000.4185)(44.1296,1003.9541)
    -- (44.1296,1012.1122);
  \path[draw=black,line join=round,line cap=round,miter limit=4.00,line
    width=0.200pt] (74.1296,1012.1122) -- (65.4196,1012.1122)(62.8942,1012.1122)
    -- (55.3181,1012.1122)(52.5401,1012.1122) -- (44.1296,1012.1122);
  \path[draw=black,line join=round,line cap=round,miter limit=4.00,line
    width=0.200pt] (34.1296,1022.1122) -- (44.1296,1012.1122);
  \path[draw=black,line join=round,line cap=round,miter limit=4.00,line
    width=0.400pt] (34.1296,992.1121) -- (64.1296,1022.1122);
  \path[draw=black,line join=round,line cap=round,miter limit=4.00,line
    width=0.400pt] (64.1296,992.1121) -- (74.1296,1012.1122);
  \path[draw=black,line join=round,line cap=round,miter limit=4.00,line
    width=0.200pt] (34.1296,992.1121) -- (44.1296,1012.1122);
  \path[draw=black,line join=round,line cap=round,miter limit=4.00,line
    width=0.200pt] (44.1296,982.1122) -- (49.3664,987.3489)(50.9736,988.9561) --
    (53.3212,991.3037)(60.0766,998.0591) -- (63.1071,1001.0896)(65.1905,1003.1730)
    -- (74.1296,1012.1122)(58.3088,996.2914) -- (55.0890,993.0715) --
    (63.1071,1001.0896);
  \path[draw=black,line join=round,line cap=round,miter limit=4.00,line
    width=0.400pt] (34.1296,1022.1122) -- (64.1296,1022.1122);
  \path[draw=black,line join=round,line cap=round,miter limit=4.00,line
    width=0.200pt] (34.1296,992.1121) -- (62.6585,1006.3766)(65.1942,1007.6445) --
    (74.1296,1012.1122);
  \path[draw=black,line join=round,line cap=round,miter limit=4.00,line
    width=0.200pt] (34.1296,1022.1122) -- (56.4661,1016.5281)(59.4072,1015.7928)
    -- (62.7266,1014.9630)(65.0585,1014.3800) -- (74.1296,1012.1122);
  \path[draw=black,line join=round,line cap=round,miter limit=4.00,line
    width=0.400pt] (34.1296,992.1121) -- (74.1296,982.1122);
  \path[fill=black] (29.53125,1021.857) node[above right] {\scalebox{0.5}{$#1$}};
  \path[fill=black] (67.326935,1022.1368) node[above right] {\scalebox{0.5}{$#2$}};
  \path[fill=black] (75.371979,1012.1121) node[above right] {\scalebox{0.5}{$#3$}};  
  \path[fill=black] (39.129608,1014.1121) node[above right] {\scalebox{0.5}{$#4$}};
  \path[fill=black] (30.174454,992.84271) node[above right] {\scalebox{0.5}{$#5$}};
  \path[fill=black] (66.245728,992.97577) node[above right] {\scalebox{0.5}{$#6$}};
  \path[fill=black] (75.03595,984.97638) node[above right] {\scalebox{0.5}{$#7$}};  
  \path[fill=black] (38.771088,984.84949) node[above right] {\scalebox{0.5}{$#8$}};
\end{tikzpicture}

}

\usepackage{varioref}

\newcommand{\tII}{\text{II}}
\newcommand{\tIII}{\text{III}}
\newcommand{\tIV}{\text{IV}}
\newcommand{\ti}{\mathrm{i}}
\newcommand{\cH}{H}
\usepackage[colorlinks,citecolor=blue,linkcolor=blue,urlcolor=blue]{hyperref}
\newcommand{\blue}[1]{\textcolor{black}{#1}}

\begin{document}

\title{Twisted Gauge Theory Model of Topological Phases in Three Dimensions}

\author{Yidun Wan}
\email{ywan@perimeterinstitute.ca}
\affiliation{Perimeter Institute for Theoretical Physics, Waterloo, ON N2L 2Y5, Canada}
\author{Juven C. Wang}
\email{juven@mit.edu} 
\affiliation{Department of Physics, Massachusetts Institute of Technology, Cambridge, MA 02139, USA}
\affiliation{Perimeter Institute for Theoretical Physics, Waterloo, ON N2L 2Y5, Canada}
\author{Huan He}
\email{huanh@princeton.edu}
\affiliation{Department of Physics, Princeton University, Princeton, NJ 08544, USA}
\affiliation{Perimeter Institute for Theoretical Physics, Waterloo, ON N2L 2Y5, Canada}

\begin{abstract}
We propose an exactly solvable lattice Hamiltonian model of topological phases in $3+1$ dimensions, based on a generic finite  group $G$  and a $4$-cocycle $\omega$ over $G$. We show that our model has topologically protected degenerate ground states and obtain the formula of its ground state degeneracy on the $3$-torus. In particular, the ground state spectrum implies the existence of purely three-dimensional looplike quasi-excitations specified by two nontrivial flux indices and one charge index. We also construct other nontrivial topological observables of the model, namely the $SL(3,\mathbb{Z})$ generators as the modular $S$ and $T$ matrices of the ground states, which yield a set of topological quantum numbers classified by $\omega$ and quantities derived from $\omega$. Our model fulfills a Hamiltonian extension of the $3+1$-dimensional Dijkgraaf-Witten topological gauge theory with a gauge group $G$. This work is presented to be accessible for a wide range of physicists and mathematicians.
\end{abstract}
\maketitle
\tableofcontents
\makeatletter
\let\toc@pre\relax
\let\toc@post\relax
\makeatother

\section{Introduction}\label{sec:intro}
Because of their potential applications---in particular to topological quantum computation---phases of matter with intrinsic topological order\cite{Wen1989,Wen1989a,Wen1990a,Wen1990c,Kitaev2003a,Levin2004,Kitaev2006,Chen2012a,Levin2012,Hung2012,Hu2012,Hu2012a,Mesaros2011,Lin2014,Kong2014} that are realizable in two-dimensions have received substantial attention. Celebrated candidates of two-dimensional topological phases include chiral spin liquids\cite{Kalmeyer1987,Wen1989a}, $\Z_2$ spin liquids\cite{Read1991,Wen1991,Moessner2001}, Abelian quantum Hall states\cite{Klitzing1980,Tsui1982,Laughlin1983}, and non-Abelian fractional quantum Hall states\cite{TaoWu1984,Moore1991,Wen1991b,Willett1987,Radu2008}.

Symmetry plays a central role in two-dimensional topological phases: A large class of two-dimensional topological phases have an underlying $2+1$-dimensional effective gauge theory description. On top of the gauge symmetry, the degenerate ground states and hence the quasi-excitations---the anyons---respect a much larger hidden symmetry, usually described by a  quantum group or modular tensor category based on the gauge group\cite{Bais2002,Rowell2009}.  That is, the anyons carry representations of the quantum group. This is evident, for example, in the Kitaev toric code model\cite{Kitaev2006} and the twisted quantum double (TQD) model\cite{Propitius1995,Hu2012a,Mesaros2011}, where the induced quantum group symmetry is the (twisted) quantum double of the finite gauge group of the underlying Dijkgraaf-Witten (DW) topological gauge theory\cite{Dijkgraaf1990}. It is then natural to ask if the same physics applies to three dimensions---the physical spatial dimension. Can one build a model of three-dimensional topological phases based on a $3+1$-dimensional topological gauge theory? Would there still be topologically protected ground states that respect a larger hidden symmetry based on the gauge symmetry? What mathematical structure describes the larger symmetry? Would there be new types of quasi-excitations, and would they remain in one-to-one relationship with the ground states? What quantum numbers would characterize the ground states and the excitations? How would such excitations connect to the two-dimensional ones? These are some crucial questions to deepen our understanding of topological phases in all dimensions and the role of symmetry.

As an attempt to answer some of the questions above, in this paper, we propose an exactly solvable Hamiltonian extension of the $3+1$-dimensional DW gauge theory with general finite gauge groups on a lattice. We shall name our model the twisted gauge theory (TGT) model, in the sense that the usual gauge transformations in the underlying gauge theory with a gauge group $G$ is twisted by a $U(1)$ \fc $\omega$ in the fourth cohomology group of $G$, $H^4[G,U(1)]$. We rigorously derive the ground state degeneracy (GSD) on the $3$-torus, the topological quantum numbers of the ground states, e.g., their topological spins as the modular $T$ matrix, and their modular $S$ matrix, which are expected to be shared by the quasi-excitations corresponding to the ground states. This model also naturally extends the TQD model based on the DW gauge theory in $2+1$ dimensions twisted by a $3$-cocycle over the gauge group. Indeed, the $3+1$-dimensional TGT model, when dimensionally-reduced to two dimensions, reproduces the $2+1$-dimensional TQD model\cite{Wang2014c}.

The meaning of Hamiltonian extension can be understood in this way: As a topological field theory, the DW theory does not have a non-vanishing Hamiltonian from Legendre transform; When the DW theory is placed on a $3+1$-dimensional spacetime with boundaries, the boundary terms ensuring the gauge invariance of the theory induce gauge invariant boundary degrees of freedom; which are in one-to-one correspondence with the ground states of the corresponding TGT model. Consequently, the partition function of the DW theory coincides with the GSD of the corresponding TGT model.

We note that recently there are also studies\cite{Walker2011,VonKeyserlingk2013,Wang2014b,Jiang2014a,Moradi2014,Kong2014,Wang2014c,Jian2014,Bi2014} of certain aspects of three-dimensional topological phases. But these studies either are based only on untwisted gauge theories, or restrict to Abelian gauge groups, or focus merely on the braiding or fusion properties of quasi-excitation but do not examine the detailed properties of the lattice Hamiltonian that yields the excitations being studied. In this work, however, we  tackle this challenge.      

We derive our results in this order: Section \ref{sec:model} proposes our TGT model. Section \ref{sec:topoOb} lays down the general setting for the topological observables. Section \ref{sec:GSD} deduces our first topological observable, the ground state degeneracy (GSD) on a $3$-torus and the topological degrees of freedom. Section \ref{sec:fractionaltopologicalnumbers} fabricates two more topological observables, the modular $\mathcal{S}$ and $\mathcal{T}$ matrices on a $3$-torus. Section \ref{sec:classification} derives the explicit formulae of the topological (fractional) quantum numbers associated with the topological observables and classifies them generally. Section \ref{sec:examples} exemplify our model concretely with a number of finite groups and different types of the $4$-cocycles of these groups. Section \ref{sec:DW} explains why out model is a Hamiltonian extension of the DW theory in $3+1$ dimensions. Section \ref{sec:disc} concludes and outlines some future directions. The Appendices collect the definitions, derivations, and proofs that are too detailed to appear in the main text.
\section{Model construction}\label{sec:model}
We establish the generic Hamiltonian of our model on graphs consisting of tetrahedra embedded in $3$-spatial dimensions. A Hilbert space of our model is comprised of all possible assignments of a group element of a finite group to each edge of the graph. Our model is exactly solvable on such Hilbert spaces.

\subsection{The defining data}\label{subsec:basic}

\begin{figure}[!ht]\centering
  \includegraphics[scale=0.8]{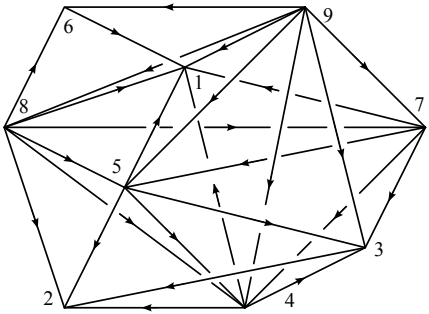}
  \caption{A crop of a graph that represents the basis vectors in the Hilbert space. Each edge $ab$, with $a<b$, is oriented from the larger enumeration to the smaller and is assigned a group element, $ab\in G$.}
  \label{fig:GraphConfiguration}
\end{figure}

We define our model on a graph embedded in a closed, orientable, $3$-dimensional manifold, e.g., a 3-sphere or a $3$-torus. Such a graph $\Gamma$  consists of solely tetrahedra (e.g., Fig. \ref{fig:GraphConfiguration}) and is free of any open edges. It may be taken as the simplicial triangulation of the closed $3$-manifold it is embedded in.  In this regard, a tetrahedron is also known as a $3$-simplex. We assign ordered labels to the vertices of $\Gamma$ and call such labels enumerations. The model is independent of the vertex enumerations as long as we keep their relative order consistent in the calculation.

We denote our model by $H_{G,\omega}$, as indicating its characterizing data---a triple $(H,G,\omega)$. The element $H$ in the triple is the Hamiltonian to be defined later. And $G$ is a finite group, Abelian or non-Abelian. The third element $\omega$ is a \emph{normalized} 4-cocycle to be explained below. We assign to each edge of $\Gamma$, on which $H_{G,\omega}$ is defined, a group element of $G$ and  orient it from the vertex with a larger enumeration to the one smaller (see Fig. \ref{fig:GraphConfiguration}). All possible configurations of the group elements on the edges of $\Gamma$ comprise the Hilbert space of the model on $\Gamma$:
\begin{equation}
  \label{IndexByLink}
  \Hil_\Gamma=\mathrm{span}\left\{\left\{g_1,g_2,...,g_E\right\}|g_e\in G\right\},
\end{equation}
where $E$ counts the number
of edges in $\Gamma$. A generic tetrahedron $abcd$ with ordered vertices $a<b<c<d$ has the following natural orientation by handedness. 
\begin{convention}\label{conv:TetOrientation}
One can grab the triangle of $abcd$ that does not contain the largest vertex, i.e., the triangle $abc$, along the boundary of the triangle, such that the three vertices are in ascending order, while the thumb points to the rest vertex $d$ of the tetrahedron. If one must use one's right hand to achieve this, $abcd$'s orientation is $+$, or $\epsilon(abcd)=1$, otherwise $-$, or $\epsilon(abcd)=-1$.    
\end{convention}

Conveniently, we denote by $ab$ for both an edge from $b$ to $a$ with $a<b$ and the group element on the edge. We let $\overline{ab}$ be the inverse element of $ab$, and $ba=\overline{ab}$ is understood. The $\Hil_\Gamma$ has a natural inner product:
\begin{align}
  \label{eq:innerProduct}
  \Blangle
  \tetrahedron{a'}{b'}{c'}{d'}{}{}{}{0.4}
  \Bvert
  \tetrahedron{a}{b}{c}{d}{}{}{}{0.4}
  \Brangle
  =& \delta_{ab,a'b'}\delta_{bc,b'c'}\delta_{ac,a'c'}\nonumber\\
  &\dots\; ,
\end{align}
where the \textquotedblleft$\dots$" neglects the $\delta$--functions on all the rest triangles that are not depicted but should be understood likewise. Generically, on the three sides of any triangle, e.g., the $ab,bc$ and $ac$ on the LHS of Eq. (\ref{eq:innerProduct}), the three corresponding group elements are independent of each other, i.e., $ab\cdot bc\neq ac$. Our notations and convention make it unnecessary to draw the group elements explicitly in a basis graph.
\begin{figure}[h!]
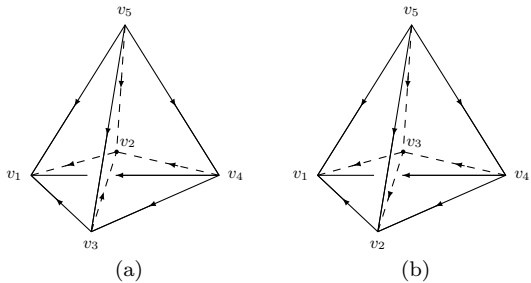

\centering
\subfigure[]{\fourTet{v_1}{v_3}{v_5}{v_4}{2}{v_2}{1}{1} \label{fig:4cocycleA}}
\subfigure[]{\fourTet{v_1}{v_2}{v_5}{v_4}{3}{v_3}{1}{1} \label{fig:4cocycleB}}
\caption{(a) The defining graph of the \fc $[v_1v_2, v_2v_3,v_3v_4,v_4v_5]$. (b) For $[v_1v_2, v_2v_3,v_3v_4,v_4v_5]^{-1}$.}
\label{fig:4cocycle}
\end{figure}

The aforementioned third ingredient, a \textit{normalized \fc} $\omega\in H^4[G,U(1))]$, is a function
$\omega:G^4\rightarrow U(1)$ that satisfies the \textit{4-cocycle
condition}
\be\label{4CocycleCondition}
  \frac{[g_1,g_2,g_3,g_4][g_0,g_1\cdot g_2,g_3,g_4][g_0,g_1,g_3,g_3\cdot g_4]}{[g_0\cdot g_1,g_2,g_3,g_4][g_0,g_1,g_2\cdot g_3,g_4][g_0,g_1,g_2,g_3]}
=1
\ee
for all $g_i \in G$, where we simplify the notation by 
\[\omega(g_1,g_2,g_3,g_4):=[g_1,g_2,g_3,g_4],\]
and satisfies the \textit{normalization condition}
\be
  \label{NormalizationCondition}
  [1,g,h,k]=[g,1,h,k]=[g,h,1,k]=[g,h,k,1]=1,
\ee
for any $g,h,k\in G$. A basic introduction to cohomology groups $H^n[G,U(1)]$ of finite groups is organized in Appendix \ref{app:HnGU1}. We stress that this normalization condition is not an \textit{ad hoc} extra condition imposed on the \fcs; instead, any group \fc naturally satisfies this condition. The reason is, any \fc $\omega$ is an equivalence class of the \fcs that differ by merely a $3$--coboundary $\delta\alpha$, where $\alpha$ is a $3$--cochain, a function from $G^3$ to $U(1)$, and $\delta$ is the coboundary operator defined in Appendix \ref{app:HnGU1}. Any \fc equivalence class can be shown to possess a representative always satisfying the normalization condition (\ref{NormalizationCondition}).

Notice that every group has a trivial 4-cocycle $\omega_0\equiv 1$ for the whole $G$. A \fc can be defined on any subgraph---a $3$-complex---composed of four tetrahedra, which share a vertex and any two of which share a triangle. Consider Fig. \ref{fig:4cocycleA} for instance: There are four tetrahedra $v_1v_2v_3v_5$, $v_1v_2v_4v_5$, $v_1v_2v_3v_4$, and $v_2v_3v_4v_5$, and five vertices all are arranged in the order $v_1<v_2<v_3<v_4<v_5$; we define the \fc for this subgraph by taking the four variables from left to right to be the four group elements, $v_1v_2$, $v_2v_3$, $v_3v_4$, and $v_4v_5$, which are along the path from the least vertex $v_1$ to the greatest vertex $v_5$ passing $v_2$, $v_3$, and $v_4$ in order; thus, the \fc reads $[v_1v_2,v_2v_3,v_3v_4,v_4v_5]$.
Note that the apparent tetrahedron $v_1v_3v_4v_5$ in Fig. \ref{fig:4cocycleA} does not exist because the figure is purely $3$-dimensional. Had one insisted upon the existence of $v_1v_3v_4v_5$, one would have to place it in a fourth dimension, such that the complex in Fig \ref{fig:4cocycleA} comprises the boundary of a $4$-simplex $v_1v_2v_3v_4v_5$. In this sense, we can also associate the \fc $[v_1v_2,v_2v_3,v_3v_4,v_4v_5]$  with a $4$-simplex. We will come back to this viewpoint later.

Such a $3$-complex is the simplest triangulation of a closed, oriented $3$-manifold. Physically, a  \fc\ defined on such a $3$-complex can be regarded as a probabilistic weight, or just a wave-function, assigned to the state of the system on the $3$-complex. This is a physical reason why the \fcs considered here are $U(1)$ numbers. Same $4$-cocycles also appear in the $3+1$-dimensional DW theory, as the fundamental building blocks of the partition function of the theory.
We shall get to this point in Section \ref{sec:DW}.

As opposed to Fig. \ref{fig:4cocycleA}, if we consider Fig. \ref{fig:4cocycleB}, which differs from Fig. \ref{fig:4cocycleA} by only the positions of the vertices $v_2$ and $v_3$,  the corresponding \fc should be an inverse one, $[v_1v_2,v_2v_3,v_3v_4,v_4v_5]^{-1}$.

One hence notices that a complex $v_1v_2v_3v_4v_5$ like those in Fig \ref{fig:4cocycle} defines a \fc $\omega^{\epsilon(v_1v_2v_3v_4v_5)}$, where $\epsilon(v_1v_2v_3v_4v_5)=\pm 1$ is the orientation of the $3$-complex that is determined by the following convention.
\begin{convention}\label{conv:4cocycle4Tet}
One first chooses any of the four tetrahedra in the defining graph of the complex and determine its orientation using handedness as described earlier, e.g., $\epsilon(v_1v_2v_3v_4)=1$ from Fig. \ref{fig:4cocycleA} and $\epsilon(v_1v_2v_3v_5)=1$ from Fig. \ref{fig:4cocycleB}. One then append the remaining vertex to the beginning of the ordered list of four vertices of the chosen tetrahedron,  e.g., $(v_5,v_1,v_2,v_3,v_4)$ from Fig. \ref{fig:4cocycleA} and $(v_4,v_1,v_2,v_3,v_5)$ from Fig. \ref{fig:4cocycleB}. If the list can be turned into ascending order by even permutations, such as $(v_5,v_1,v_2,v_3,v_4)$ from Fig. \ref{fig:4cocycleA}, one has $\epsilon(v_1v_2v_3v_4v_5) =\epsilon(v_1v_2v_3v_4)=1$, and otherwise, $\epsilon(v_1v_2v_3v_4v_5)=-\epsilon(v_1v_2v_3v_5)=-1$ from Fig. \ref{fig:4cocycleB}.
The result is independent of the choice of the tetrahedron.
\end{convention}

The reader may notice some abuse of language in the sequel. For instance, ``a \fc" $\omega$ may stand for a class $[\omega]$, a representative $\omega$, or the evaluation of $\omega$ on a $3$-complex. Also, although Eq. \eqref{4CocycleCondition} is the only \fc condition in an abstract sense, from time to time, we may refer \fc conditions to the evaluations of the condition \eqref{4CocycleCondition} on different $4$-simplices or $3$-complexes. These however should not mislead contextually.
\subsection{The lattice Hamiltonian}\label{subsec:Hamiltonian}
The Hamiltonian of our model takes the form  
\be\label{eq:Hamiltonian}
  H=-\sum_v A_v-\sum_f B_f.
\ee
Here $A_v$ is the vertex operator defined on each vertex $v$, and $B_f$ is the face operator defined at each triangular face $f$. As to be seen, the $4$--cocycles introduced in the previous subsection will constitute the matrix elements of the operators $A_v$ and $B_f$. As in the TQD model in $(2+1)$-d, an operator $A_v$ facilitates a gauge transformation of the group element on each edge incident at $v$, and a $B_f$ imposes the zero flux condition on the face $f$. Thus, the ground states of such a Hamiltonian are gauge invariant and flux-free on all faces. We now elucidate these operators.

The operator $B_f$ acts  on a basis vector as
\be\label{eq:actionOfBf}
  B_f\BLvert \oneTriangle{v_1}{v_2}{v_3} \Brangle
  =\delta_{v_1v_2\cdot v_2v_3\cdot v_3v_1}
  \BLvert \oneTriangle{v_1}{v_2}{v_3}\Brangle 
\ee
and yield a phase factor, a delta function $\delta_{v_1v_2\cdot v_2v_3\cdot v_3v_1}$, which is unity if ${v_1v_2\cdot v_2v_3\cdot v_3v_1=1 }$, the identity element of $G$, and 0 otherwise. Again, the ordering of $v_1,v_2$, and $v_3$ is irrelevant because 
$\delta_{v_1v_2\cdot v_2v_3\cdot v_3v_1}
=\delta_{v_3v_1\cdot v_1v_2\cdot v_2v_3}$ and
$\delta_{v_1v_2\cdot v_2v_3\cdot v_3v_1}
=\delta_{\overline{v_1v_2\cdot v_2v_3\cdot v_3v_1}}
=\delta_{\overline{v_3v_1}\cdot \overline{v_2v_3}\cdot \overline{v_1v_2}}
=\delta_{v_1v_3\cdot v_3v_2\cdot v_2v_1}$.
Namely, on the three sides of any triangle $f$ with $B_f=1$, the three group elements obey the \textit{chain rule}:
\begin{equation}
\label{ChainRule}
v_1v_3=v_1v_2\cdot v_2v_3
\end{equation}
for any enumerations
$v_1,v_2,v_3$ of $f$'s vertices.

The operator $A_v$ is is more involved. It is an averaged sum over the group $G$,
\begin{equation}
  \label{eq:Av}
  A_v=\frac{1}{|G|}\sum_{[vv']=g\in G}A_v^g,
\end{equation} 
where $|G|$ is the order of $G$. The finer operator $A_v^g$ acts on a vertex $v$ by a group element $g\in G$; it replaces $v$ by another enumeration $v'$ such that $v'v=g$. In our convention, the new enumeration $v'$ must be ``slightly" less than $v$ but greater than all the enumerations that are less than $v$ in the original set of enumerations before $A^g_v$ acts. In a dynamical picture of Hamiltonian evolution, $A^g_v$ evolves $v$ from one ``time' to $v'$ at a later time, which results in an timelike edge $v'v\in G$ in the $3+1$-dimensional \textquotedblleft spacetime". Let us consider the following simplest subgraph---a single tetrahedron---of some large $\Gamma$ to illustrate how an $A_v$ acts on $\Gamma$: 
\be
A_{v_4}^g : \BLvert \fourTet{v_1}{v_2}{v_4}{v_3}{0}{}{0.5}{1} \Brangle \mapsto \BLvert\fourTet{v_1}{v_2}{v'_4}{v_3}{0}{}{0.5}{1}\Brangle,
\ee
where $v'_4v_4=g$. Now put together the two tetrahedra before and after the action in the above equation as two spatial slices and the edge $v'_4v_4$, which is not shown, as along the `time' (the fourth) dimension, we obtain a 4-simplex. This is a path integral picture, which motivates us to attribute the amplitude of $A^g_{v_4}$ to an evaluation of the 4-simplex, which is naturally given by the \fc associated with the 4-simplex (recall our earlier discussion). That is,
\be\label{eq:AvAmpSingleTet}
\begin{aligned}
&\Blangle\fourTet{v_1}{v_3}{v'_4}{v_3}{0}{}{0.5}{1}\BRvert A^g_{v_4}\BLvert  \fourTet{v_1}{v_2}{v_4}{v_3}{0}{}{0.5}{1} \Brangle\\
 =\ & \delta_{v'_4v_4,g}\Blangle\fourTet{v_1}{v_2}{v_4}{v_3}{4}{v'_4}{0.6}{1} \Brangle\\
=\ &\delta_{v'_4v_4,g}[v_1v_2,v_2v_3,v_3v'_4,v'_4v_4]^{\epsilon(v_1v_2v_3v'_4v_4)}\\
=\ &\delta_{v'_4v_4,g}[v_1v_2,v_2v_3,v_3v'_4,v'_4v_4]^{-1},
\end{aligned}
\ee   
where the big bracket in the second row means an evaluation of the 4-simplex in the bracket, which gives rise to the \fc in the third row. The orientation appearing is understood this way: 
\begin{convention}\label{conv:4cocyleAv1Tet}
Since the new vertex $v'_4$ is set to be slightly off the 3d slice made of the tetrahedron $v_1v_2v_3v_4$, and since every newly created vertex bears a label slightly less than that of the original vertex acted on by the vertex operator, one can always choose the convention such that $\epsilon(v_1v_2v_3v'_4v_4) =\epsilon(v_1v_2v_3v_4)\sgn{v'_4,v_1,v_2,v_3,v_4}$. And $\sgn{v'_4,v_1,v_2,v_3,v_4}$ is the sign of the permutation that takes the list of vertices in the argument to purely ascending as $(v_1,v_2,v_3,v'_4,v_4,v_5)$, which embraces the 4-simplex $v_1v_2v_3v'_4v_4$. 
\end{convention}

By Convention \ref{conv:TetOrientation} of tetrahedral orientation, we have $\epsilon(v_1v_2v_3v_4)=1$ and clearly $\sgn{v'_4,v_1,v_2,v_3,v_4}=-$, leading to the last equality in Eq. \eqref{eq:AvAmpSingleTet}. 

Note that the result above does not depend on how one actually projects the 4-simplex on the plane. For example, one may place the vertex $v'_4$ completely outside of the tetrahedron and should obtain the same orientation. It is however crucial that any singular projection, e.g., placing $v'_4$ right on a face or edge, is forbidden. Moreover, if the ordering of the vertices in the above example is changed, one may obtain a different orientation accordingly. 

There is a coordinate-based method of determining the orientation of a $4$-simplex or the $3$-complex bounding a $4$-simplex. Take the $3$-complex on the RHS of Eq. \eqref{eq:AvAmpSingleTet} for an example. One places the $3$-complex in an Euclidean frame. Then the sign of the determinant $\det[v_1v_2,v_2v_3,v_3v_4',v_4'v_4]$ is the orientation of the complex. Here we also use $v_iv_j$ for the vector from $v_i$ to $v_j$ in the Euclidean frame. In the action by the vertex operator $A_v^g$, however, the newly created vertex $v'$ is assumed to be slightly smaller the $v$. That is, $v'$ is just slightly off the $3$-dimensional manifold of the complex, with a infinitesimal coordinate in the $4$-th dimension. Back to the current example, the $\det[v_1v_2,v_2v_3,v_3v_4',v_4'v_4]$ is simply proportional to $\det[v_1v_2,v_2v_3,v_4]$. And the proportionality factor the permutation sign of $v_4'$ in the ordered list of the five vertices, if we take the assumption that the $4$-th coordinate of $v'_4$ is positive. This is precisely the meaning of our Convention \ref{conv:4cocyleAv1Tet}. The virtue of our method is that we no longer need to refer to concrete coordinates of the vertices to compute the orientation of a $3$-complex or $4$-simplex. In fact, our method naturally generalizes to higher dimensions. 

\begin{figure}[h!]
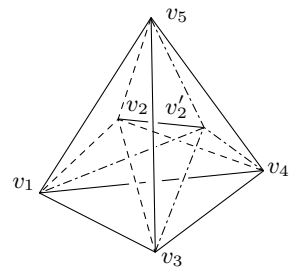

\centering
\AvQuard{v_5}{v_4}{v_3}{v_1}{v_2}{v_2'}
\caption{The topology of the action of $A_{v_2}^g$. The arrows are omitted for simplicity because the vertex ordering is explicit.}
\label{fig:AvFourTet}
\end{figure}

Because we consider closed graphs only, there are not any boundary vertices; hence, unlike demonstrated in the simplest example above, each $A_v$ actually acts on a vertex $v$ that is shared by more than one tetrahedra, and it should create more than one $4$-simplices, each of which contributes a \fc to the amplitude. Bearing the convention introduced via the simplest example above, we are now ready to show a typical example in Eq. (\ref{eq:AvgFourTet}). We assume, without losing generality, that the the five vertices are enumerated as $v_1<v_2<v_3<v_4<v_5$. The basis vector on the LHS of \eqref{eq:AvgFourTet} is specified by ten group elements,
$v_1v_2$, $v_1v_3$, $v_1v_4$, $v_1v_5$, $v_2v_3$, $v_2v_4$, $v_2v_5$, $v_3v_4$, $v_3v_5$, and $v_4v_5$.
On this basis vector, $A_{v_2}^g$ acts as
\begin{align}
  \label{eq:AvgFourTet}
  &A_{v_2}^g\BLvert \fourTet{v_1}{v_3}{v_5}{v_4}{2}{v_2}{0.6}{1} \Brangle
  \nonumber\\
  =&\frac{[v_1v'_2,v'_2v_2,v_2v_3,v_3v_5][v'_2v_2,v_2v_3,v_3v_4,v_4v_5]} {[v_1v'_2,v'_2v_2,v_2v_3,v_3v_4] [v_1v'_2,v'_2v_2,v_2v_4,v_4v_5]}
  \nonumber\\
  &\times\delta_{v'_2v_2,g}
  \BLvert \fourTet{v_1}{v_3}{v_5}{v_4}{2}{v_2'}{0.6}{1} \Brangle,
\end{align}
where the new enumerations are $v_1<v'_2<v_2< v_3<v_4<v_5$. The action of $A^g_{v_2}$ also imposes on the newly created triangles the following \textit{chain rules}.
\be
\begin{aligned}
\label{eq:ChainRuleInAv}
&[v_1{v_2}]=[v_1v_2']\cdot[v_2'{v_2}],\\
&[v'_2{v_3}]=[v'_2v_2]\cdot[v_2{v_3}],\\
&[{v'_2}v_4]=[{v'_2}v_2]\cdot[v_2v_4],\\
&[{v'_2}v_5]=[{v'_2}v_2]\cdot[v_2v_5].
\end{aligned}
\ee
The amplitude of this $A^g_{v_2}$ can be read off from the `spacetime' complex composed of tetrahedra before and after the action and the tetrahedra sharing the ``timelike'' edge $v_2'v_2$, as depicted in Fig \ref{fig:AvFourTet}. In this figure, there are four $4$-simplices, namely $v_1v'_2v_2v_3v_4$, $v_1v'_2v_2v_3v_5$, $v_1v'_2v_2v_4v_5$, and $v_2'v_2v_3v_4v_5$, whose orientations are respectively $-1$, $+1$, $+1$, and $-1$, as determined by Convention \ref{conv:4cocyleAv1Tet}. This explains the amplitude of the $A^g_{v_2}$ in Eq. \eqref{eq:AvgFourTet}. We emphasize again that Fig. \ref{fig:AvFourTet} is merely one possible projection of the $4$-complex on the plane but our result of the amplitude is independent of the projection.    

Fig. \ref{fig:AvFourTet} also has a precise topological meaning. Had one imagined the original $3$-complex $v_1v_2v_3v_4v_5$ as a $4$-simplex (although it is not one because $v_1v_3v_4v_5$ is not a tetrahedron is the original setting), then Fig. \ref{fig:AvFourTet} would be an illustration of a $4$-dimensional Pachner move\cite{Pachner1978,Pachner1987}, the $1\rightarrow 5$ move, which splits $v_1v_2v_3v_4v_5$ into five $4$-simplices, $v_1v'_2v_2v_3v_4$, $v_1v'_2v_2v_3v_5$, $v_1v'_2v_2v_4v_5$, $v_2'v_2v_3v_4v_5$, and $v_1v'_2v_3v_4v_5$ that do not exist either because again $v_1v_3v_4v_5$ is not a tetrahedron. The actual $1\rightarrow 5$ move should take place in five dimensions. We will come back to this point in the next section.  

If $v'_2v_2=g=1$, the unit element of $G$, is true in Eq. \eqref{eq:AvgFourTet}, then the amplitude becomes
\be\label{eq:Avg=1}
\frac{[v_1v'_2,1,v_2v_3,v_3v_5][1,v_2v_3,v_3v_4,v_4v_5]} {[v_1v'_2,1,v_2v_3,v_3v_4] [v_1v'_2,1,v_2v_4,v_4v_5]}\equiv 1
\ee
by the \textit{normalization condition} \eqref{NormalizationCondition}. That is, $A_v^{g=1}=\mathbb{I}$, identity operator. In this case, the $A_v^1$ simply imposes the usual gauge transformations in lattice gauge theories via the chain rules \eqref{eq:ChainRuleInAv}.

In Appendix \ref{app:algAvBf}, we show that all $B_f$ and $A_v$ are mutually-commuting projection operators. Consequently, the ground states of the Hamiltonian \eqref {eq:Hamiltonian} are common $+1$ eigenvectors of all these local projection operators. It is clear that the ground states of the Hamiltonian \eqref {eq:Hamiltonian} have the exactly the same energy and thus are \emph{degenerate}, typically when the Hamiltonian is defined in some non-trivial spatial topology, e.g., a $3$-torus. As we will show later, the degeneracy of these ground states is a topological invariant that partially characterizes our model. Moreover, the Hamiltonian \eqref {eq:Hamiltonian} also indicates that any excitation of the model has a finite gap of energy above the degenerate ground states; hence, each such Hamiltonian describes a \emph{gapped phase}. Such gapped phases exhibiting topologically protected degenerate ground states are phases with intrinsic topological order, or simply put, \emph{topological phases}. In two dimensions, a celebrated example would be the  $\Z_2$ spin liquid\cite{Kitaev2006}, whose $\mathrm{GSD}=4$ on a $2$-torus. Another significant $2$-dimensional example is the Fibonacci phase\cite{Nayak2008}, whose GSD is 2 on a $2$-torus. The latter is known to be the best candidate of realizing topological quantum computation in two-dimensions. Therefore, we are justified to claim that our model describes topological phases in three spatial dimensions.
It deserves future effort to look for the potential applications of the $3$-dimensional topological phases yielded by our model. 

Because $A_v$ imposes gauge transformations twisted by a $4$-cocycle, and because of another physical reason to be revealed in Section \ref{subsec:topoDof}, we shall tentatively name our model the \textbf{twisted gauge theory (TGT) model}.
\subsection{Equivalent models due to equivalent \fcs}\label{subsec:equivModel}
Since we now have a TGT model defined by a \fc,
and since a \fc les in an equivalence class of \fcs, one may ask whether two equivalent \fc\ in an equivalence class define two equivalent models. To answer this question, we begin with two TGT models $H_{G,\omega}$ and $H_{G,\omega'}$, respectively defined by two equivalent \fcs $\omega$ and $\omega'$. We assume that $\omega$ and $\omega'$ defer by merely a $4$--coboundary $\delta\alpha$ of $3$--cochain $\alpha:G^3\mapsto U(1)$ normalized by $\alpha(1,y,z)=\alpha(x,1,z) =\alpha(x,y,1)=1$ for all $x,y,z \in G$. 
\be\label{eq:equiv4cocycle}
\begin{aligned}
&\omega'(g_0,g_1,g_2,g_3)\\
=&\delta\alpha(g_0,g_1,g_2,g_3)\omega(g_0,g_1,g_2,g_3)\\
=&\frac{\alpha(g_1,g_2,g_3)\alpha(g_0,g_1g_2,g_3)}{\alpha(g_0g_1,g_2,g_3) \alpha(g_0,g_1,g_2g_3)\alpha(g_0,g_1,g_2)} \omega(g_0,g_1,g_2,g_3),
\end{aligned}
\ee
where $g_i\in G$. Since each \fc is defined on a $3$-complex made of four tetrahedra, such as those in Fig \ref{fig:4cocycle}, each $3$-cochain $\alpha(g_1,g_2,g_3):=[g_1,g_2,g_3]$ can be defined on a tetrahedron. Therefore, one can view Eq. (\ref{eq:equiv4cocycle}) as a local gauge transformation of the \fc $\omega$.

Because the amplitudes of the operators $B_f$ are just $\delta$-functions  immune to the transformation \eqref{eq:equiv4cocycle}, to see how $H_{G,\omega'}$ and $H_{G,\omega}$ are related, one needs only to study the operators $A_v^g(\omega')$ and $A_v^g(\omega)$. There is no loss of generality to revisit the vertex operators on the common vertex of four tetrahedra, similar to  that in Eq. (\ref{eq:AvgFourTet}). Equation (\ref{eq:equiv4cocycle}) makes the following derivation straightforward. 
\begin{align}
  \label{eq:AvEquiv4Cocycle}
  &A_2^g(\omega')\BLvert \fourTet{1}{3}{0}{4}{2}{2}{0.6}{0} \Brangle
  \nonumber\\
  =&\frac{[01,12',2'2,23]'[02',2'2,23,34]'} {[01,12',2'2,24]' [12',2'2,23,34]'}   \BLvert \fourTet{1}{3}{0}{4}{2}{2'}{0.6}{0} \Brangle\nonumber\\
  =&\frac{[01,12,23][12,23,34]}{[02,23,34][01,12,24]}\times \frac{[01,12',2'2,23][02',2'2,23,34]} {[01,12',2'2,24] [12',2'2,23,34]}
  \nonumber\\
  &\times
  \frac{[02',2'3,34][01,12',2'4]}{[01,12',2'3][12',2'3,34]}
  \BLvert \fourTet{1}{3}{0}{4}{2}{2'}{0.6}{0} \Brangle,
\end{align}
where for simplicity the $\delta$--function $\delta_{[2'2],g}$ is implicit.
In the second equality above, the second term containing four  $\omega$'s is exactly the amplitude of $A^g_2(\omega)$. Moving the first fraction of $\alpha$'s in the second equality above to the LHS indicates the action of $A^g_3(\omega')$ on the rescaled state
\[
  \frac{[02,23,34][01,12,24]}{[01,12,23][12,23,34]} 
  \BLvert \fourTet{1}{3}{0}{4}{2}{2}{0.6}{0} \Brangle,
\]
which agrees with the that of $A^g_3(\omega)$ on the original state. Being just a local $U(1)$ phase, such rescaling can factorize into the local $U(1)$ transformations on the basis tetrahedral states:
\be\label{eq:basisTransformEquiv3Cocycle}
\BLvert\fourTet{a}{b}{d}{c}{0}{}{0.5}{1}\Brangle\mapsto[ab,bc,cd]^{-\epsilon(abcd)} \BLvert \fourTet{a}{b}{d}{c}{0}{}{0.5}{1}\Brangle,
\ee
where $\epsilon(abcd)$ is the orientation of the tetrahedron $abcd$. The amplitude of the $A^g_v(\omega')$ in the new basis coincides with that of the $A^g_v(\omega)$ in the original basis; hence, the two Hamiltonians $H_{G,\omega}$ and $H_{G,\omega'}$ would have the same spectrum.

Any two \fcs related by $\omega'=\omega\delta\alpha$ can be continuously deformed into each other. To this end, let us define a 3--cochain $\alpha^{(t)}(x,y,z)=\alpha(x,y,z)^t$, with $0 \leq t \leq 1$ a continuous parameter. Then, for all $0 \leq t \leq 1$, $\omega^{(t)}=\omega\delta\alpha^{(t)}$ is equivalent to $\omega$,  with $\omega^{(0)}=\omega$ and $\omega^{(1)}=\omega'$. As a result, in Eq. (\ref{eq:basisTransformEquiv3Cocycle}), with $\alpha=[ab,bc,cd]$ replaced by $\alpha^{(t)}$, the local $U(1)$ transformation becomes continuous, such that \emph{the Hamiltonian remains gapped} for any $t$; hence, no phase transition occurs in the one--parameter passage with the Hamiltonian $H_{G,\omega^{(t)}}$ from $0\leq t \leq1$. Therefore, \emph{the gapped Hamiltonians $H_{G,\omega'}$ and $H_{G,\omega}$ defined by two equivalent \fcs $\omega'$ and $\omega$ do describe the same topological order.}

\section{From symmetries to Topological observables }\label{sec:topoOb}
To understand the observables and symmetries of topological phases, we start by drawing an analogy between Hydrodynamics and topological phases. The diffeomorphism group acting on the fluid offers a systematic way to examine the topological properties of hydrodynamical fluid, such as the stability and interactions of currents and fluxes\cite{ArnoldHydrobook1998}. On the other hand, the topological properties of a discrete model of topological phases can be systematically examined by the discrete diffeomorphisms. These properties include the GSD and the braiding and fusion of the quasi-excitations.

The discrete diffeomorphisms we shall consider are the mutations transformations of the graph. These transformations alter the local graph structure but preserve the topology of the $3$-manifold in which the graph is embedded. A \textit{topological observable} is then a Hermitian operator invariant under the these mutation transformations.

Although many condensed-matter and other physical systems may not possess the mutation (or diffeomorphism) symmetry, certain discrete models of topological phases, such as the Kitaev model\cite{Kitaev2003a,Kitaev2006}, the Levin-Wen model\cite{Levin2004}, and the Walker-Wang model\cite{Walker2011}, do respect this kind of symmetry, in the sense that their ground state Hilbert spaces are invariant under the mutation transformations pertinent to their constructions. Therefore, we can adopt any topological observable, e.g., the GSD, of these models to characterize them, at least partially.

That said, we now devise the mutation transformations in our model. They form a unitary symmetry of the ground states of our model. As such, the GSD to be derived  is indeed a topological observable of our model.

As proven in Appendix \ref{app:algAvBf}, $A_v$ and $B_f$ are projectors and commute with each other. Thus each ground state is a $+1$ eigenvector shared by all the operators $A_v$ and $B_f$. We begin with the ground state projector:
\begin{equation}
  \label{eq:GSDprojector}
  P^0_{\Gamma}=(\prod_{f\in\Gamma}B_f)(\prod_{v\in\Gamma} A_v),
\end{equation}
which projects an arbitrary state to a ground state. Thus, the Hilbert subspace of the ground states is
\begin{equation}
  \label{GroundStateSubspace}
  \mathcal{H}^0_{\Gamma}=\left\{|\Phi\rangle\Bigl|\Bigr. P_{\Gamma} |\Phi\rangle=|\Phi\rangle\right\}.
\end{equation}

Symmetry transformations in a lattice model are normally defined for a fixed lattice and thus should not change the lattice structure. On the contrary, the mutation moves in our model can send one graph to another. Accordingly, our Hamiltonian (\ref{eq:Hamiltonian}) and hence the the Hilbert space are seemingly uninvariant under  the mutation moves.
This is however not an issue, as we shall show later in this section, the physical content of the model, namely the ground states and the spectra of the topological observables are indeed invariant under the mutation transformations.

We first lay down the mutation moves on the graphs and then define the corresponding mutation transformations. The mutation moves at our disposal are those generated by the Pachner moves that relate any two triangulations $\Gamma$ and $\Gamma'$ of a $3$-manifold\cite{Pachner1978,Pachner1987}:
\begin{align}
& f_1:\; \bmm\twoTet{}{}{}{}{}{0.25}\emm\mapsto\bmm\threeTet{}{}{}{}{}{0.25}\emm,\;\\
& f_2:\; \bmm\threeTet{}{}{}{}{}{0.25}\emm\mapsto\bmm\twoTet{}{}{}{}{}{0.25}\emm,\;\\
& f_3:\; \bmm\fourTet{}{}{}{}{0}{}{0.5}{0}\emm\mapsto\bmm\fourTet{}{}{}{}{1}{}{0.5}{0}\emm,\\
& f_4:\; \bmm\fourTet{}{}{}{}{1}{}{0.5}{0}\emm\mapsto\bmm\fourTet{}{}{}{}{0}{}{0.5}{0}\emm.
\end{align}
Associated with each mutation move generator $f_i:\Gamma\rightarrow\Gamma'$ is a linear transformation, the mutation transformation $T_i:\Hil_{\Gamma}\rightarrow\Hil_{\Gamma'}$:
\begin{align}
  &T_1\BLvert \twoTet{v_0}{v_1}{v_2}{v_3}{v_4}{0.25}\Brangle\label{eq:T1move}\\
  =&\sum_{v_2v_4\in G}[v_0v_1,v_1v_2,v_2v_3,v_3v_4]^{\epsilon(v_4|v_0v_1v_2v_3)}
  \BLvert \threeTet{v_0}{v_1}{v_2}{v_3}{v_4}{0.25} \Brangle\nonumber\\
  =&\sum_{v_2v_4\in G}[v_0v_1,v_1v_2,v_2v_3,v_3v_4]^{-1} 
  \BLvert \threeTet{v_0}{v_1}{v_2}{v_3}{v_4}{0.25} \Brangle.\nonumber
\end{align}
Before moving on to the other three mutation transformations, let us explain the principles that determine the linear properties $T_1$ because they will determine that of $T_2$ through $T_4$ as well. 

In Eq. \eqref{eq:T1move} for $T_1$, and in the subsequent equations for the other three transformation, we assume, without losing generality, $v_0<v_1<v_2<v_3<v_4$. 
$T_1$ turns the original two tetrahedra on the LHS of Eq. \eqref{eq:T1move} to the three tetrahedra on the RHS of the equation. Topologically, this cannot be directly realized in $3$d but only in $4$d, in the sense that the five tetrahedra in total before and after the transformation form the boundary of a $4$-simplex that is usually also drawn on the plane as the picture on the RHS of Eq. \eqref{eq:T1move}. This motivates the \fc as the amplitude of $T_1$ in Eq. \eqref{eq:T1move}. The \fc comes with a sign exponent. To determine this sign, one can think of the original two tetrahedra and the three new tetrahedra are on different $3$d hypersurfaces in $4$d. And following our convention taken in defining the vertex operators, we always assume the new hypersurface is "lower" than the original one, such that the orientation of the $4$-simplex bounded by the two hypersurfaces share the orientation of the original hypersurface, namely the $3$-complex made of the original tetrahedra. In Eq. \eqref{eq:T1move}, to be precise, the original $3$-complex is made of the tetrahedra $v_0v_1v_2v_3$ and $v_0v_1v_3v_4$, and its orientation reconciles Convention \ref{conv:4cocyleAv1Tet}. That is, one can choose either of the two tetrahedra in this case and determine its orientation by Convention \ref{conv:TetOrientation}, then use the relative order of the fifth vertex with respect to the vertices of the chosen tetrahedron to decide the orientation of the $3$-complex, which is independent of the choice of the tetrahedron in the complex. As such, in Eq. \eqref{eq:T1move}, we take $v_0v_1v_2v_3$ and denote the orientation of the original $3$-complex on the LHS by $\epsilon(v_4|v_0v_1v_2v_3)$, which is obviously $-1$ by our conventions.

Besides, since $T_1$ creates a new edge $v_2v_4$, the corresponding group element $v_2v_4\in G$ must be summed over to remove the arbitrariness. Note that $T_1$ and the other three mutation transformations to be defined do not alter the existing edges of the original graph, neither the group elements on these edges. 

Bearing the above explanation in mind, one  is ready  to understand the construction of $T_2$ through $T_4$ as follows.       
\begin{align}
&T_2\BLvert \threeTet{v_0}{v_1}{v_2}{v_3}{v_4}{0.25} \Brangle\label{eq:T2move}\\
  =&[v_0v_1,v_1v_2,v_2v_3,v_3v_4]^{\epsilon(v_3|v_0v_1v_2v_4)}
  \BLvert \twoTet{v_0}{v_1}{v_2}{v_3}{v_4}{0.25}\Brangle\nonumber\\
  =&[v_0v_1,v_1v_2,v_2v_3,v_3v_4] 
  \BLvert \twoTet{v_0}{v_1}{v_2}{v_3}{v_4}{0.25} \Brangle.\nonumber
\end{align}
\begin{align}
  &T_3\BLvert \fourTet{v_1}{v_2}{v_4}{v_3}{0}{}{0.5}{1}\Brangle \nonumber\\
  =&\sum_{\substack{v_0v_1,v_0v_2,\\ v_0v_3,v_0v_4\in G}} 
   [v_0v_1,v_1v_2,v_2,v_3,v_3v_4]^{\epsilon(v_1v_2v_3v_4)}\nonumber \\
&\x\BLvert\fourTet{v_1}{v_2}{v_4}{v_3}{1}{v_0}{0.5}{1}\Brangle\label{eq:T3move}\\
  =&\sum_{\substack{v_0v_1,v_0v_2,\\ v_0v_3,v_0v_4\in G}}
  [v_0v_1,v_1v_2,v_2,v_3,v_3v_4] \BLvert \fourTet{v_1}{v_2}{v_4}{v_3}{1}{v_0}{0.5}{1} \Brangle. \nonumber
\end{align}
\begin{align}
&T_4 \BLvert \fourTet{v_0}{v_1}{v_4}{v_3}{3}{v_2}{0.5}{1} \Brangle\nonumber\\
=& [v_0v_1,v_1v_2,v_2,v_3,v_3v_4]^{\epsilon(v_4|v_0v_1v_2v_3)}
\BLvert \fourTet{v_0}{v_1}{v_4}{v_3}{0}{}{0.5}{1} \Brangle\nonumber\\
=& [v_0v_1,v_1v_2,v_2,v_3,v_3v_4]^{-1} \BLvert \fourTet{v_0}{v_1}{v_4}{v_3}{0}{}{0.5}{1} \Brangle. \label{eq:T4move}
\end{align}

A generic mutation transformation $T$ is a composition of the transformations $T_1$ through $T_4$ defined above. A crucial step now is to prove that \textit{the mutation transformations generated by $T_1$, $T_2$, $T_3$ , and $T_4$ form a unitary symmetry on the ground-state Hilbert space.}  We denote by $\Hil^0_{\Gamma}$ the ground-state subspace of the Hilbert space $\Hil_{\Gamma}$ on a graph $\Gamma$. We divide the proof into two logical steps.

\noindent (i). \textit{Mutation transformations keep $\Hil^0_{\Gamma}$ invariant.}

Namely, for any mutation transformation $T:\Hil_{\Gamma}\rightarrow \Hil_{\Gamma'}$, if $|\Phi\rangle\in\Hil^0_{\Gamma}$, then $T|\Phi\rangle\in\Hil^0_{\Gamma'}$.

Equivalently, we can show that $T P^0 _{\Gamma}=P^0_{\Gamma'}T$ holds for any mutation transformation $T$ over the entire subspace $\Hil^{B_f=1}$, where $P_{\Gamma}$ and $P^0_{\Gamma'}$ project respectively onto $\Hil^0_{\Gamma}$ and $\Hil^0_{\Gamma'}$. We detail the proof in Appendix \ref{app:Tunitarity}. As a remark, in general, however, $H' T \neq TH$.

\noindent (ii). \textit{Mutation transformations are unitary on $\Hil^0_{\Gamma}$.}

In other words, for any mutation transformation $T:\Hil_{\Gamma}\rightarrow \Hil_{\Gamma'}$,  
\begin{equation}
  \label{unitary}
  \left\langle T\Phi\right.\left|T\Psi\right\rangle
  =\left\langle \Phi\right.\left|\Psi\right\rangle,\quad\forall |\Phi\rangle, |\Psi\rangle\in\Hil^0_{\Gamma}.
\end{equation}

We need only to verify Eq. (\ref{unitary}) for $T_1$ through $T_4$ but save the proof in Appendix \ref{app:Tunitarity}.
 
The above implies a bijection between the ground-state Hilbert spaces on any two graphs connected by the mutation moves. Because two such graphs triangulate the same $3$-manifold, the dimension of the ground-state Hilbert space---the GSD---is a topological observable, whose expectation value is a topological invariant. The GSD of our model on some $3$-manifold is thus justified to be the trace of the ground state projector (\ref{eq:GSDprojector}) over the $B_f=1$ subspace of the Hilbert space on any graph $\Gamma$ triangulating the manifold, namely,
\be\text{GSD}=\text{tr}(P^0_{\Gamma}),\ee
which is Hermitian and invariant under the mutations.

\section{Degenerate ground states}\label{sec:GSD}
As pointed out in the end of Section \ref{subsec:Hamiltonian}, our model describes topological phases with topologically protected degenerate ground states. And as shown in the previous section, the GSD of a topological phase described by our model is a topological invariant. We thus may partially characterize a topological phase described by our model by its GSD in certain spatial topology. In the same spatial topology, two topological phases with different GSDs must be distinct.   

Another key property of a topological phase is the topological quantum numbers of the quasi-excitations in the phase, such as the fractional self and mutual statistics of these excitations. Here is an important remark on the difference between two dimensions and three dimensions, regarding the relation between ground states and excitations. In the TQD model of $2$-dimensional topological phases and in fact in all $2$-dimensional phases to date, the GSD on a $2$-torus counts quasi-excitation types\cite{Hu2012}. In our TGT model, however, the GSD on a $3$-torus is \emph{not} equal to the number of types of loop and particle excitations. Typically, the $3$d GSD \emph{over-counts} the excitation types.\cite{Wang2014c} We will come back to this issue in Section \ref{subsec:GSbasis}.

The topology-dependent GSD roots in the long-range entanglement and the global degrees of freedom in the  ground states. But how to characterize these global degrees of freedom? The answer of the question will help to (1) distinguish between different topological phases with identical topology-dependent GSD, and (2) comprehend the relationship between the global degrees of freedom in the degenerate ground states and the topological properties of the quasi-excitations.

Below we place our model on a $3$-torus  and derive the corresponding GSD formula. We also study the global degrees of freedom in the ground states.

\subsection{GSD on a 3-torus}\label{subsec:GSDtorus}
Thanks to the topological invariance of the GSD of our model, we can derive the GSD on the simplest graph that triangulates the manifold on which the model is defined.

On a $3$-sphere, whose topology is trivial, namely its fundamental group is trivial, the GSD of our model is identical to one. This feature is in fact shared by all models of  topological phases to date. 

\begin{figure}[h!]
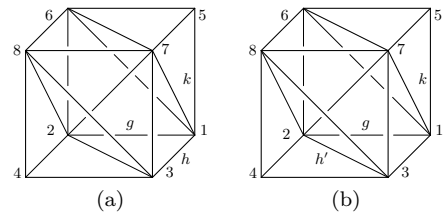

\centering
\subfigure[]{
\torusCube{4}{3}{2}{1}{8}{7}{6}{5}{2}
\label{fig:3torusA}}
\subfigure[]{
\torusCubeNB{4}{3}{2}{1}{8}{7}{6}{5}{2}
\label{fig:3torusB}}
\caption{The simplest triangulation of a $3$-torus. Vertices are in the order of enumerations $1<2<3<4<5<6<7<8$, so the arrows are omitted for simplicity. Any two squares opposite to each other are identified along the edges with the same arrow. Restricted to the subspace $\Hil^{B_f=1}$, there are only three independent group degrees of freedom. (a) A natural presentation of the ($zxy$) basis: $k=15,g=12,h=13\in G$. (b) A physical presentation of the ($zxy'$) basis: $k$, $g$, and $h'=23=\bar gh$.}
\label{fig:3torus}
\end{figure}

In 3-dimension, the simplest closed, orientable manifold is a $3$-torus, whose simple triangulation is shown in Fig. \ref{fig:3torus}.

We first focus on Fig.  \ref{fig:3torusA}, which apparently contains six tetrahedra but in fact has a single vertex, i.e., all the eight vertices must be identified. To make it easy to construct the amplitude of  the operators $A_v^x$  in terms of the \fcs, however, we first treat the eight vertices differently as in Fig. \ref{fig:3torus}. This is justified by the periodic boundary condition that identifies all the differently labeled vertices. Note that the orientations of the boundary edges in Fig. \ref{fig:3torus} are consistent with the periodic boundary condition.

The graph in Fig. \ref{fig:3torusA} actually  supports a natural basis of the subspace $\Hil^{B_f=1}$, spanned by the basis vectors
\be\label{eq:basisVectNatural}
  \left\{\left|g,h,k\right\rangle\right.\left|g,h,k\in G,gh=hg,gk=kg,hk=kh\right\}
\ee
where the constraints of commutativity are due to the diagonals respectively on the six boundary squares of the cube in Fig. \ref{fig:3torusA}. As seen in the figure, the basis vectors \eqref{eq:basisVectNatural} can be collectively written as $\ket{15,12,13}$, without referring to the explicit group elements. This natural presentation of the basis later will turn out  physically unclear and urge we to switch to the physical presentation in Fig. \ref{fig:3torusB}. But let us stick to the natural one first within this subsection because it is intuitive to do so.

That the graph in Fig. \ref{fig:3torusA} has a single vertex enables us to take a simplified notation $A^x$ of the vertex operator by omitting the vertex subscript. Applying definition \eqref{eq:AvAmpSingleTet}, the $A^x$ with $x\in G$ acts on the basis vector
as\begin{align}
  \label{eq:torusAxEnumeration}
  &A^x \ket{15,12,13}
  \nonumber\\
  =&[23,34,48',8'8][23,37,78',8'8]^{-1}[26,67,78',8'8]\times
    \nonumber\\
  &\frac{[23,37',7'7,78'][12,26,67',7'7]}{[26,67',7'7,78'][12,23,37',7'7][15,56,67',7'7]}\times
  \nonumber\\
  &[15,56',6'6,67'][12,26',6'6,67']^{-1}[26',6'6,67',7'8']\times\nonumber\\
  &[15',5'5,56',6'7']^{-1}\x \nonumber\\
  &[23,34',4'4,48']^{-1}\x \nonumber\\
  &\frac{[12,23',3'3,37'][23',3'3,34',4'8']}{[23',3'3,37',7'8']}\times \nonumber\\
  &\frac{[2'2,23',3'7',7'8'][12',2'2,26',6'7']}{[2'2,23',3'4',4'8'][2'2,26',6'7',7'8'][12',2'2,23',3'7']}\times \nonumber\\
  &\frac{[1'1,12',2'3',3'7'][1'1,15',5'6',6'7']}{[1'1,12',2'6',6'7']}\times \nonumber\\
  &\ket{1'5',1'2',1'3'},
\end{align}
where $v'v=x$  for all $v=1,\dots 8$, and the $A^x$ acts on the eight virtually different vertices in descending order. One may obtain a differently looking amplitude by acting the $A^x$ on the vertices in alternative, say, ascending, order. Nevertheless, the topological invariance and commutativity \eqref{eq:AvCommute} enforce the new amplitude to be the same as that in Eq. \eqref{eq:torusAxEnumeration}. This can be straightforwardly  shown by using \fc conditions.

Now we can substitute the group elements $g,h, k,$ and $x$ into the above equation and get
\begin{align}
  \label{eq:torusAx}
  &A^x \ket{k,g,h}
  \nonumber\\
  =&[\bar gh,g,k\bar x,x][\bar gh,k,g\bar x,x]^{-1}[k,\bar gh,g\bar x,x]\times
    \nonumber\\
  &\frac{[\bar gh,k\bar x,x,g\bar x][g,k,\bar gh\bar x,x]}{[k,\bar gh\bar x,x,g\bar x][g,\bar gh,k\bar x,x][k,g,\bar gh\bar x,x]}\times
  \nonumber\\
  &[k,g\bar x,x,\bar gh\bar x][g,k\bar x,x,\bar gh\bar x]^{-1}[k\bar x,x,\bar gh\bar x,xg\bar x]\times\nonumber\\
  &[k\bar x,x,g\bar x,x\bar gh\bar x]^{-1}\x \nonumber\\
  &[\bar gh,g\bar x,x,k\bar x]^{-1}\x \nonumber\\
  &\frac{[g,\bar gh\bar x,x,k\bar x][\bar gh\bar x,x,g\bar x,xk\bar x]}{[\bar gh\bar x,x,k\bar x,xg\bar x]}\times \nonumber\\
  &\frac{[x,\bar gh\bar x,xk\bar x,xg\bar x][g\bar x,x,k\bar x,x\bar gh\bar x]}{[x,\bar gh\bar x,xg\bar x,xk\bar x][x,k\bar x,x\bar gh\bar x,xg\bar x][g\bar x,x,\bar g h\bar x,xk\bar x]}\times \nonumber\\
  &\frac{[x,g\bar x,x\bar gh\bar x,xk\bar x][x,k\bar x,xg\bar x,x\bar gh\bar x]}{[x,k\bar x,xg\bar x,x\bar gh\bar x]}\times \nonumber\\
  &\ket{xk\bar x,xg\bar x,xh\bar x}.
\end{align}

At this stage, one can verify the product rule $A^xA^y=A^{xy}$  using  \fc\ conditions, which we do not show here. This reconciles Eq. \eqref{eq:AvProd}. 
The ground-state projector is the average of $A^x$ over $G$: 
\be
\label{eq:torusProjector}
P^0=\frac{1}{|G|}\sum_x A^x
\ee

As aforementioned, the trace of the ground--state projector \eqref{eq:torusProjector} is the topological observable GSD:  
\begin{align}
  \label{eq:GSDomega}
  \text{GSD}=&\text{tr}(\frac{1}{|G|}\sum_x A^x)
  \nonumber\\
  =&\frac{1}{|G|}\sum_{g,h,k,x}\delta_{gh,hg} \delta_{gk,kg} \delta_{hk,kh} \bra{k,g,h}A^x\ket{k,g,h}
  \nonumber\\
  =&\frac{1}{|G|}\sum_{g,h,k,x}\delta_{gh,hg} \delta_{gk,kg} \delta_{hk,kh} \delta_{hx,xh}\delta_{xg,gx}\delta_{xk,kx}\times
    \nonumber\\
  &[\bar gh,g,k\bar x,x][\bar gh,k,g\bar x,x]^{-1}[k,\bar gh,g\bar x,x]\times
    \nonumber\\
  &\frac{[\bar gh,k\bar x,x,g\bar x][g,k,\bar gh\bar x,x]}{[k,\bar gh\bar x,x,g\bar x][g,\bar gh,k\bar x,x][k,g,\bar gh\bar x,x]}\times
  \nonumber\\
  &[k,g\bar x,x,\bar gh\bar x][g,k\bar x,x,\bar gh\bar x]^{-1}[k\bar\ x,x,\bar gh\bar x,xg\bar x]\times\nonumber\\
  &[k\bar x,x,g\bar x,x\bar gh\bar x]^{-1}[\bar gh,g\bar x,x,k\bar x]^{-1}\x \nonumber\\
  &\frac{[g,\bar gh\bar x,x,k\bar x][\bar gh\bar x,x,g\bar x,xk\bar x]}{[\bar gh\bar x,x,k\bar x,xg\bar x][x,\bar gh\bar x,xg\bar x,xk\bar x]}\times \nonumber\\
  &\frac{[x,\bar gh\bar x,xk\bar x,xg\bar x][g\bar x,x,k\bar x,x\bar gh\bar x]}{[x,k\bar x,x\bar gh\bar x,xg\bar x][g\bar x,x,\bar g h\bar x,xk\bar x]}\times \nonumber\\
  &\frac{[x,g\bar x,x\bar gh\bar x,xk\bar x][x,k\bar x,xg\bar x,x\bar gh\bar x]}{[x,k\bar x,xg\bar x,x\bar gh\bar x]}
\end{align}
where this trace is over the subspace $\mathcal{H}^{B_f=1}$.

Amazingly, the complexity in Eq. (\ref{eq:GSDomega}) can be greatly reduced in many cases because of an implicit simple mathematical structure. To this end, in the next subsection we first reveal the algebraic structure hidden in Eq. \eqref{eq:torusAx}, equipped with which we be ready to simplify the GSD expression \eqref{eq:GSDomega}.

\subsection{Topological degrees of freedom}\label{subsec:topoDof}
The algebraic structure in the expression \eqref{eq:torusAx} will pave an even path for  studying the topological degrees of freedom in the ground states and exploring the significance of the GSD {not yet} fully uncovered. We enclose the lengthy, distractive derivation in Appendix \ref{app:Rewrite3torusAx} but present only the result here: Eq. (\ref{eq:torusAx}) can be reexpressed as follows.
\be
  \label{eq:torusAxRewrite}
   A^x\ket{k,g,h}  =
  \frac{[\bar x,x\bar gh\bar x]_{k, g}}{[\bar gh,\bar x]_{k, g}}
  \ket{xkx,xg\bar x,xh\bar x}.
\ee
On the RHS, the fraction consists of two normalized \textbf{doubly-twisted} $2$-cocycles defined in Eq. \eqref{eq:Dtwisted2cocycleDef}.  A doubly-twisted $2$-cocycle $[d,e]_{a,b}$ satisfies not the usual but a twisted $2$-cocycle condition, namely, 
\be\label{eq:dT2cocycleCond}
\widetilde\delta[c,d,e]_{a,b}=\frac{[d,e]_{\bar cac ,\bar cbc}[c,de]_{a,b}}{[cd,e]_{a,b}[c,d]_{a,b}}\bigg\vert_{ab=ba}=1,
\ee
for $a,b,c,d\in G$ and $ab=ba$. The normalization reads $[1,e]_{a,b}=[d,1]_{a,b}=1.$ Because $k$ and $g$ do commute, the numerator and denominator on the RHS of Eq. \eqref{eq:torusAxRewrite} indeed satisfy the twisted $2$-cocycle condition \eqref{eq:dT2cocycleCond}. 

Eq. \eqref{eq:torusAxRewrite} also suggests that the basis presentation $\ket{k,g,h}$ does not reflect the physics precisely because $g,h$, and $k$ do not all appear in the amplitude of $A^x$ individually; rather, it is $k$, $g$, and $\bar gh$ each play an important role individually. Staring at Fig. \ref{fig:3torusA}, one sees that $23=\bar g h$. Hence, by redefining $h'=\bar gh$, it would be better to present the basis vector by $\ket{k,g,h'}$, as seen in Fig. \ref{fig:3torusB}.  Note that such a replacement of basis presentation is not any sort of basis transformation because the actual basis state is the $3$-torus in Fig. \ref{fig:3torus} that can be presented symbolically by any three independent group degrees of freedom on the edges of the torus. Clearly, $k,g$, and $h'$ are also three such group elements. Note that $(g,h')$ and $(g,h)$ span the same plane perpendicular to $k$. Furthermore, it is conventional and more convenient notation-wise in the subsequent study to change our notation of $[d,e]_{a,b}$ as
\be\label{eq:betaDef}
\beta_{a,b}(d,e)\defeq [d,e]_{a,b}
\ee
Therefore, hereafter we rewrite Eq. (\ref{eq:torusAxRewrite}) as
\be\label{eq:torusAxBeta}
\begin{aligned}
   A^x\ket{k,g,h'} =
  &\frac{\beta_{k,g}(\bar x,xh'\bar x)}{\beta_{k',g'}(\bar h',x)}
  \ket{xk\bar x,xg\bar x,xh'\bar x} \\
  =&\eta^{k,g}(h',x)^{-1}\ket{xk\bar x,xg\bar x,xh'\bar x},
\end{aligned}
\ee
where we define
the ratio\be\label{eq:eta}
\eta^{k,g}(h',x)=\frac{\beta_{k,g}(h',\bar x)}{\beta_{k,g}(\bar x,xh'\bar x)},
\ee
for any given $g\in G$ and $h'\in Z_{k,g}=\{y\in G|yg=gy,yk=ky\}$, the centralizer subgroup for $g,k\in G$. It is straightforward to show by the defining Eqs. \eqref{eq:twisted3cocycleDef} 
and \eqref{eq:Dtwisted2cocycleDef}  that
\be\label{eq:etah=g'Is1}
\eta^{k,g}(g,x)=1,
\ee
for all $g,k,x\in G$, and that
\be\label{eq:etah=k'Is1}
\eta^{k,g}(k,x)=1,
\ee
for all $x\in G$ and $kg=gk$. This is precisely our case.
Importantly, if $x\in Z_{k,g,h'}=\{x\in G|xk=kx,xg=gx,xh'=h'x\}$, the $U(1)$ numbers
\be\label{eq:rhoEta}
\rho^{k,g}(h',x)\defeq\eta^{k,g}(h',x)\Big|_{x\in Z_{k,g,h'}}
\ee
furnish a $1$--dimensional representation of the centralizer subgroup $Z_{k,g,h'}\subseteq G$. 
This is due to the fact that $\rho^{k,g}(h',x)\rho^{k,g}(h',y)=\rho^{k,g}(h',xy)$, owing to the product rule $A^x A^y=A^{xy}$ on the ground states.

Equation \eqref{eq:torusAxBeta} implies that the ground-state Hilbert space is spanned by the set of vectors:
\begin{align}
  \label{eq:GGquotConj}
  \left\{\frac{1}{|G|}\sum_{x\in G} \eta^{k,g}(h',x)
  \ket{xk\bar x,xg\bar x,xh'\bar x}\Biggl|k,g,h'\ \mathrm{commute}\right\}.
\end{align}
This result may easily lead one to the false statement that the GSD value is equal to the size of $\text{Hom}\left(\pi_1(T^3),G\right)/conj$, where the $conj$ in the quotient is the conjugacy equivalence: $(k,g,h')\sim(xkx,xg\bar x,xh'\bar x)$ for any $x$. The fallacy is ascribed to  that the set \eqref{eq:GGquotConj} generally over-counts the ground states because 
the terms therein that are summed for some $k$, $g$, and $h'$ may vanish and render the homologous states absent. 
To capture and classify the non-vanishing  ground states and hence nail down the correct GSD formula, let us take a closer look at the algebraic structure hidden among the functions $\beta_{a,b}$ defined in Eq. (\ref{eq:betaDef}).

Although the details are examined in the end of Appendix \ref{app:Rewrite3torusAx},
it is worthwhile noting here that if $\beta_{a,b}$ has all its variables restricted to $Z_{a,b}$, it would satisfy the usual $2$-cocycle condition. 
Namely, 
\be\label{eq:DtwistedUsual2cocycleCond}
\beta_{a,b}[e,f]\beta_{a,b}[de,f]^{-1}\beta_{a,b}[d,ef]\beta_{a,b}[d,e]^{-1}=1,
\ee
for all $d,e,f\in Z_{a,b}$.

Each function $\beta_{a,b}$ in fact specifies a class of projective representations
of $Z_{a,b}$. Such representations are dubbed $\beta_{a,b}$-\textbf{representations} $\widetilde{\rho}:Z_{a,b}{\rightarrow}\text{GL}\left(Z_{a,b}\right)$, defined by
\be\label{eq:betarepresentation}
\widetilde{\rho}^{a,b}(x)\widetilde{\rho}^{a,b}(y)=\beta_{a,b}(x,y)\widetilde{\rho}^{a,b}(xy).
\ee
The consequence that $\beta_{a,b}$ is normalized is $\widetilde{\rho}(e)\widetilde{\rho}(x)=\widetilde{\rho}(x)\widetilde{\rho}(e) =\widetilde{\rho}(x)$. Beside, the $2$--cocycle condition \eqref{eq:DtwistedUsual2cocycleCond} implies the associativity $\widetilde{\rho}(x)\left(\widetilde{\rho}(y)\widetilde{\rho}(z)\right) =\left(\widetilde{\rho}(x)\widetilde{\rho}(y)\right)\widetilde{\rho}(z)$. These two facts make $\widetilde{\rho}$ truly a well-defined representation. In the special case where $\beta_{a,b}=1$, $\tilde{\rho}^{a,b}$ reduces to a linear representations of $Z_{a,b}$.

We shall be interested only in the classification of  the $\beta_{a,b}$--representations of $Z_{a,b}$ for fixed $a,b\in G$. Some important and relevant properties of these representations are discovered and elucidated in 
Appendix \ref{app:centralizer&projChi}, and we simply catalogue the results as follows for convenience.

A key concept is $\beta_{a,b}$-\emph{regularity}. An element $c\in Z_{a,b}$ is $\beta_{a,b}$-\textbf{regular} if $\beta_{a,b}(c,d)=\beta_{a,b}(d,c)$, $\forall d\in Z_{a,b,c}\subseteq Z_{a,b}$. Furthermore, $c$ is $\beta_{a,b}$--regular if and only if the entire conjugacy class $[c]$ of $c$ is. This reiterates the Theorem \ref{theo:betaRegClass} established in Appendix \ref{app:centralizer&projChi}. Similarly, $[c]$ is called a $\beta_{a,b}$-\textbf{regular conjugacy class} if $c$ is $\beta_{a,b}$-regular. Specifically, for a given $\beta_{a,b}$, $a$ and $b$ are readily $\beta_{a,b}$-regular.

Let $r(G)$ be the number of conjugacy classes in $G$, and $C^A$ with $A=1,\dots,r(G)$ denote the classes. Because $Z_a\cong Z_b$ $\forall a,b\in C^A$, one needs not to distinguish between these isomorphic centralizers in many cases but simply can work with a generic one of them, denoted by $Z^A$, as the one associated with any representative $g^A$ of $C^A$. To cope with this fact, we write any set of representatives of all classes $C^A$ as $R_C=\{g^A\in C^A|A=1\dots r(G)\}$.

Likewise, we let  $r(Z^A)$ be the number of conjugacy classes in the centralizer subgroup $Z^A$, and $C^B_{Z^A}$ with $B=1,\dots,r(Z^A)$ the conjugacy classes in $Z^A$. For any $a\in C^A$ and $b\in C^B_{Z^A}$,  we denote the number of $\beta_{a,b}$--regular conjugacy classes in $Z_{a,b}$ by $r(Z_{a,b},\beta_{a,b})$. Then, the following inequality is obvious:
\be\label{eq:rGbetaLrG}
r(Z_{a,b},\beta_{a,b})\leq r(Z_{a,b}).
\ee
For finite groups, there are precisely $r(Z_{a,b},\beta_{a,b})$ inequivalent irreducible $\beta_{a,b}$--representations  of $Z_{a,b}$. Typically, in the case where $\beta_{a,b}=1$, we recover the familiar equality between the number of irreducible linear representations and that of the conjugacy classes of a finite group. Eq. \eqref{eq:rGbetaLrG} indicates that the irreducible $\beta_{a,b}$-regular representations of $Z_{a,b}$ are fewer than the irreducible liner representations.

To relate this classification of projective representations of $Z^{A,B}$ with the global degrees of freedom in the ground states, let us rewrite the GSD expression \eqref{eq:GSDomega} as
\begin{align}
  \text{GSD}=&
  \frac{1}{|G|}\sum_{\substack{g',h,k'\in G\\ x\in Z_{g',h,k'}}}\delta_{kh',h'k} \delta_{gk,kg}\delta_{h'k,kh'}\eta^{k,g}(h',x)^{-1}\nonumber \\
  =&  \frac{1}{|G|}\sum_{k\in G}\sum_{g\in Z_{k}}\sum_{h\in Z_{k,g}}\sum_{x\in Z_{k,g,h'}}\rho^{k,g}(h',x).\label{eq:GSDbeta}
\end{align}
To further simply the above, we acknowledge the identity: 
\begin{align}
  \label{eq:SumBetaToDelta}
  \frac{1}{|Z_{k,g,h'}|}\sum_{x\in Z_{k,g,h'}}
  \rho^{k,g}(h',x)
  =
  \left\{
  \begin{array}{ll}
    1, & h'\text{ is }\beta_{k,g}\text{-regular}.\\
    0, & \text{otherwise}.
  \end{array}
  \right.
\end{align}
Clearly, $|Z_{k,g,h'}|$ is the order of $Z_{k,g,h'}$ for given $k,g,h'\in G$.
We prove Eq. \eqref{eq:SumBetaToDelta} as follows. The phase $\rho^{k,g}(h',x)$ above is defined in Eq. (\ref{eq:rhoEta}), which is a $1$--dimensional representation of $Z_{k,g,h'}$. This representation is trivial, namely $\rho^{k,g}=\rho^0=\mathds{1}$ if $h'$ is $\beta_{k,g}$-regular; otherwise it is a non--trivial irreducible representation. 
 Equation \eqref{eq:SumBetaToDelta} is then an immediate result of the orthogonality condition
\be
\frac{1}{|Z_{k,g,h'}|}\sum_{x\in Z_{k,g,h'}}{\rho^{k,g}_{(j)}}(h',x)
  =\delta_{j,0},
\ee
where $j=0$ and $j\neq 0$ respectively label the trivial representation and the non-trivial irreducible representations. 

Plugging Eq. \eqref{eq:SumBetaToDelta} into \eqref{eq:GSDbeta} yields
\begin{align}
  \text{GSD}=&\sum_{\substack{k\in G,\\g\in Z_{k},h'\in Z_{k,g}}}\frac{|Z_{k,g,h'}|}{|G|}\times
  \left\{
  \begin{array}{ll}
    1, & h\text{ is }\beta_{k,g}\text{-regular}\\
    0, & \text{otherwise}
  \end{array}
  \right.
  \nonumber\\
  =&\sum_{k\in G}\frac{|Z_{k}|}{|G|}\sum_{g\in Z_{k}}\frac{|Z_{k,g}|}{|Z_{k}|} \sum_{h'\in Z_{k,g}}\frac{|Z_{k,g,h'}|}{|Z^{k,g}|}\nonumber\\ 
&\quad\quad \times
  \left\{
  \begin{array}{ll}
    1, & h'\text{ is }\beta_{k,g}\text{--regular}\\
    0, & \text{otherwise}
  \end{array}
  \right.
  \nonumber\\
  =&\sum_{k\in G}\frac{1}{|C^{k}|}\sum_{g\in Z_{k}}\frac{1}{|C^{g}_{Z_{k}}|} \sum_{h'\in Z_{k,g}}\frac{1}{|C^{h'}_{Z_{k,g}}|}\nonumber\\ 
&\quad\quad \times
  \left\{
  \begin{array}{ll}
    1, & h'\text{ is }\beta_{k,g}\text{--regular}\\
    0, & \text{otherwise}
  \end{array}
  \right.
  \nonumber\\ 
=&\sum_A\sum_{k\in C^A}\frac{1}{|C^{A}|}\sum_{g\in Z^A}\frac{1}{|C^{g}_{Z^A}|} \sum_{h'\in Z_{k^A,g^B}}\frac{1}{|C^{h'}_{Z_{k^A,g^B}}|}\nonumber\\ 
&\quad\quad \times
  \left\{
  \begin{array}{ll}
    1, & h'\text{ is }\beta_{k^A,g^B}\text{-regular}\\
    0, & \text{otherwise}
  \end{array}
  \right.
  \nonumber\\   
  =&\sum_{A=1}^{r(G)}\sum_{B=1}^{r(Z^A)} r(Z_{k^A,g^B},\beta_{k^A,g^B}),  \label{eq:GSDconjugacy}
\end{align}
where the final line is independent of the choice of the representatives of $C^A$ and $C^B_{Z^A}$. 

The aforementioned equality between the number of $\beta_{k^A,g^B}$-regular conjugacy classes of $Z_{k,g}$ and the number of $\beta_{k^A,g^B}$-regular representations of $Z_{k^A,g^B}$ 
 also sets the GSD formula \eqref{eq:GSDconjugacy} in an alternative but equivalent form,
\be\label{eq:GSDrepresentations}
  \text{GSD}
  =\sum_{A=1}^{r(G)}\sum_{B=1}^{r(Z^A)}\#(\beta_{k^A,g^B}\text{-regular irreps of }Z^{A,B}),
\ee
where $\#$ denotes \textquotedblleft the number of".

\subsection{Bases of the ground states}\label{subsec:GSbasis}

As preluded in the earlier subsection, we have a simple GSD formula from the sophisticated expression (\ref{eq:GSDomega}). The GSD of our model on a $3$-torus thus amounts to summing the number of irreducible projective $\beta_{k^A,g^B}$-regular representations of the centralizer subgroup $Z_{k^A,g^B}$ for each pair of conjugacy classes $C^A$ and $C^B_{Z^A}$ in $G$.

This encourages us to label the ground states of our model on a $3$-torus by {triples} $(k^A,g^B,h')$, where the three elements $g^A$, $g^B$, and $h'$ respectively run over $R_{C_G}$, $R_{C_{Z^A}}$, and the set of $\beta_{k^A,g^B}$-regular conjugacy class representatives of $Z_{k^A,g^B}$. A Fourier transform (defined below) can turn the triples $(k^A,g^B,h')$ into the equivalent triples $(A,B,\mu)$ with $A=1\dots r(G)$, $B=1,\dots,r(Z^A)$, and $\mu$ labeling $\widetilde{\rho}^{k^A,g^B}_{\mu}$, the irreducible $\beta_{k^A,g^B}$-representations of $Z_{k^A,g^B}$. We claim and prove later that the basis vectors $\ket{A,B,\mu}$ can be defined via the Fourier transform:
\be
  \label{eq:AmuBasis}
  \ket{A,B,\mu}
  =\frac{1}{\sqrt{|G|}}\sum_{\substack{k\in C^A,g\in C^B_{Z_{k}}\\ h'\in Z_{k,g}}}\,\widetilde{\chi}^{k,g}_{\mu}(h')\,\ket{k,g,h'},
\ee
where the \emph{projective characters} $\widetilde{\chi}^{k,g}_{\mu}$ are defined by \emph{the trace of the representations $\widetilde{\rho}^{k,g}_{\mu}$}:
\be
\widetilde{\chi}^{k,g}_{\mu}(h) \equiv \text{tr}\widetilde{\rho}^{k,g}_{\mu}(h).
\ee  

It might appear somewhat awkward that the index $B$ can be fixed while the conjugacy class $C^A$ has an internal space, as for two $k,k'=xk\bar x\in C^A$ with $x\in G$, $g\in Z_k$ is in general different from $g'=xg\bar x\in Z_{k'}$. Nevertheless, according to Proposition \ref{prop:isoConjClass}, the isomorphism $Z_k\cong Z_{k'}$ induces a bijection between the conjugacy class $C^B_{Z_k}$ of $g$ and $C^{B'}_{Z_{k'}}$   of $g'$. Hence, as far as topologically and physically invariant properties are concerned, we need not to distinguish the index $B$ and $B'$. Likewise, since the centralizers $Z_{k,g}$ are isomorphic for all $k\in C^A$ and $g\in C^B_{Z_{k}}$, so are the set of irreducible $\beta_{k,g}$-regular representations of $Z_{k,g}$ for all $k\in C^A$ and $g\in C^B_{Z_{k}}$. This statement is proven in Appendix \ref{app:centralizer&projChi}. Therefore, the same label $\mu$ works for all $\beta_{k,g}$-representations. We save 
the construction of the isomorphism among the irreducible $\beta_{k,g}$-representations for $g\in C^A$ for Appendix \ref{app:centralizer&projChi}. 

The projective characters $\widetilde{\chi}^{k,g}_{\mu}(h')$ transform under simultaneous conjugation of $k,g$, and $h'$ as
\be
\label{eq:CharacterInConjugacyClasses}
\widetilde{\chi}^{xk\bar x,xg\bar x}_\mu(xh'\bar x) = \eta^{k,g}(h',x)^{-1} \widetilde{\chi}^{k,g}_{\mu}(h'),
\ee
for all $x\in G$. For each conjugacy class $C^A$ with its representative element $k^A$ and each conjugacy class $C^B_{Z_{k^A}}$ with some representative $g^B$, if we find a $\beta_{k^A,g^B}$-representation $\widetilde{\rho}^{k^A,g^B}_{\mu}$ of $Z_{k^A,g^B}$, in principle we can construct the $\beta_{k,g}$-representations for any other elements $k\in C^A$ and $g\in C^B_{Z_{k^A}}$. Throughout this work, the representations being considered all meet the relation \eqref{eq:CharacterInConjugacyClasses}.

In general, the projective characters can not be functions of conjugacy classes because of the relation  $\widetilde{\chi}^{k,g}_{\mu}(xh'\bar x)= \left[\beta_{k,g}(h'\bar x,x)/\beta_{k,g}(x,h\bar 'x)\right]\widetilde{\chi}^{k,g}_{\mu}(h')$, which is a result of definition \eqref{eq:betarepresentation}. Nonetheless, the following completeness and orthogonality relations of these projective characters still hold: 
\be
\begin{aligned}
  \label{eq:characterrelation}
  &\frac{1}{|Z_{k,g}|}\sum_{h'\in Z_{k,g}}
  \overline{\widetilde{\chi}^{k,g}_{\mu}}(h')
  \widetilde{\chi}^{k,g}_{\nu}(h')
  =\delta_{\mu\nu},\\
  &\frac{|C^A|}{|Z_{k,g}|}
  \sum_{\mu}\,
  \overline{\widetilde{\chi}^{k,g}_{\mu}}(a)\,
  \widetilde{\chi}^{k,g}_{\mu}(b)
  =
  \left\{
  \begin{array}{ll}
    1, & a\in [b]\\
    0, & \text{otherwise}
  \end{array}
  \right.,
\end{aligned}
\ee
for all $\beta_{k,g}$-regular elements $h',a,b\in G$. Equation \eqref{eq:characterrelation} makes the basis \eqref{eq:AmuBasis} orthonormal. Importantly, for any $h'$ not $\beta_{k,g}$-regular, $\widetilde{\chi}^{k,g}(h')=0$. This is 
just Proposition \ref{prop:chiIs0} proven in Appendix \ref{app:centralizer&projChi}.

Now, to prove that $\ket{A,B,\mu}$ is a ground state, we simply need to corroborate its invariance under the action of the ground-state projector \eqref{eq:torusProjector}:
\begin{align}
& P^0\ket{A,B,\mu}\nonumber\\
=&\frac{1}{|G|}\sum_{x\in G}A^x\ket{A,B,\mu}\nonumber\\
=&\frac{1}{\sqrt{|G|^3}}\sum_{x\in G}\sum_{\substack{k\in C^A,\\g\in C^B_{Z_{k}}\\ h'\in Z_{k,g}}}\widetilde{\chi}^{k,g}_{\mu}(h')\eta^{k,g}(h',x)^{-1} \ket{xk\bar x, xg\bar x,xh'\bar x}\nonumber\\
=&\frac{1}{\sqrt{|G|^3}} \sum_{x\in G} \sum_{\substack{k\in C^A,\\g\in C^B_{Z_{k}}\\ h'\in Z_{k,g}}} \widetilde{\chi}^{xk\bar x,xg\bar x}_\mu(xh'\bar x)\ket{xk\bar x, xg\bar x,xh'\bar x}\nonumber\\
=&\frac{1}{\sqrt{|G|^3}}\sum_{\substack{k\in C^A,\\g\in C^B_{Z_{k}}\\ h'\in Z_{k,g}}}\widetilde{\chi}^{k,g}_{\mu}\ket{k, g,h'}\sum_{x\in G}1\nonumber\\
=&\frac{1}{\sqrt{|G|}}\sum_{\substack{k\in C^A,\\g\in C^B_{Z_{k}}\\ h'\in Z_{k,g}}}
  \widetilde{\chi}^{k,g}_{\mu}(h')\ket{k, g,h'}=\ket{A,B,\mu}.\label{eq:P0Amu}
\end{align}
In the derivation above, the second and third equalities follows respectively Eqs. \eqref{eq:torusAxBeta} and \eqref{eq:CharacterInConjugacyClasses}, whereas the fourth equality relies on Proposition \ref{prop:isoZab} in Appendix \ref{app:centralizer&projChi}. The inverse transformation of Eq. \eqref{eq:AmuBasis} reads
\be\label{eq:ghToAmu}
  \ket{k,g,h'}
  =\frac{1}{\sqrt{|G|}}\sum^{r(Z^{A,B},\beta_{k,g})}_{\nu=1}\,
  \overline{\widetilde{\chi}^{k,g}_{\nu}(h')}\,\ket{A,B,\nu},
\ee
which is defined in $\Hil^0$ only, with $k\in C^A$ and $g\in C^B_{Z_k}$ assumed.

We therefore conclude that the set of $\ket{A,B,\mu}$ does span an orthonormal basis of the ground states, i.e.,
\be\label{eq:H0basis}
\begin{aligned}
\Hil^0=\mathrm{span}\big\{&\ket{A,B,\mu}:A=1\dots r(G), B=1,\dots,r(Z^A), \\ 
&\mu=1\dots r(Z_{k^A,g^B},\beta_{k^A,g^B})\big\}.
\end{aligned}
\ee
The mathematical structure that classifies the topological degrees of freedom in the ground states via representation theory is now uncovered. Starting with our model specified by a 4--cocycle $\omega$ over $G$, we obtain the result that the topological degrees of freedom are dictated by the doubly-twisted $2$-cocycles $\beta_{k,g}$ over $Z_{k^A,g^B}$. 

Recall the ground state basis $\ket{A,\mu}$ in the $2+1$-dimensional TQD model\cite{Hu2012a}, where $A$ labels fluxes while $\mu$ labels the charges as irreducible representations of $Z^A$. So we may, in the current case of $3+1$ dimensions, also associate the labels $A$ and $B$ in $\ket{A,B,\mu}$ with fluxes, whereas $\mu$ with charges. The obvious and major difference here is that $3+1$-dimensional ground states and hence the quasi-excitations carry two flux labels. Recall that a flux ground state in $2+1$ dimensions physically is an untraced Wilson loop. In $3+1$ dimensions, a ground state with two nontrivial flux labels would be an untraced Wilson membrane, indicating that 
\emph{the corresponding quasi-excitation would be a loop in general instead of just a point-like particle}. 
Pointlike excitations would still be those bearing a single nontrivial flux label. 
Besides, in $3+1$ dimensions, the two flux labels are not independent of each other except for Abelian groups.  

According to Fig. \ref{fig:3torus}, the three group elements $k,g$, and $h$ are on an equal footing, the basis $\ket{A,B,\mu}$ cannot be the sole eigen-basis of the ground state projector $P^0$. 
Indeed, the expression \eqref{eq:AxTorusInIxabc} that rewrites the amplitude \eqref{eq:torusAx} implies that there are six possible eigenbases, depending on which two of $k, g$, and $h'$ and in which order they are Fourier transformed into flux labels. All these bases are related by linear transformations. To be more specific, according to Eq. \eqref{eq:etaPermuteID}, the other five bases should read
\be\label{eq:5otherBases}
\begin{aligned}
&\ket{B,A,\mu}
  =\frac{1}{\sqrt{|G|}}\sum_{\substack{g\in C^B,k\in C^A_{Z_{g}}\\ h'\in Z_{g,k}}}\,\overline{ \widetilde{\chi}^{g,k}_{\mu}}(h')\,\ket{k,g,h},\\
&\ket{A,\nu,C}
  =\frac{1}{\sqrt{|G|}}\sum_{\substack{k\in C^A,h'\in C^C_{Z_{k}}\\ g\in Z_{k,h'}}}\,\overline{ \widetilde{\chi}^{k,h'}_{\nu}}(g)\,\ket{k,g,h'},\\
&\ket{C,\nu,A}
  =\frac{1}{\sqrt{|G|}}\sum_{\substack{h\in C^C,k'\in C^A_{Z_{h}}\\ g'\in Z_{h,k'}}}\,\widetilde{\chi}^{h,k'}_{\nu}(g')\,\ket{k,g,h'},\\
&\ket{\rho,B,C}
  =\frac{1}{\sqrt{|G|}}\sum_{\substack{g'\in C^B,h\in C^C_{Z_{g'}}\\ k'\in Z_{g',h}}}\, \widetilde{\chi}^{g',h}_{\rho}(k')\,\ket{k,g,h'},\\
&\ket{\rho,C,B}
  =\frac{1}{\sqrt{|G|}}\sum_{\substack{h\in C^C,g'\in C^B_{Z_{g'}}\\ k'\in Z_{h,g'}}}\,\overline{ \widetilde{\chi}^{h,g'}_{\rho}}(k')\,\ket{k,g,h'}.
\end{aligned}
\ee
Equation \eqref{eq:DtwistedIDabba} shows that $\beta_{a,b}(c,d)=\beta_{b,a}(c,d)^{-1}$, $\forall ab=ba, c,d\in Z_{a,b}$, which implies that 
\be\label{eq:rhoRhoDag}
\widetilde\rho^{a,b}_\lambda=\big(\widetilde{\rho}^{b,a}_\lambda\big)^\dag.
\ee
We then infer that among the six bases, any basis and the one with the two flux labels exchanged are dual to each other. For an explicit example,
\be
\begin{aligned}
&\braket{B',A',\nu}{A,B,\mu}\\
=&\frac{1}{|G|}\sum_{\substack{g_1\in C^{B'},k_1\in C^{A'}_{Z_{g}}\\ h_1'\in Z_{g,k}}} \sum_{\substack{k\in C^A,g\in C^B_{Z_{k}}\\ h'\in Z_{g,k}}}\\
&\times\widetilde{\chi}^{g_1,k_1}_{\nu}(h_1')\widetilde{\chi}^{k,g}_{\mu}(h') \braket{k_1,g_1,h'_1}{k,g,h'}\\
=&\delta_{B'A}\delta_{A'B}\delta_{\nu\mu}\sum_{\substack{k\in C^A,g\in C^B_{Z_{k}}\\ h\in Z_{g,k}}}\frac{1}{|G|}\\
=&\delta_{B'A}\delta_{A'B}\delta_{\nu\mu}\sum_{k\in C^A}\frac{|Z_{k}|}{|G|} \sum_{g\in C^B_{Z_{k}}}\frac{|Z_{k,g}|}{|Z_{k}|}\sum_{h'\in Z_{k,g}}\frac{1}{|Z_{k,g}|}\\
=&\delta_{B'A}\delta_{A'B}\delta_{\nu\mu},
\end{aligned}
\ee
where the second equality acknowledges Eq. \eqref{eq:rhoRhoDag}. The linear relations between the six bases can be worked out similarly. Physically, these bases correspond to a kind of dimensional reduction of the $(3+1)$d topological phase into certain $(2+1)$d ones along different $3$d axes. Such dimensional reduction for Abelian $G$ has been studied\cite{Jiang2014,Wang2014}.

Back to $2+1$ dimensions again, the ground--state basis $\ket{A,\mu}$ labels the set of all inequivalent irreducible representation spaces of the TQD model defined by a $3$-cocycle $\alpha$.  The TQD also plays a central role in the orbifolds by a symmetry group $G$ of a holomorphic conformal field theory. The term quantum double reflects the charge-flux duality of the ground states in the $(2+1)$d model. Topologically, this duality is implied by the vertex-face duality  between the graph $\Gamma$ considered and the dual graph, as charge excitations live on the vertices of $\Gamma$, while the fluxes live on the faces. We may understand the term ``twisted'' as twisting the gauge transformation in the underlying topological gauge theory or as twisting the linear representations to projective representations of $Z^{A,B}\subseteq G$. 

In $3+1$ dimensions, however, there is a much richer variety of ground states and the excitations on top of them. A ground state $\ket{A,B,\mu}$ now has three indices. When either $A$ or $B$ is trivial, the corresponding excitation reduces to a dyon in $2+1$ dimensions. When both $A$ and $B$ are nontrivial, the excitation is loop instead of point like because $A$ and $B$ are bases of fluxes in two direction, such that the ground state is a membrane with fluxes integrated over the membrane surface. When the membrane is cut open to create a pair of excitations, the excitations are indeed loops bounding the open surfaces. There then lacks the familiar duality between charges and fluxes in $(3+1)$-d. Thus, the ground states $\ket{A,B,\mu}$ would cease to form a quantum-double like algebra. Although the precise mathematical structure of the ground states is important, it is not necessary for our understanding of the topological and certain physical properties of the states; hence, we shall not dwell on this problem but will report our study on it elsewhere.

It is time to answer another important question raised in the introduction section. That is, do the ground states and quasi-excitations still remain in one-to-one correspondence? They do not, in general. Typically, for any TGT with a finite Abelian group, the ground state basis $\ket{A,B,\mu}$ \emph{over-counts} the excitation types for the following reason. The ground state basis $| A, B, \mu \rangle$ on 3-torus has two flux indices $A,B$ and one charge index $\mu$. Nevertheless, in a DW theory with an Abelian gauge group, the number of pure string or loop excitations is determined by the flux type---only one flux index, and the number of pure particles is determined by the charge type---also one charge index. For example, the usual $\Z_{N}$ gauge theory has $N$ types of pure fluxes (string) and $N$ types of pure charges (particles), so the theory has $N^2$ distinct excitations in total. But the GSD of our model on 3-torus with $G=\Z_{N}$, by counting the independent labels in the basis, has GSD=$N^3$. For non-Abelian gauge groups, the situation is more subtle, in which we suspect the ground states are still more than the excitation species. But we shall seek for an affirmative answer in future work.

\section{Topological quantum numbers}\label{sec:fractionaltopologicalnumbers}
We have studied the GSD---the simplest topological observable---of our model; however,  GSD alone is not able to uniquely specifies a topological phase. Two models defined by two inequivalent $4$-cocycles may happen to own the same GSD but still describe two distinct topological phases.

An immediate challenge is how to distinguish two distinct topological phases with the same GSD. We thus should study other emergent topological quantum numbers that together with the GSD, can facilitate a finer classification such topological phases.

To realize the goal above, we shall first dwell on the subspace $\Hil^{B_f=1}$ to discover the topological observables pertinent to this space. Then we shall tackle the eigenvalue problems of these observables to obtain the expected topological quantum numbers. In 3d TGT model, these quantum numbers are also related to the braiding statistics of quasi-excitations, loops and/or particles. 

\subsection{$\text{SL}(3,\Z)$ generators as Topological observables}\label{sub:TopoObservableSL2Z}
As a reminder, we have formulated the mutation transformations, which are local unitary transformations  that can modify the local structure of the graph while preserving the global topology the graph triangulates. The mutation transformations thus leave the global degrees of freedom in the ground states intact. 

In contrast, our purpose here is to search for the large transformations that globally  vary the graph, without, however, affecting the topology. These transformations will provide us a richer class of topological observables. To seek for these large, global transformations, it suffices to work on the simplest triangulations of a $3$-torus, like those in Fig. \ref{fig:3torus}.

In three dimensions, the transformations we shall focus on are known to be the modular transformations forming the modular group $SL(3,\Z)$. This group is generated by
\be  \label{eq:SL2Z}
  \str=\bpm
     0 & 0 & 1 \\
     1 & 0 & 0 \\
     0 & 1 & 0
    \epm,
    \quad
  \ttr=\bpm
     1 & 1 & 0 \\
     0 & 1 & 0 \\
     0 & 0 & 1
    \epm.
\ee
Note that our generator $\ttr$ preserves the third dimension, or $\mathbf{z}$-direction by our convention. 
Any pure $2$d modular transformation can also be generated by the two matrices above. For example, 
\[
\str_{\mathbf{xy}}=\bpm
     0 & -1 & 0 \\
     1 & 0 & 0 \\
     0 & 0 & 1
    \epm=(\ttr^{-1}\str)^3(\str\ttr)^2\str\ttr^{-1}.
\]  

\begin{figure}[h!]
\centering
\includegraphics[scale=2]{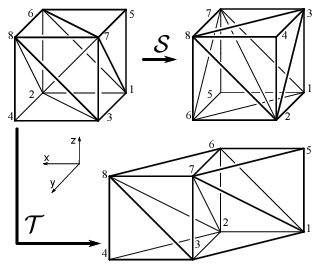}
\caption{$\str$ and $\ttr$ transformations of a $3$-torus.}
\label{fig:SL3Ztransformation}
\end{figure}

To evaluate the modular transformations in our ground states, such that their matrix elements are functions of the $4$--cocycle that defines the model, we redraw the 3-torus in Fig. \ref{fig:3torus} in the coordinate frame in Fig. \ref{fig:SL3Ztransformation}. Clearly, the coordinate system is accordant with the vertex eumerations in the graph. Figure \ref{fig:SL3Ztransformation} illustrates the $\str$ and $\ttr$ transformations on the 3-torus on the upper left portion of the figure.
We remark that Fig. \ref{fig:SL3Ztransformation} can be thought as the $\str$ and $\ttr$ transformations of the wavefunction of the model; hence, the $\str$ and $\ttr$ transformations on the corresponding basis vectors would appear to be opposite to those in the figure. 

We establish the $\str$ and $\ttr$ transformations on the subspace $\Hil^{B_f=1}$ in the following
(The details of the construction are saved to Appendix \ref{app:modularTrans}).   
Similar to the vertex operators, the action of $\str$ or $\ttr$ should be an average over its actions by different group elements of $G$, as in Eq. \eqref{eq:STequalSumSTx}. We claim that $\str^x$ acts as
\begin{align}
  \label{eq:SxtransformationEnumerations}
  &\str^x \BLvert\torusBeforeS{1}{5}{2}{6}{3}{7}{4}{8}{1.75}\Brangle
  \nonumber\\
  = &[8'2,23,34,48]^{-1}[8'2,23,37,78][8'2,26,67,78]^{-1}\x \nonumber \\
&\frac{[6'1,12,23,37][6'8',8'2,26',67][6'1,15,56,67]}{[6'8',8'2,23,37][6'1,12,26,67]}\x \nonumber \\
&\frac{[4'6',6'1,12,26]}{[4'6',6'8',8'2,26][4'6',6'1,15,56]}\x \nonumber \\
&[2'4',4'6',6'1,15] [7'8',8'2,23,34] \x \nonumber \\
&\frac{[5'6',6'8',8'2,23]}{[5'7',7'8',8'2,23][5'6',6'1,12,23]}\x \nonumber \\
&\frac{[3'5',5'7',7'8',8'2][3'4',4'6',6'8',8'2]}{[3'5',5'6',6'8',8'2][3'4',4'6',6'1,12]}\x \nonumber \\
&\frac{[3'5',5'6',6'1,12][1'3',3'4',4'6',6'1]}{[1'3',3'5',5'6',6'1][1'2',2'4',4'6',6'1]}\x \nonumber\\
  &\frac{[1'2',2'3',3'4',4'6']^2[1'2',2'3',3'6',6'7']}{[2'3',3'4',4'6',6'8']^2[2'3',3'6',6'7',7'8']} \nonumber\\
&\BLvert \torusAfterS{1'}{2'}{3'}{4'}{5'}{6'}{7'}{8'}{1.75}\Brangle.
\end{align}
Here we set the order of the enumerations by $1'<2'<3'<4'<5'<6'<7'<8'<1<2<3<4<5<6<7<8$, in order that the boundary edges are oriented consistently with the periodic boundary condition. 
Clearly, the vector in Eq. \eqref{eq:SxtransformationEnumerations} transforms in a way opposite to that in Fig. \ref{fig:SL3Ztransformation}. We remark that the triangulation on the LHS of the above equation is merely a sheared view of Fig. \ref{fig:3torusA}, simply for a better visualization of the $\str$ transformation.    

Substituting $8'8=6'7=4'6=2'5=7'4=5'3=3'2=1'1=x$ and the other relevant group elements in the above equations, and after a lengthy calculation, the action of $\str^x$ on the physical basis can be expressed as
\begin{align}
  \label{eq:SxPhysBasis}
&\str^x\ket{k,g,h'}\nonumber \\
=&\eta^{xk\bar x,xg\bar x}(xh'\bar x,h'gk\bar x)\nonumber \\
&\x\frac{\beta_{xh'g\bar x,xk\bar x}(x\bar kg\bar x,xk\bar x)}{\beta_{xk\bar x,xg\bar x}(xh'\bar x  ,xg\bar x)}\ket{xh'g\bar x,xk\bar x,x\bar kg\bar x},
\end{align}
which simply recasts Eq. \eqref{eq:SxReduced} in the physical basis.

Likewise, we find that $\ttr^x$ acts as
\begin{align}
  \label{eq:Txtransformation}
  & \ttr^x \BLvert\torusCubeNB{4}{3}{2}{1}{8}{7}{6}{5}{1.75}\Brangle
  \nonumber\\
  = &[7'2,23,34,48]^{-1}[7'2,23,37,78][7'2,26,67,78]^{-1}\x \nonumber\\
&\frac{[8''7',7'2,26,67][8''1,12,23,37][8''1,15,56,67]}{[8''7',7'2,23,37][8''1,12,26,67]}\x \nonumber \\
&\frac{[6'8'',8''1,12,26]}{[6'8'',8''1,15,56][6'8'',8''7',7'2,26]}\x \nonumber \\
&[5'6',6'8'',8''1,15] \x \nonumber \\
&[3'7',7'2,23,34] \x \nonumber \\
&\frac{[4''8'',8''7',7'2,23]}{[4''3',3'7',7'2,23][4''8'',8''1,12,23]}\x \nonumber \\
&\frac{[2'4'',4''3',3'7',7'2][2'6',6'8'',8''7',7'2]}{[2'4'',4''8'',8''7',7'2][2'6',6'8'',8''1,12]}\x \nonumber \\
&\frac{[2'4'',4''8'',8''1,12][1'2',2'6',6'8'',8''1]}{[1'2',2'4'',4''8'',8''1][1'5',5'6',6'8'',8''1]}\x \nonumber \\
 &\frac{[1'2',2'4'',4''8'',8''7']}{[1'2',2'6',6'8'',8''7']} \x\nonumber \\
&[1'5',5'6',6'8'',8''7']^2 \BLvert \torusAfterT{3'}{4''}{1'}{2'}{7'}{8''}{5'}{6'}{1.75}\Brangle,
\end{align}
where we order the enumerations as $1'<2'<4''<3'<5'<6'<8''<7'<1<2<3<4<5<6<7<8$, also consistent with the boundary condition. With $x=1'1=2'2=4''3=3'4 =5'5=6'6=8''7=7'8$, Eq. \eqref{eq:Txtransformation} takes the following form in terms of the group elements.
\be\label{eq:TxPhysBasis}
\begin{aligned}
&\ttr^x\ket{k,g,h'}\\
=&\eta^{xk\bar x,xg\bar x}(xh'\bar x,h'gk\bar x)\\ 
&\x\beta_{xk\bar x,xg\bar x}(xh'\bar x,xg\bar x)\ket{xkx,xg\bar x,xgh'\bar x},
\end{aligned}
\ee
which rewrites the very Eq. \eqref{eq:TxNatBasis} in the physical basis.

The $\str$  and $\ttr$ transformations are then defined as
\be
\label{eq:STequalSumSTx}
\str=\frac{1}{|G|}\sum_x \str^x,
\quad \ttr=\frac{1}{|G|}\sum_x \ttr^x.
\ee
The two operators above represent the modular $\str$ and $\ttr$ matrices in Eq. \eqref{eq:SL2Z} on the 
subspace $\mathcal{H}^{B_f=1}$ of our model. 
To show that the $\str$ and $\ttr$ defined above are indeed topological observables and symmetry transformations on the ground states, one need to first sum the expressions \eqref{eq:SxPhysBasis} and \eqref{eq:TxPhysBasis} over $x\in G$ and then study their actions on the ground states. Fortunately, two properties of the ground states save our labor. First, Eq. \eqref{eq:GGquotConj} indicates that a ground state basis vector is an average over all possible simultaneous conjugation of all the group elements involved. Second, the Fourier transformed basis $\ket{A,B,\mu}$ is also invariant under simultaneous conjugation of all the group elements in the Fourier transform, which is a consequence of Proposition \ref{prop:isoZab}. Now that $\str$ or $\ttr$ not only introduce simultaneous conjugation of all the group elements in a ground state basis vector but also average over all such conjugation, this average can be absorbed into a redefinition of the basis vector. Therefore, as far as the ground states on a $3$-torus are concerned, the action of $\str$ or $\ttr$  should be the same as the action of $\str^x$ or $\ttr^x$ for an arbitrary $x\in G$. It is then legal and sufficient to choose simply $x=1\in G$ to serve our purposes. Precisely, this means
\be\label{eq:STsimple}
\begin{aligned}
&\str\ket{A,B,\mu}=\str^1\ket{A,B,\mu},\\
&\ttr\ket{A,B,\mu}=\ttr^1\ket{A,B,\mu}.
\end{aligned}
\ee      

On the ground-state basis comprised of the eigenvectors $\{\Phi_K\}$ of $\ttr$,
we have\be
\label{eq:GSintermsofT}
\ttr\ket{\Phi_K}=\theta_K \ket{\Phi_K},\quad K=1,2,...,\text{GSD,}
\ee
where $\theta_K$ wil be shown to be a $U(1)$ phase. Later, we will be able to identify the eigenvectors $\Phi_k$  with the ground-state basis vectors $\ket{A,B,\mu}$ we discovered earlier.

It turns out that other eigenvectors of $\ttr$ also have vanishing eigenvalue. These eigenvectors in fact correspond to the excited states in our model. We shall report our study of them elsewhere.

The eigenvalues $\theta_K$ of $\ttr$ thus sever as a set of topological numbers of our model. Staring at Fig. \ref{fig:SL3Ztransformation}, one can see that $\ttr$ performs a global twisting of the graph, so it is reasonable to regard the eigenvalues $\theta_K$ as the topological spins of the topological states $\ket{\Phi_k}$.

Similarly, the modular $\str$ operator also offers a set of topological quantum numbers, which are its matrix elements evaluated on the topological states, namely,\be
\label{eq:smatrix}
S_{IJ}=\left\langle\Phi_I \right| \str\ket{\Phi_J},\quad I,J=1,2,...,\text{GSD}.
\ee
The matrix above is orthonormal:
\be \label{eq:Smatrixcondition}
  \sum_{J}S_{IJ}\overline{S_{JK}}=\delta_{IK}.
\ee
Remarkably, we have derived a representation of the $3$-dimensional modular $\str$ and $\ttr$ matrices, purely based on our model and in terms of $4$-cocycles of $G$, without, however, relying on any group representation theory. 

Putting the pieces together, we now have at our disposal a rich set of topological quantum numbers,  $\{\text{GSD},\theta_K,S_{IJ}\}$, to classify the topological phases described by our model more accurately.
Yet, we do not have the explicit expressions of the quantum numbers $\theta_K$ and $S_{IJ}$. This is our next task.
\subsection{$3$-torus modular $\str$ and $\ttr$ matrices}
We now explicitly solve for the topological quantum numbers $\{\theta_K,S_{IJ}\}$. We stress that although the topological observables $\str$ and $\ttr$ are defined on the subspace $\Hil^{B_f=1}$, we shall solve their eigenvalue problems within $\Hil^0\subset \Hil^{B_f=1}$. The reason is that the eigenvectors on $\ttr$ with finite eigenvalues all lie in the ground-state space. 

Firstly, we diagonalize the $\ttr$ matrix in Eq. \eqref{eq:STequalSumSTx} on the ground states. In Section \ref{subsec:GSbasis}, we have learned that the ground states are superpositions of the orthonormal basis vectors $\ket{A,B,\mu}$ defined in Eq. (\ref{eq:AmuBasis}). It turns out that fortunately, the basis vectors $\ket{A,B,\mu}$ are the sought-after eigenvectors of $\ttr$. Appendix \ref{app:solModularTrans} offers a step-by-step proof of this claim. Below, we simply present the result. Namely, the modular $\ttr$ acts on the $\ket{A,B,\mu}$ basis  as
\begin{align}
  \label{eq:TonAmu}
  &\ttr\ket{A,B,\mu}
  \nonumber\\
  =&\ttr^1\ket{A,B,\mu}
  \nonumber\\
  =&\frac{\overline{\widetilde{\chi}^{k^A,g^B}_{\mu}}(g^B)}{\text{dim}_{\mu}} \ket{A,B,\mu},
\end{align}
where $\text{dim}_{\mu}$ is the dimension of the representation $\mu$, and $g^B$ is an arbitrary representative of $C^B_{Z_k}$. Each basis vector $\ket{A,B,\mu}$ is thus an eigenvector of $\ttr$, with eigenvalue
\be\label{eq:Teigenvalue}
\theta^{A,B}_{\mu}=\frac{\overline{\widetilde{\chi}^{k^A,g^B}_{\mu}}(g^B)}{\text{dim}_{\mu}} .
\ee
This hints a physical meaning of the projective characters as the ground-state wave-functions of our model that are collapsed in $\ttr$'s eigenstates.
The formula \eqref{eq:Teigenvalue} also leads to interpreting $\theta^{A,B}_{\mu}$ mathematically
as an invariant that characterizes the representation $\widetilde{\rho}^{k,g}_\mu$. 
This precisely reconciles Eq. \eqref{eq:rhogg} and our proof of it. We thus name $\theta^{A,B}_{\mu}$ \emph{topological spins}. 

The formula \eqref{eq:Teigenvalue} of the topological spins may appear to be abstract still. Interestingly, they topological spins $\theta^{A,B}_{\mu}$ can be expressed directly in terms of the twisted $3$-cocycles $\alpha_{AB}$ as
\be\label{eq:thetaomegaA}
(\theta^{A,B}_{\mu})^{p_B}=\alpha_{AB}
\ee
defined by
\begin{align}
  \label{eq:alphaAB}
  \alpha_{AB}\defeq\prod_{n=0}^{p_B-1}[g^B,(g^B)^n,g^B]_{k^A}
\end{align}
for conjugacy classes $C^A$ of $G$, $C^B_{Z^A}$ of $Z^A$. The number $p_B$ is the degree of $g^B$. It is the least integer such that the $p_B$-th power of $g^B$, $(g^B)^{p_B}=1$. The twisted $3$-cocycle $\alpha_{AB}$ is independent of the the representative $g^B\in C^B_{Z^A}$ and is thus a function of conjugacy classes. The formulae \eqref{eq:thetaomegaA} and \eqref{eq:alphaAB} are a result of repeated applications of Eq. \eqref{eq:betarepresentation} to $(\widetilde{\rho}^{A,B}_{\mu}(g^B))^{p_B}=(\theta^{A,B}_{\mu})^{p_B}\mathds{1}$. The topological spins $\theta^{A,B}_{\mu}$ therefore take value in the $p_B$--th roots of $\alpha_{AB}$. 
Furthermore, each of the $p_B$ distinct $p_B$--th roots appears exactly $r(Z^{A,B,}\beta_{{k^A},g^B})/p_B$ times in $\{\theta^{A,B}_{\mu}\}$ for all representations $\mu$. To verify that the ratio $r(Z^{A,B,}\beta_{{k^A},g^B})/p_B$ is indeed an integer, we first observe that each element of $Z_{k,g}$ can be uniquely written as ${g^n}h$ with integer $n$ between $0$ and $p_B$ for some $h\in Z_{k,g}$. Second, $Z_{k,g}$ has $p_B$ 1--dimensional representations, namely, $\rho^j({g^n}h)=\exp(2\pi\ii{j}{n}/p_B)$. Hence, for each representation $\mu$, there exists a representation $\mu'$, such that $\widetilde{\rho}^{A,B}_{\mu'}(g^nh)=\rho^j(g^nh)\widetilde{\rho}^{A,B}_{\mu}(g^nh)$. Therefore, the relation $\theta^{A,B}_{\mu'}=\exp(2\pi\ii{n}/p_B)\theta^{A,B}_{\mu}$ holds.

Appendix \ref{app:solModularTrans} evaluates the matrix elements of the modular $\str$ operator in the eigenvectors of the $\ttr$ matrix and expresses them in terms of the projective characters as well. 
While leaving the details to the appendix, here we show the final formula of the $\str$ matrix elements:\be  \label{eq:SonAmuBnu}
\begin{aligned}
& S_{(AB\mu)(A'B'\nu)}\\
=&\left\langle A,B,\mu\right|\str \ket{A',B',\nu}\\
=&\frac{{\theta^{A,B}_{\mu}}^*\theta^{A',B'}_{\nu}} {|G|}
\sum_{\substack{k\in C^A\cap Z_{g,h'},\\ g\in C^B_{Z_k}\cap C^{A'}\\h'\in Z_{k,g}\cap C^{B'}_{Z_g}}} \widetilde{\chi}^{g,h'}_{\nu}(k)\overline{\widetilde\chi^{k,g}_\mu}(h').
\end{aligned}
\ee

Similar to the $\ttr$ matrix eigenvalues, the $\str$ matrix elements are also invariants carried by the projective representations of $Z^{A,B}$ for any $A$ and $B$. Since these invariants are functions of the projective characters, which are specified by the doubly-twisted $2$-cocycles $\beta_{k,g}$, the role of these $2$-cocycles cannot be overestimated. Unfortunately, as aforementioned, we still lack the knowledge whether the projective $\beta_{k^A,g^B}$-representations of all $Z^{A,B}$ in $G$ and any given \fc of $G$ can be lifted to certain irreducible linear representations of certain algebraic structure. In other words, do the ground states of our model carry irreducible linear representations of the algebraic structure? But we conjecture that they do, and the algebraic structure may be a $2$-category\footnote{Private communications with Liang Kong and Yuting Hu.}.

We expect that these irreducible representations would still classify the particle and loop quasi-excitations. 
The invariants of each irreducible representation identifies the topological quantum numbers of the corresponding quasi-excitation. In $2$-dimensions, the $\str$--matrix has the origin as a braiding operation that exchanges any two of these quasi-excitations, while the $\ttr$--matrix contains the statistical spins of the corresponding quasiparticles which are determined by the braiding operation. We expect this to hold for $3$-dimensions as well. This is partially verified via a dimensional-reduced approach.\cite{Wang2014c,Jiang2014a} In the sense that for those $3$-dimensional quasi-excitations that can be viewed as closed loops of $2$-dimensional quasi-excitations, their $\str$ and $\ttr$ matrices can be block-diagonalized into the $\str$ and $\ttr$ matrices of the reduced $2$-dimensional quasi-excitations. Nevertheless, there exist purely $3$-dimensional quasi-excitations in our model that admits no canonical dimensional-reduction into $2$-dimensional anyons. The topological and physical meanings of the $\str$ and $\ttr$ matrices of such quasi-excitations are an important problem to solve, and we shall report our results elsewhere.
\section{Classify the topological phases}\label{sec:classification}
We believe that the topological phases are classified by the set of topological quantum numbers $\{\mathrm{GSD},\theta^{A,B}_{\mu},s_{(AB\mu),(A'B'\nu)}\}$. In all examples to be discussed in Section \ref{sec:examples}, naively one may think those TGT models are classified by the group $G$ and the fourth cohomology classes of $\omega$. This is wrong, however. Even if $\omega$ and $\omega'$ are \emph{inequivalent}, the two models $H_{G,\omega}$ and $H_{G,\omega'}$ may still yield the same topological observables and hence the same topological phase and.\cite{Wang2014c} On the other hand, as seen in Section \ref{subsec:equivModel}, if $\omega$ and $\omega'$ are equivalent, so must be the two models $H_{G,\omega}$ and $H_{G,\omega'}$.

In this section, we offer to some extent the general classifications of the topological quantum numbers.

\subsection{When the $4$--Cocycle is cohomologically trivial}\label{subsec:trivial3cocycle}
We have demonstrated in Section \ref{subsec:equivModel} that two TGT models defined by two equivalent $4$--cocycles describe the same topological phase. We now examine such topological phases closely.

The simplest case arises when the 4--cocycle $\omega$ defining our model is cohomologically trivial. That is, the $\omega$ is equivalent to the trivial $4$--cocycle $\omega^0=1$. This enables us to write the $\omega$ as a 4--coboundary:
\be
\label{eq:4couboundary}
\begin{aligned}
\omega(a,b,c,d)=&\delta\alpha(a,b,c,d)\\
=&\frac{\alpha(b,c,d)\alpha(a,bc,d)\alpha(a,b,c)}{\alpha(ab,c,d)\alpha(a,b,cd)},
\end{aligned}
\ee
where $\alpha(x,y,z)$ is any normalized $3$-cochain, a function $\alpha:G\x G\x G\rightarrow U(1)$ that satisfies $\alpha(1,y,z)=\alpha(x,1,z)=\alpha(x,y,1)=1$ for all $x,y,z\in G$. In this case, as we will show, the TGT model is untwisted, or in other wors, it reduces to a usual gauge theory model.

Induced by the $\omega$ in Eq. \eqref{eq:4couboundary}, the twisted $3$-cocycle $\alpha_a$, defined in Eq. \eqref{eq:twisted3cocycleDef}, is automatically trivial because it turns out to be a twisted $3$-coboundary:
\be
\alpha_a(b,c,d)=\widetilde{\delta}\beta_a(b,c,d),
\ee
where
\be\label{eq:betaDue2Equiv4cocycle}
\beta_a(x,y)=\frac{[a,x,y][x,y,\overline{xy}axy]}{[x,\bar x ax,y]}
\ee
is a twisted $2$-cochain, as the slant product of certain $3$-cocycle of $G$. The twisted $3$-coboundary $\widetilde{\delta}\beta_a$ of $\beta_a$ reads
\be\label{eq:twisted3coboundaryDue2Equiv4cocycle}
\widetilde{\delta}\beta_a(b,x,y)=\frac{\beta_{\bar b ab}(x,y)\beta_a(b,xy)}{\beta_a(bx,y)\beta_a(b,x)},
\ee
for all $a,b,x,y \in G$. This confirms that $\alpha_a(b,c,d)\sim\ 1$. Clearly, the $\beta_a$ defined in Eq. \eqref{eq:betaDue2Equiv4cocycle} is normalized. 

As far as ground states are concerned, we need only to focus on the doubly-twisted $2$-cocycle $\beta_{a,b}$ defined in Eq. \eqref{eq:Dtwisted2cocycleDef} that plays the key role in the topological quantum numbers of the ground states. Restricted to $\Hil^0$, since we have $ab=ba$ for $\beta_{a,b}$, the triviality of $\alpha_a(b,c,d)$ implies that $\beta_{a,b}$ is also automatically trivial because it takes the form of a doubly-twisted $2$-coboundary:
\be\label{eq:Dtwisted2cocyleDue2Equiv4cocycle}
\beta_{a,b}(c,d)=\widetilde\delta\epsilon_{a,b}(c,d),
\ee 
where
\be\label{eq:Dtwisted1cochainDue2Equiv4cocycle}
\epsilon_{a,b}(x)=\frac{\beta_a(x,\bar xbx)}{\beta_a(b,x)}
\ee
is the doubly-twisted $1$-cochain derived from the twisted $2$-cochain in Eq. \eqref{eq:betaDue2Equiv4cocycle}. The corresponding doubly-twisted $2$-coboundary reads
\be\label{eq:Dtwisted2coboundaryDue2Equiv4cocycle}
\widetilde\delta\epsilon_{a,b}(x,y)=\frac{\epsilon_{\bar xax,\bar xbx}(y)\epsilon_{a,b}(x)}{\epsilon_{a,b}(xy)}.
\ee
Since $\beta_a$ is normalized, so is $\epsilon_{a,b}$. Equations. \eqref{eq:betaDue2Equiv4cocycle} and \eqref{eq:Dtwisted1cochainDue2Equiv4cocycle} also imply the following identities
\be\label{eq:epsilonConst}
\begin{aligned}
\epsilon_{a,b}(a)=\epsilon_{a,b}(b)=1,\, &\forall ab=ba,\\
\epsilon_{a,b}(c)\epsilon_{a,c}(b)=1,\, &\forall cb=bc,\\
\epsilon_{a,b}(c)\epsilon_{c,b}(a)=1,\, &\forall c\in Z_{a,b},\\
\epsilon_{a,b}(c)\epsilon_{b,c}(a)^{-1}=1,\, &\forall c\in Z_{a,b}.
\end{aligned}
\ee 
Note that the above identities are clearly independent of which $\omega$ is chosen in its equivalent class, and thus are constantly true for a given model in this case.

Substituting the $\beta_{a,b}$ in Eq. \eqref{eq:Dtwisted2cocyleDue2Equiv4cocycle} into definition \eqref{eq:betarepresentation}, we immediately see that the irreducible projective $\beta_{a,b}$-representations $\widetilde\rho^{a,b}$ of $Z_{a,b}$ become proportional to linear representations $\rho^{a,b}$, namely 
\be
\label{eq:BetaRepresToLinearRepres}
\widetilde{\rho}^{k,g}_{\mu}(h')=\epsilon_{k,g}(h')\rho^{k,g}_{\mu}(h'),\,\forall h'\in Z_{k,g}.
\ee

The ground-state basis (\ref{eq:AmuBasis}) then becomes
\be
\label{eq:AmubaisTrivialAlpha}
\ket{A,B,\mu}=\frac{1}{\sqrt{|G|}}\sum_{\substack{k\in C^A,g\in C^B_{Z_k}\\h'\in Z_{k,g}}}\epsilon_{k,g}(h')\chi^{k,g}_{\mu}(h')\ket{k,g,h'},
\ee
where $\chi^{k,g}_{\mu}=\mathrm{tr}\rho^{k,g}_{\mu}$ is the usual linear character.

Eq \eqref{eq:Dtwisted2cocyleDue2Equiv4cocycle} and \eqref{eq:Dtwisted2coboundaryDue2Equiv4cocycle} indicate that $\beta_{a,b}(x,y)=\beta_{b,a}(y,x)$ for all $x,y\in Z_{a,b}$ satisfying $xy=yx$. Consequently, all elements in $Z_{k,g}$ are $\beta_{k,g}$-regular, and $\eta^{k,g}(h',x)\equiv 1$ holds for all $x\in Z_{k,g,h'}$. The GSD formula (\ref{eq:GSDbeta}) therefore reduces to
\begin{align}
\label{eq:GSDforTrivialAlpha}
  \text{GSD}=\sum_{\substack{k\in G\\g\in Z_k}}\sum_{\substack{h'\in Z_{k,g}\\x\in Z_{k,g,h'}}}\frac{1}{|G|} =\left|\frac{\text{Hom}(\pi_1(T^3),G)}{conj}\right|,
\end{align}
where we quotient by the equivalence relation $(k,g,h')\sim (xk\bar x,xg\bar x,xh\bar x)$, for any $x\in G$.

Using \eqref{eq:BetaRepresToLinearRepres} and the identities \eqref{eq:epsilonConst} of the $\epsilon_{a,b}$, we can rewrite the topological numbers $\theta^A_{\mu}$ and the $\str$ matrix 
as
\be
\label{eq:TTrivialAlpha}
\theta^{A,B}_{\mu}
=\frac{\widetilde{\chi}^{k^A,g^B}_{\mu}(g^B)}{\text{dim}_{\mu}}
=\frac{\chi^{k^A,g^B}_{\mu}(g^B)}{\text{dim}_{\mu}}
\ee
and
\be  \label{eq:STrivialAlpha}
\begin{aligned}
s_{(AB\mu)(A'B'\nu)}=\frac{{\theta^{A,B}_{\mu}}^*\theta^{A',B'}_{\nu}}{|G|}\sum_{\substack{k\in C^A\cap Z_{g,h'},\\ g\in C^B_{Z_k}\cap C^{A'}\\h'\in Z_{k,g}\cap C^{B'}_{Z_g}}}
  \chi^{g,h'}_{\nu}(k)\overline{\chi^{k,g}_{\mu}}(h').
\end{aligned}
\ee

If the \fc\ $\omega$  is precisely the trivial $\omega^0=1$, the ground states happen to carry the irreducible \textit{linear} representations of all the centralizers $Z^{A,B}\subseteq G$.
If $\omega\in [\omega^0]$ but $\omega\neq \omega^0$, since $\beta_{k,g}\neq 1$, the ground states carry irreducible \textit{projective} representations of all the centralizers $Z^{A,B}\subseteq G$. But because these projective representations are proportional to the linear representations up to a phase, all topological quantum numbers are the same as those in the case of $\omega^0$. This complies with the fact that equivalent \fcs define equivalent TGT models.  So effectively in this case, only linear representations appear, and the TGT models are equivalent to the model without being twisted by any \fc.

\subsection{When the doubly twisted $2$--cocycle is cohomologically trivial}\label{subsec:trivialBeta}

A $4$-cocycle $\omega\notin [\omega^0]$ could still be regarded ``trivial" at a deeper level. This is understood in the case where the doubly twisted $2$-cocycle $\beta_{a,b}$ induced by the $\omega$ is cohomologically trivial by accident. That is, the induced $\beta_{a,b}$ is merely a twisted $2$--coboundary:
\be\label{eq:BetaToEpsilon}
  \beta_{a,b}(c,d)=\widetilde\delta\epsilon_{a,b}(c,d),
\ee
for all $a,b,c,d\in G$. Beware that however, the doubly twisted $1$--cochain $\epsilon_{a,b}$ here ceases to have the closed form in Eq. \eqref{eq:Dtwisted1cochainDue2Equiv4cocycle} in general because $\omega$ is not cohomologically trivial. The identities \eqref{eq:epsilonConst} thus do not hold in general either. By Eqs. \eqref{eq:Dtwisted2cocyleDue2Equiv4cocycle} and \eqref{eq:Dtwisted2coboundaryDue2Equiv4cocycle}, it is then easy to show that in the current case, the $\eta^{k,g}$ defined in Eq. \eqref{eq:eta} turns out to be
\be\label{eq:EtaForTrivialBeta}
  \eta^{k,g}(h',x)=\frac{\epsilon_{xk\bar x,xg\bar x}(xh'\bar x)}{\epsilon_{k,g}(h')}
\ee
for all $h'\in Z_{k,g}$ and $x \in G$. We obviously have $\eta^{k,g}(h',x)=1$, $\forall x\in Z_{k,g,h'}$.

The ground--state subspace in the current case are also spanned by the basis vectors $\ket{A,B,\mu}$ of the form in Eq. \eqref{eq:AmubaisTrivialAlpha}, where $\mu$ labels the $\beta_{k,g}$-representations $\widetilde{\rho}^{k,g}_{\mu}$ of $Z_{k,g}$. These projective representations are yet proportional to the linear representations $\rho^{k,g}_{\mu}$, as in Eq. \eqref{eq:BetaRepresToLinearRepres}.

Because Eq. \eqref{eq:EtaForTrivialBeta} makes all the elements of $Z_{k,g}$ $\beta_{k,g}$-regular, the GSD formula here simply copies Eq. \eqref{eq:GSDforTrivialAlpha}.

By Eq. \eqref{eq:BetaRepresToLinearRepres}, the topological spins $\theta^{A,B}_{\mu}$ and the $\str$ matrix elements are related to $\epsilon_{k,g}(h')$ by,
\be
\label{eq:TTrivialBeta}
\theta^{A,B}_{\mu}
=\epsilon_{k^A,g^B}(g^B)\frac{\chi^{k^A,g^B}_{\mu}(g^B)}{\text{dim}_{\mu}}
\ee
and
\begin{align}  
s_{(AB\mu)(A'B'\nu)}=&\frac{{\theta^{A,B}_{\mu}}^*\theta^{A',B'}_{\nu}}{|G|}\sum_{\substack{k\in C^A\cap Z_{g,h'},\\ g\in C^B_{Z_k}\cap C^{A'}\\h'\in Z_{k,g}\cap C^{B'}_{Z_g}}}\label{eq:STrivialBeta}\\ &\x  \chi^{g,h'}_{\nu}(k)\overline{\chi^{k,g}_{\mu}}(h')\,\epsilon_{g,h'}(k)\epsilon_{k,g}(h')^{-1}.\nonumber
\end{align}

As opposed to the GSD formula that coincides with that of the $[\omega^0]$-model, the topological spinss $\theta^{A,B}_{\mu}$ and the $\str$ matrix differentiate the current model from the $[\omega^0]$-model. Equation \eqref{eq:TTrivialBeta} implies the physical significance of the phases $\epsilon$: They endow each ground-state basis vector $\ket{A,B,\mu}$ with an extra spin factor $\epsilon_{k^A,g^B}(g^B)$, other than that in Eq. \eqref{eq:TTrivialAlpha}.

\subsection{Classifying the topological numbers}\label{subsec:GeneralCase}
Because our model is defined by a $4$-cocycles $\omega$, all the topological quantum numbers of the model should depend on $\omega$ in one way or another. Indeed, seen in the special cases discussed in the previous subsection, some topological quantum numbers may not depend on $\omega$ directly or explicitly but via the quantities derived from $\omega$, such as the equivalence class of $\omega$ and the induced doubly-twisted $2$-cocycles. Below, we remark three general characteristics of the topological quantum numbers $\{\text{GSD},\theta^{A,B}_{\mu},s_{(AB\mu)(A'B'\nu)}\}$ of our model on a $3$-torus.
\begin{enumerate}
\item
The set $\{\text{GSD},\theta^{A,B}_{\mu},s_{(AB\mu)(A'B'\nu)}\}$ of a TGT model defined by an $\omega$ depends on the equivalence class $[\omega]\in H^4(G,U(1))$.
\item The GSD of a TGT model depends only on the equivalence classes $[\beta_{k^A,g^B}]\in H^2(Z^{A,B},U(1))$, independent of the representatives $k^A$ of $C^A$ and $g^B$ of $C^B_{Z_{k^A}}$.
\item The topological spins $\{\theta^{A,B}_{\mu}\}$ are classified by $\{r(Z^{A,B},\beta_{k^A,g^B}),\alpha_{AB}\}$, where $\alpha_{AB}$ is defined in Eq. \eqref{eq:alphaAB}.
\end{enumerate}
As follows, we elucidate these characteristics in order.

To show characteristic 1, one can verify that any two equivalent $4$-cocycles $\omega'$ and $\omega$ related by by Eq. \eqref{eq:4couboundary} induce two equivalent doubly-twisted $2$-cocycles $\beta'_{a,b}$ and $\beta_{a,b}$, which differ by a doubly-twisted $2$-coboundary: 
\be\label{eq:equivTbetaDue2Equiv3cocycle}
\beta'_{a,b}(c,d)=\beta_{a,b}(c,d)\widetilde{\delta}\epsilon_{a,b}(c,d),
\ee
where the doubly-twisted $2$-coboundary $\widetilde{\delta}\epsilon$ appears to be either that in Eq. \eqref{eq:Dtwisted1cochainDue2Equiv4cocycle} or the one in Eq. \eqref{eq:Dtwisted2coboundaryDue2Equiv4cocycle}.
Equations \eqref{eq:Dtwisted2coboundaryDue2Equiv4cocycle} and \eqref{eq:epsilonConst} lead to the equalities $\text{GSD}'=\text{GSD}$, ${\theta'}^{A,B}_{\mu}={\theta}^{A,B}_{\mu}$, and $s'_{(AB\mu)(A'B'\nu)}=s_{(AB\mu)(A'B'\nu)}$.

Characteristic 2 is a result of Eq. (\ref{eq:GSDconjugacy}), where the GSD sums the numbers $r(Z^{A,B},\beta_{k^A,g^B})$ over all $C^A$ of $G$ and $C^B$ of $Z^A$. This can also be corroborated in an alternative way. We first observe that in Eq. (\ref{eq:GSDbeta}), the GSD sums the phases $\rho^{k,g}$, $1$-dimensional representations of $Z_{k,g,h'}$ with $h'\in Z_{k,g}$. Each $\rho^{k,g}$ is a function of the $\beta_{k,g}$ defined in Eq. (\ref{eq:rhoEta}).    Assuming there exists two \fcs $\omega$ and $\omega'$ that may be inequivalent but still induce $\beta_{a,b}$ and $\beta'_{a,b}$, which are equivalent for all $a,b\in G$ in the following sense. There exists a normalized doubly-twisted $1$-cochain $\epsilon_{a,b}:G\rightarrow U(1)$ indexed by $a,b\in G$, which obeys $\epsilon_{a,b}(1)=1$, such that $\beta_{a,b}$ and $\beta'_{a,b}$ differ by just the doubly-twisted $2$--coboundary of $\epsilon_{a,b}$, i.e.,
\be\label{eq:BetaBetaEpsilon}
  \beta'_{a,b}(x,y)=\frac{\epsilon_{a,b}(x)\epsilon_{\bar xax,\bar xbx}(y)}{\epsilon_{a,b}(xy)} \beta_{a,b}(x,y),
\ee
for all $a,b,x,y \in G$. Restricting the above equation to $x\in Z_{a,b}$ and $y\in Z_{a,b,x}$, $\beta'_a$ and $\beta_a$ become equivalent up to a usual $2$-coboundary over $Z_{a,b}$. By Eq. (\ref{eq:BetaBetaEpsilon}) and Eq. \eqref{eq:GSDbeta}, we have
\[
\rho'^{k,g}(h',x)=\rho^{k,g}(h',x),
\]
which confirms that $\mathrm{GSD}'=\mathrm{GSD}$. Moreover, according to Eq. (\ref{eq:SumBetaToDelta}), the sum of $\rho^{k,g}(h',x)$ over $Z_{k,g,h'}$ is either one or zero, regardless of the representatives $k\in C^A$ and $g\in C^B_{Z_k}$. Hence, {as expected from the analysis in Section \ref{subsec:equivModel}, characteristic 2} is true.

Based on Sec.\ref{sec:fractionaltopologicalnumbers}, the topological spins $\theta^{A,B}_{\mu}$ for given $A$ and $B$ are the distinct roots of the $3$-cocycle $\alpha_{AB}$ in Eq. \eqref{eq:alphaAB} that is a function of only the classes $C^A$ and $C^B_{Z^A}$. Besides, each of the $p_B$-distinct $p_B$-th roots occurs exactly $r(Z^{A,B},\beta^{k^A,g^B})/p_B$ times in  the set $\{\theta^{A,B}_{\mu}\}$ for all representations $\mu$. Therefore, characteristic 3 holds.
\section{$3+1$-dimensional examples} \label{sec:examples}
This section provides various examples of our model with finite Abelian groups, in particular a systematic study of 4-cocycle for a product of cyclic groups, $G=\Z_{N_1} \times \Z_{N_2} \times \Z_{N_3} \times \Z_{N_4} \times \dots $. Here the 4-cocycle is denoted by $\omega=\omega(a,b,c,d)$ with $a,b,c,d\in G$. Since any finite Abelian group is isomorphic to a product of cyclic groups, this example is generic for finite Abelian groups.

In Sec. \ref{subsec:Zn1n2n3}, we consider the Abelian TGT model with a finite Abelian $G$ that still exhibits
{\bf Abelian statistics}. This happens when there are two or three (less than four) cyclic groups $\Z_{N_1}$, $\Z_{N_2}$, $\Z_{N_3}$ each contributing 
one or two group elements to the 4-cocycle.
Such Abelian statistics are examined recently.\cite{Wang2014b,Jiang2014a,Wang2014c}  
In Sec. \ref{subsec:Zn1n2n3n4}, we study the lattice TGT model, when the Abelian TGT model of a finite Abelian $G$ produces
{\bf non-Abelian statistics}. This occurs when there are four cyclic groups $\Z_{N_1}$, $\Z_{N_2}$, $\Z_{N_3}$, $\Z_{N_4}$ each contributing 
a group element to the 4-cocycle.
Such phenomena are first examined in Ref.\onlinecite{Wang2014c}.
In Sec. \ref{subsec:Zmn}, we discuss the cohomology group and cocycles from $G=\Z_m^n$ or more generally $G=\prod_i \Z_{N_i}$.

We would remind the reader that the irreducible projective $\beta_{a,b}$-representations 
are dictated by the doubly-twisted 2-cocycle defined in Eq. \eqref{eq:betaDef}. Its dependence on the \fc $\omega$ is detailed in Appendix \ref{app:Rewrite3torusAx}.
The irreducible projective $\beta^{}_{a,b}$-representation means Eq. (\ref{eq:betarepresentation}): 
\begin{align}
\label{eq:betaabRep}
\widetilde{\rho}_{\alpha}^{a,b}(c)\widetilde{\rho}_{\alpha}^{a,b}(d)=\beta^{}_{a,b}(c,d)\widetilde{\rho}_{\alpha}^{a,b}(c  d),   
\end{align}
where we reiterate $\widetilde{\rho}_{\alpha}^{a,b}(c)$, the irreducible projective $\beta^{}_{a,b}$-representation.

As studied in Section \ref{sec:classification}, when $\beta^{}_{a,b}(c,d)$ is a 2-coboundary, each corresponding projective representation $\widetilde{\rho}^{a, b}_{\alpha}(c)$ is proportional to a linear-representation ${\rho}^{a, b}_{\alpha}(c)$. According to the proportionality \eqref{eq:BetaRepresToLinearRepres}, we can divide the projective representation $\widetilde{\rho}^{a, b}_{\alpha}(c)$ into two parts:
\begin{eqnarray} \label{eq:projectTOlinear}
\text{projective Rep} &:& {\epsilon}_{a, b}^{}(c)  {\epsilon}_{a, b}^{}(d)= \beta^{}_{a, b}(c,d) {\epsilon}_{a, b}^{}(c d), \nonumber\\
\text{linear Rep} &:& {\rho}_{\alpha}^{a, b}(c) {\rho}_{\alpha}^{a, b}(d)= {\rho}_{\alpha}^{a, b}(c d).  
\end{eqnarray}
We repeat here that $\epsilon_{a, b}(c)$ is a doubly-twisted $1$-cochain. The $\epsilon_{a, b}(c)$ has a closed form \eqref{eq:Dtwisted1cochainDue2Equiv4cocycle} when the \fc $\omega$ is trivial; otherwise, it does not. As we find: Type II and III \fcs are cohomologically nontrivial; however, the induced $\beta_{a,b}(c,d)$ is yet a $2$-coboundary, i.e., cohomologically trivial. On the other hand, Type IV \fcs\ are also cohomologically nontrivial but the induced $\beta_{a,b}(c,d)$ is nontrivial and thus not a $2$-coboundary. As such, Eq. \eqref{eq:projectTOlinear} works for Type II and Type III 4-cocycle to be introduced in Sec. \ref{subsec:Zn1n2n3}.
But it does not work for Type IV 4-cocycle to be introduced in Sec. \ref{subsec:Zn1n2n3n4}.

\subsection{$G=\Z_{N_1} \times \Z_{N_2} \times \Z_{N_3}$} \label{subsec:Zn1n2n3}

Let us consider $G=\Z_{N_1} \times \Z_{N_2} \times \Z_{N_3}$, with which the TGT model of a finite Abelian G has only Abelian statistics and $1$--dimensional projective representation.
The cohomology group can be derived as $\cH^4[ \Z_{N_1} \times \Z_{N_2} \times \Z_{N_3},U(1)] 
=     \Z_{N_{12}}^2
         \times \Z_{N_{23}}^2 
         \times  \Z_{N_{13}}^2
         \times \Z_{N_{123}}^2 $.
Here we define $N_{12\dots j} \equiv \gcd(N_1,N_2,\dots,N_j)$, with $gcd$ as their greatest common divisor.

The generators of the 4-cocycle $\omega$ for the $(\Z_{N_{ij}})^2$ elements are (here $(i,j)=(1,2),(2,3)$ or $(1,3)$):
\begin{align}
&&{\omega_{{ \tII}}^{(1st,ij)} (a,b,c,d)=  e^{\big( \frac{2 \pi \ii  }{ (N_{ij} \cdot N_j  )   }    (a_i b_j )( c_j +d_j - [c_j+d_j  ]) \big)} },\\
&& {\omega_{{ \tII }}^{(2nd,ij)} (a,b,c,d) =  e^{\big( \frac{2 \pi \ii  }{ (N_{ij} \cdot N_i  )  }   (a_j b_i )( c_i +d_i - [c_i+d_i  ])  \big)} }.
\end{align}

The generators of 4-cocycle $\omega$ for the $\Z_{N_{ijl}}$ elements are (here $(i,j,l)=(1,2,3)$):
\begin{align}
&& {\omega_{{\tIII}}^{(1st,ijl)}(a,b,c,d) = e^{\big( \frac{2 \pi \ii  }{ (N_{ij} \cdot N_l )  }  (a_i b_j )( c_l +d_l - [c_l+d_l  ]) \big)} }, \\
&& {\omega_{{\tIII}}^{(2nd,ijl)}(a,b,c,d) = e^{\big( \frac{2 \pi \ii  }{ (N_{li} \cdot N_j )  }  (a_l b_i )( c_j +d_j - [c_j+d_j  ]) \big)} }.
\end{align}

The 4-cocycle $\omega$ can be written correspondingly from the generators, for example:
\begin{align}
\omega =&\prod_{ \underset{(i,j,l)=(1,2,3)}{(i,j)=(1,2),(2,3),(1,3)} }\omega_{{ \tII}}^{(1st,ij)} \omega_{{ \tII}}^{(2nd,ij)} \omega_{{\tIII}}^{(1st,ijl)} \omega_{{\tIII}}^{(2nd,ijl)} \\
=&\prod_{ \underset{(i,j,l)=(1,2,3)}{(i,j)=(1,2),(2,3),(1,3)} } (\omega_{{ \tII}}^{(1st,ij)})^{p_{{ \tII(ij)}}^{(1st)} } (\omega_{{ \tII}}^{(2nd,ij)})^{p_{{ \tII(ij)}}^{(2nd)} }  \cdot \nonumber \\
&  (\omega_{{\tIII}}^{(1st,ijl)})^{p_{{ \tIII(ijl)}}^{(1st)} } (\omega_{{\tIII}}^{(2nd,ijl)})^{p_{{ \tIII(ijl)}}^{(2nd)} } 
\end{align}
Here ${p_{{ \tII(ij)}}^{(1st)} } \in \Z_{N_{ij}}$,  ${p_{{ \tII(ij)}}^{(2nd)} } \in \Z_{N_{ij}}$, ${p_{{ \tIII(ijl)}}^{(1st)} } \in \Z_{N_{ijl}}$, ${p_{{ \tIII(ijl)}}^{(2nd)} } \in \Z_{N_{ijl}}$ are the group elements
in $\cH^4( \Z_{N_1} \times \Z_{N_2} \times \Z_{N_3},U(1))$.
We organize the following data in Table \ref{4cocycleinducedN123table}, (1) the group elements in a cohomology group $\cH^4[G,U(1)]$, (2) the notations of different types of $\omega$, (3)
explicit forms of the \fcs $\omega$, (4) the induced doubly-twisted 2-cocycle $\beta^{}_{a,b}(c,d)$ by a 4-cocycle $\omega$. 
 Table \ref{4cocycleEpProjRep3DN123table} presents other data: (5) A projective scalar representation ${\epsilon}_{a, b}^{}(c)$, 
but without the irreducible linear scalar representation ${\rho}^{a, b}_{\text{}\alpha}(c)$. (6) The full projective scalar representation $\widetilde{\rho}^{a, b}_{\text{}\alpha}(c) ={\rho}^{a, b}_{\text{}\alpha}(c)  {\epsilon}_{a, b}^{}(c)$
endorsed with a charge-flux irreducible linear scalar representation ${\rho}^{a, b}_{\text{}\alpha}(c)=\exp \big( \sum_k \frac{2 \pi \ti  }{ N_k   }  \; \alpha_k c_k  \big) $.

For  Types II and III 4-cocycles, the induced $\beta^{}_{a,b}(c,d)$ are cohomologically trivial as 2-coboundary. Thus, we can apply Eq.(\ref{eq:projectTOlinear})
to obtain Table \ref{4cocycleEpProjRep3DN123table}. 

\begin{widetext}

\begin{center}
\begin{table} [!h]
\begin{tabular}{|c||c|c|c|c| }
\hline
$\scriptstyle\cH^4[G,U(1)]$ & $\omega$ type  & 4-cocycle $\omega$ as $\omega(a,b,c,d)$ &  $\beta^{}_{a, b}(c,d)$  \\[0mm]  \hline \hline 
$\Z_{N_{ij}}$ & II 1st   & ${\omega_{{ \tII}}^{(1st,ij)} =  \exp \big( \frac{2 \pi \ii p_{{ \tII(ij)}}^{(1st)} }{ (N_{ij} \cdot N_j  )   }    (a_i b_j )( c_j +d_j - [c_j+d_j  ]) \big)}$   & 
$\exp \big( 
\frac{2 \pi i p_{{ \tII(ij)}}^{(1st)} }{ (N_{i j} \cdot N_j  )   }  \; 
(a_j b_i-a_i b_j )( c_j +d_j - [c_j+d_j  ]) \big) $  \\[2mm] \hline
$\Z_{N_{ij}}$ & II 2nd   & ${\omega_{{ \tII }}^{(2nd,ij)}  =  \exp \big( \frac{2 \pi \ii p_{{ \tII(ij)}}^{(2nd)} }{ (N_{ij} \cdot N_i  )  }   (a_j b_i )( c_i +d_i - [c_i+d_i  ])  \big) }$ & 
$\exp \big( 
\frac{2 \pi i p_{{ \tII(ij)}}^{(2nd)} }{ (N_{i j} \cdot N_i  )  }  \; 
(a_i b_j -a_j b_i )( c_i +d_i - [c_i+d_i  ])  \big) $  \\[2mm] \hline
$\Z_{N_{ijl}}$ & III 1st   & ${\omega_{{\tIII}}^{(1st,ijl)} = \exp \big( \frac{2 \pi \ii p_{{ \tIII(ijl)}}^{(1st)} }{ (N_{ij} \cdot N_l )  }  (a_i b_j )( c_l +d_l - [c_l+d_l  ]) \big) }$ & 
$\exp \big( 
\frac{2 \pi i p_{{ \tIII(ijl)}}^{(1st)} }{ (N_{i j} \cdot N_l )  }  \; 
(a_j b_i-a_i b_j )( c_l +d_l - [c_l+d_l  ]) \big)$  \\[2mm] \hline
$\Z_{N_{ijl}}$ & III 2nd   & ${\omega_{{\tIII}}^{(2nd,ijl)} = \exp \big( \frac{2 \pi \ii p_{{ \tIII(ijl)}}^{(2nd)} }{ (N_{li} \cdot N_j )  }  (a_l b_i )( c_j +d_j - [c_j+d_j  ]) \big) }$  & 
$\exp \big( 
\frac{2 \pi i p_{{ \tIII(ijl)}}^{(2nd)} }{ (N_{j i} \cdot N_l )  }  \; 
(a_i b_l -a_l b_i )( c_j +d_j - [c_j+d_j  ]) \big)$   \\[2mm] \hline
\end{tabular}
\caption{In the first column, we list the group elements in the cohomology group $\cH^4[G,U(1)]$ with $G=\Z_{N_1} \times \Z_{N_2} \times \Z_{N_3}$. In the second and the third columns, we record the name and the form of corresponding 4-cocycles. In the fourth column, we provide the induced doubly-twisted 2-cocycle $\beta_{a,b}(c,d)$ by the 4-cocycle $\omega$ in the same row.}
\label{4cocycleinducedN123table}
\end{table}
\end{center}

\begin{center}
\begin{table} [!h]
\begin{tabular}{|c|c|c| }
\hline
$ {\epsilon}_{a, b}^{}(c)$ &  $\widetilde{\rho}^{a, b}_{\text{}\alpha}(c) ={\rho}^{a, b}_{\text{}\alpha}(c) \cdot {\epsilon}_{a, b}^{}(c)$  \\[0mm]  \hline \hline 
 $  { {\epsilon}_{\tII\;a, b}^{(1st)}(c)=  \exp \big( \frac{2 \pi \ti p_{{ \text{II}(ij)}}^{(1st)} }{ (N_{ij} \cdot N_j  )   }  \;  (a_j b_i-a_i b_j ) c_j  \big)  }$   & 
$\widetilde{\rho}^{(1st)a, b}_{\text{II},\alpha}(c) =\exp \big( \sum_k \frac{2 \pi \ti  }{ N_k   }  \; \alpha_k c_k  \big) \cdot \exp \big( \frac{2 \pi \ti p_{{ \text{II}(ij)}}^{(1st)} }{ (N_{ij} \cdot N_j  )   }  \;  
(a_j b_i -a_i b_j ) c_j  \big)$  \\[2mm] \hline
$ { {\epsilon}_{\tII\; ab}^{(2nd)}(c) = \exp \big( \frac{2 \pi \ti p_{{ \text{II}(ij)}}^{(2nd)} }{ (N_{ij} \cdot N_i  )   }  \;  (a_i b_j-a_j b_i ) c_i  \big)  }$ & 
$\widetilde{\rho}^{(2nd)a, b}_{\text{II},\alpha}(c) =\exp \big( \sum_k \frac{2 \pi \ti  }{ N_k   }  \; \alpha_k c_k  \big) \cdot 
\exp \big( \frac{2 \pi \ti p_{{ \text{II}(ij)}}^{(2nd)} }{ (N_{ij} \cdot N_i  )   }  \;  (a_i b_j-a_j b_i ) c_i  \big)$  \\[2mm] \hline
${ \epsilon_{\tIII\; a, b}^{(1st)} =\exp \big(  \frac{2 \pi \ti p_{{ \text{III}(ijl)}}^{(1st)} }{ (N_{ij} \cdot N_l )  }  \;  (a_j b_i-a_i b_j )  c_l \big)  }$ & 
${\widetilde{\rho}^{(1st)a, b}_{\text{III},\alpha}(c) =\exp \big( \sum_k \frac{2 \pi \ti  }{ N_k   }  \; \alpha_k c_k  \big) \cdot  \exp \big(  \frac{2 \pi \ti p_{{ \text{III}(ij l)}}^{(1st)} }{ (N_{ij} \cdot N_l )  }  \;  (a_j b_i-a_i b_j )  c_l \big)}$  \\[2mm] \hline
${ \epsilon_{\tIII\;a, b}^{(2nd)}=\exp \big(  \frac{2 \pi \ti p_{{ \text{III}(ij l)}}^{(2nd)} }{ (N_{l i} \cdot N_j )  }  \;  (a_i b_l-a_l b_i )  c_j \big) }$  & 
$\widetilde{\rho}^{(2nd)a,b}_{\text{III},\alpha}(c) =\exp \big( \sum_k \frac{2 \pi \ti  }{ N_k   }  \; \alpha_k c_k  \big) \cdot   \exp \big(  \frac{2 \pi \ti p_{{ \text{III}(ij l)}}^{(2nd)} }{ (N_{l i} \cdot N_j )  }  \;  
(a_i b_l-a_l b_i )  c_j \big)$   \\[2mm] \hline
\end{tabular}
\caption{This table continues from Table \ref{4cocycleinducedN123table}, here $\widetilde{\rho}^{ab}_{\text{}\alpha}(c)$ for a (3+1)d TGT model with
$G=\Z_{N_1} \times \Z_{N_2} \times \Z_{N_3}$ of $\cH^4[G,U(1)]$.
We derive $\widetilde{\rho}_{\alpha}^{a,b}(c)$ from the equation,
$\widetilde{\rho}_{\alpha}^{a,b}(c)\widetilde{\rho}_{\alpha}^{a,b}(d)=\beta_{a,b}(c,d)\widetilde{\rho}_{\alpha}^{a,b}(cd)$,
which presents the {projective representation}, 
because the induced 2-cocycle belongs to the second cohomology group $\cH^2[G,U(1)]$.
The $\widetilde{\rho}_{\alpha}^{a,b}(c)$: $(Z_a, Z_b)$ ${\rightarrow}$ $\text{GL}\left(Z_a,Z_b\right)$ can be written
as a general linear matrix. For $G=\Z_{N_1} \times \Z_{N_2} \times \Z_{N_3}$, it is only a $1$--dimensional projective representation.}
\label{4cocycleEpProjRep3DN123table}
\end{table}
\end{center}

\end{widetext}

\subsection{$G=\Z_{N_1} \times \Z_{N_2} \times \Z_{N_3} \times \Z_{N_4}$}  \label{subsec:Zn1n2n3n4}

Here we discuss an example when the TGT model of a finite Abelian G has non-Abelian statistics and $N$--dimensional projective representation.
This is the case when the finite Abelian $G$ contains four cyclic group, $G=\Z_{N_1} \times \Z_{N_2} \times \Z_{N_3} \times \Z_{N_4}$,
and when $\gcd(N_1,N_2,N_3,N_4)>1$. Its cohomology group is 
$\cH^4[\prod_i \Z_{N_i},U(1)] 
=  \underset{1 \leq i < j < l \leq 4}{\prod} 
         \Z_{N_{ij}}^2
         \times \Z_{N_{ijl}}^2 
         \times \Z_{N_{1234}}$.
The previous Type II and Type III 4-cocycles can generate all group elements in Sec.\ref{subsec:Zn1n2n3n4} except the $\Z_{N_{1234}}$ subgroup.
The generator for $\Z_{N_{1234}}$ subgroup for the full cohomology group $\cH^4[\prod_i \Z_{N_i},U(1))]$ is,
\begin{align}
\omega_{ \tIV}^{(1234)} (a,b,c,d)=\exp \big( \frac{2 \pi \ii }{ N_{1234} }  a_1 b_2 c_3 d_4 \big).
\end{align}
And the corresponding ``Type IV 4-cocycle'' is with an exponent $p_{{ \tIV}{(1234)}}^{} \in \Z_{N_{1234}}$. 
\begin{align}
&\omega_{{ \tIV}}^{(1234)} (a,b,c,d)=(\omega_{{ \tIV}}^{(1234)} (a,b,c,d) )^{ p_{{ \tIV}{(1234)}} }\\
&=\exp \big( \frac{2 \pi \ii p_{{ \tIV}{(1234)}}^{}}{ N_{1234} }  a_1 b_2 c_3 d_4 \big).
\end{align}

The induced 2-cocycle $\beta^{}_{a,b}(c,d)$ by a Type IV 4-cocycle $\omega_{{ \tIV}}^{(1234)}$ is:
\begin{eqnarray}
&\beta^{}_{a,b}(c,d)=\exp \Big(  \frac{2 \pi i
p_{{ IV}{(1234)}}^{}}{ N_{1234} } \; 
\big( (a_4 b_3 -a_3 b_4 ) c_1 d_2\nonumber \\
&+(a_2 b_4 -a_4 b_2 ) c_1 d_3 +(a_4 b_1 -a_1 b_4 ) c_2 d_3\nonumber \\ 
&+(a_3 b_2 -a_2 b_3 ) c_1 d_4 \nonumber \\
&+ (a_1 b_3 -a_3 b_1 ) c_2 d_4+(a_2 b_1 -a_1 b_2 ) c_3 d_4 \big) \Big), 
\end{eqnarray}
not a 2-coboundary, thus cohomologically nontrivial.

Indeed, such a Type IV 4-cocycle has a \emph{higher-dimensional} irreducible projective representation $\widetilde{\rho}^{a, b}_{\text{}\alpha}(c)$ 

The higher dimensional representations and non-Abelian statistics from a Type IV 4-cocycle imply that the GSD is different from those in the Abelian cases. In particular, on a $3$-torus, the GSD for the Abelian case is simply GSD=$|G|^3$ for an Abelian $G$. The cocycle twist of Type IV 4-cocycles promotes a twisted Abelian model $G=(\Z_m)^4$ model to be an intrinsic non-Abelian one, and we compute its GSD numerically in Table \ref{tab:GSDZmfourth}.

\onecolumngrid

\begin{widetext}

\begin{table}[!h]
\begin{center}
\begin{tabular}{lc|ccccc}
\hline\hline
  GSD     & & ${ p_{{ \tIV}{(1234)}} }=0 $ & ${ p_{{ \tIV}{(1234)}} }=1$ &  ${ p_{{ \tIV}{(1234)}} }=2$ &  ${ p_{{ \tIV}{(1234)}} }=3$ &  ${ p_{{ \tIV}{(1234)}} }=4$ \\\hline
$m=2$  & & $2^{12}=4096$ & 1576 & & &\\
$m=3$  & & $3^{12}=531441$ & 82161 & 82161 & &\\
$m=4$  & & $4^{12}=16777216$ & 1939456 & 6455296 & 1939456 &\\
$m=5$  & & $5^{12}=244140625$ & 12012625 & 12012625 & 12012625 & 12012625\\ \hline
\hline
\end{tabular}\caption{ GSD on a $3$-torus $\mathbb{T}^3$ with $G=\Z_m^4$}
\label{tab:GSDZmfourth}
\end{center}
\end{table}

\begin{table}[!h]
\begin{center}
\begin{tabular}{lc|c|c}
\hline\hline
  GSD     & & ${ p_{{ \tIV}{(1234)}} }=0 $ &   ${ p_{{ \tIV}{(1234)}} } \neq 0$ (if $\gcd({ p_{{ \tIV}{(1234)}} },m)=1$)       \\\hline
$m=${prime}  &  & $(m^4)^3$ &  $\text{GSD}_{\mathbb{T}^3,\tIV}$($m_{\text{prime}}$) = $m^3-m^5-m^6-m^7+m^8+m^9+m^{10}$   \\ \hline
\hline
\end{tabular}\caption{ GSD on a $3$-torus $\mathbb{T}^3$ with $G=\Z_m^4$.
If $q \neq 0$ and if $\gcd(q,m)=1$; it is truncated to,
$\text{GSD}_{\mathbb{T}^3,\tIV}(m_{\text{prime}})$.
}
\label{tab:GSDZmfourth2}
\end{center}
\end{table}
\end{widetext}

\twocolumngrid

Indeed, when $m$ is a prime number, for $G=\Z_m^4$ TGT model with Type IV 4-cocycle, such as $m=2,3,5,\dots$, we obtain an analytic formula of its GSD:
\be
\begin{aligned}
&\text{GSD}_{\mathbb{T}^3,\tIV}(m_{\text{prime}}) \\
=&m^3-m^5-m^6-m^7+m^8+m^9+m^{10},
\end{aligned}
\ee
which is consistent with Table \ref{tab:GSDZmfourth}. We can derive this GSD in two alternative ways, either by the conservation of total quantum dimensions or by the dimensional reduction of a three-dimensional topological phase to degenerate states of several sectors of two-dimensional topological phases\cite{Wang2014c}. To a representation theory point of view, part of the GSD is due to Abelian excitations carrying scalar representations, part of  GSD is due to non-Abelian excitations carrying $m$-dimensional representations (especially for prime $m$'s). This way, we can decompose the GSD to its Abelian and non-Abelian sectors with their representations' dimensions ($|G|=m^4$),
\begin{align}
& \text{GSD}_{\mathbb{T}^3,\tIV}(m_{\text{prime}}) \label{eq:T3TypeIVdec}
 \equiv \text{GSD}^{Abel}_{\mathbb{T}^3,\tIV} + \text{GSD}^{nAbel}_{\mathbb{T}^3,\tIV}  \nonumber\\
&= \big(m^8+m^9-m^5\big) +\big(m^{10} -m^{7} -m^{6} +m^{3}\big)
\end{align}
The conservation of total quantum dimensions are constrained by a cubic equality:
\begin{align}
& |G|^3={ \text{GSD}^{Abel}_{\mathbb{T}^3,\tIV} \cdot  \dim_{1}^2 +\text{GSD}^{nAbel}_{\mathbb{T}^3,\tIV} \cdot \dim_{m}^2} \nonumber \\
& ={ \text{GSD}^{Abel}_{\mathbb{T}^3,\tIV} + \text{GSD}^{nAbel}_{\mathbb{T}^3,\tIV} \cdot m^2}=m^{12},
\end{align}
where we use $\dim_{m}=m$ to denote the dimension of the $m$-dimensional representation. If the corresponding ground state has an excited counterpart, $\dim_m$ would also be the {\it quantum dimension} of that excitation.

When $m$ is not a prime (such as $m=4$), the dimensions of the representations have more choice: $\dim_{m'}$,
where $m'$ is a divisor of $m$. Namely, $m/m' \in \mathbb{Z}$.
For example, when $m=4$, we have dimensions of representations as $\dim_{1}$, $\dim_{2}$ and $\dim_{4}$.
Because the conservation of quantum dimensions still holds in the form of a cubic equality:
\begin{align}
|G|^3={ \text{GSD}^{Abel}_{\mathbb{T}^3,\tIV} \cdot  \dim_{1}^2 +\sum_{m'} \text{GSD}^{nAbel}_{\mathbb{T}^3,\tIV(m')} \cdot \dim_{m'}^2}.
\end{align}
To satisfy this equality, we learn that the factorization of non-prime $m$ to smaller $m'$ makes the overall sum of GSD becomes larger.
Indeed, we see it is the case that for $m=4$, namely,
\begin{align}
&|G|^3 >\text{GSD}_{\mathbb{T}^3,\tIV}(m_{\text{non-prime}})|_{\gcd({ p_{{ \tIV}{(1234)}} },m)\neq 1} \nonumber \\
&>\text{GSD}_{\mathbb{T}^3,\tIV}(m_{\text{non-prime}})|_{\gcd({ p_{{ \tIV}{(1234)}} },m)=1} \nonumber \\ 
& > m^3-m^5-m^6-m^7+m^8+m^9+m^{10}.
\end{align}
The relation should work for any positive integer $m$. 
The general formula of the GSD on $\mathbb{T}^3$ for any $m$ is expected to relate to the factorization of $m$.

\subsection{$G=(\Z_m){}^n$ or $G=\prod_i \Z_{N_i}$} \label{subsec:Zmn}
We end this section by considering the finite Abelian group $\Z_m^n$ and $\prod_i \Z_{N_i}$for some integer $m$ and $n$. A special case with $n\leq 3$ were investigated in previous sub sections in 
Sec.\ref{subsec:Zn1n2n3}. 
For $n>3$, things are similar to $G=\Z_m^4$ case in Sec.\ref{subsec:Zn1n2n3n4}. The second, the third
 and the fourth cohomology groups are
\begin{align}
\label{ZmnCohomologyGroups}
&H^2[(\Z_m){}^n,U(1)]\simeq (\Z_m){}^{n \choose 2},
\nonumber\\
&H^3[(\Z_m){}^n,U(1)]\simeq (\Z_m){}^{{n \choose 1}+{n \choose 2}+{n \choose 3}},
\nonumber\\
&H^4[(\Z_m){}^n,U(1)]\simeq (\Z_m){}^{2 {n \choose 2}+2 {n \choose 3}+{n \choose 4}}.
\end{align}

More generally $G=\prod_i \Z_{N_i}$, we have
\begin{align}
& \cH^2[\prod_i \Z_{N_i},U(1))]=  \prod_{1 \leq i < j \leq k} 
          \Z_{N_{ij}},
           \;\;\;  \label{eq:H2}\\
& \cH^3[\prod_i \Z_{N_i},U(1)] 
=  \prod_{1 \leq i < j < l \leq k} 
         \Z_{N_{i}}\times \Z_{N_{ij}}
         \times \Z_{N_{ijl}}, 
           \;\;\;  \label{eq:H3}\\
&\cH^4[\prod_i \Z_{N_i},U(1)] 
=  \prod_{1 \leq i < j < l<m \leq k} 
         \Z_{N_{ij}}^2
         \times \Z_{N_{ijl}}^2 
         \times \Z_{N_{ijlm}}.\;\;\;  \label{eq:H4}
\end{align}
The 4-cocycles for finite Abelian groups are limited to Type II, III, and IV 4-cocycles discussed in Sec. \ref{subsec:Zn1n2n3} and \ref{subsec:Zn1n2n3n4}. The data above is sufficient to exemplify our general studies of the TGT model in previous sections.
\section{Correspondence with the Dijkgraaf-Witten theory}\label{sec:DW}
In this section, we briefly demonstrate the relation between our TGT model and the Dijkgraaf-Witten (DW) topological gauge theory.\cite{Dijkgraaf1990} The $d+1$-dimensional discrete DW gauge theory with a finite gauge group $G$ is defined on a triangulation $\mathbf{T}$ of a $d+1$-manifold $M$. The triangulstion $\mathbf{T}$ consists of $d+1$-simplices. Each edge, or $1$-simplex, of $\mathbf{T}$ is oriented and graced with a group element of $G$ (see Fig. \ref{fig:4cocycle} for a $3+1$-d example). The topological partition function of this gauge theory reads
\be\label{eq:Zdw}
\begin{aligned}
\mathcal{Z} =&\sum_\gamma \e^{\ii S[\gamma]}=\sum_\gamma \e^{\ii 2\pi \langle \omega_{d+1}, \gamma(\mathbf{T})\rangle\pmod {2\pi}} \\
=&\frac{1}{|G|^{V_\mathbf{T}}}  \sum_{\{[v_av_b]\}} \prod_i \omega_{d+1}(\{[v_av_b]\})^{\epsilon_i} \Big\vert_{v_av_b\in T_i}.
\end{aligned}
\ee
Here we sum over all embedding maps $\gamma: M \mapsto BG$, from the spacetime manifold $M$ to $BG$, the classifying space of $G$. In the second equality, we triangulate $M$ to $\mathbf{T}$, where each edge $v_a v_b$ connects the two vertices $v_a$ and $v_b$. The action $ \langle \omega_{d+1}, \gamma(\mathbf{T})\rangle$ evaluates the cocycles $\omega_{d+1}$ on the spacetime $(d+1)$-complex $\mathbf{T}$. 
Based on the relation between the topological cohomology class of $BG$ and the cohomology group of $G$: 
$H^{d+2}[BG,\Z] =H^{d+1}[G,\R/\Z]=H^{d+1}[G,U(1)]$,\cite{Dijkgraaf1990,Hung2012b} 
we can view $\omega_{d+1}$ as the $(d+1)$-cocycles in the cohomology group $H^{d+1}[G,U(1)]$. 
Each edge $v_a v_b$ carries a group element $[v_av_b]\in G$. The $1/|G|^{V_\mathbf{T}}$ is the normalization factor, where $V_\mathbf{T}$ counts the number of vertices in $\mathbf{T}$. 
The cocycle $\omega_{d+1}$ is evaluated on all the $d+1$-simplices $T_i$ of $\mathbf{T}$. 

In $3+1$ dimensions, the $\omega_4$ is in fact the \fc $\omega$ we have been dealing with in the entire paper. Restricted to this case, the partition function \eqref{eq:Zdw} is invariant under the Pachner moves that connect two simplicial triangulations of $M$. It is also gauge invariant under the lattice gauge transformation, whose definition is the same as that of our vertex operator $A^g_v$ (see Eq. \eqref{eq:AvgFourTet} for an example). 

If the manifold $M$ is open, one has to add boundary terms to Eq. \eqref{eq:Zdw} but the bulk gauge transformation ceases to work on the boundary. Here is the reason. The partition function \eqref{eq:Zdw} sums over the embedding maps of $M$ into the classifying space $BG$. When $M$ is open, the boundary condition on $M$ nevertheless fixes the boundary values of the all the embedding maps. As such, the Pachner moves involving the boundary simplices would be disallowed because they can vary the boundary value of an embedding. Consequently, on the three--dimensional boundary $\partial M$, degrees of freedom that cannot be gauged away by the bulk gauge transformation emerge, forming the boundary states. In this scenario, the four--dimensional partition function can be regarded as the transition amplitude between the boundary states, or simply a wavefunction of the boundary degrees of freedom.
Being a topological gauge theory, however, the DW theory does not have a finite Hamiltonian via Legendre transform; hence, on the boundary, the notion of ground and excited states is missing. Nonetheless, the notion of gauge-invariant and uninvariant states still exists. In fact, if $\partial M$ is closed, only gauge-invariant states survive the boundary. The size of the Hilbert space of the boundary gauge-invariant states can be obtained by the standard technique of "gluing" and "sewing" in topological field theories. Particularly in our case, the DW partition function turns out to count the dimension of the Hilbert space of the gauge-invariant states on the closed boundary $\partial M$.

A natural but challenging question arises: can one construct a Hamiltonian on the closed boundary $\partial M$, such that the ground states of the Hamiltonian are in one-to-one correspondence with the gauge-invariant boundary states of a DW theory on $M$? In this work, we have actually tackled this challenge.  One of the authors of the current paper has shown in Ref.\onlinecite{Hu2012a} the correspondence between the TQD model and the $2+1$-dimensional DW theory. The entire analysis in the $2+1$ case straightforwardly applies to our $3+1$ case by just going one dimension higher, so we will not repeat it. To be precise, let us place the DW theory on the triangulation of a $3+1$-dimensional manifold $X\x S^1$, where $S^1$ is the circle, and $\partial X$, the boundary of $X$, is a torus, such that $\partial X\x S^1$ is a $3$-torus. Then we place our TGT model on the triangulation of $\partial X\x S^1$. Knowing that the gauge transformation on the boundary of the DW theory is the same as the vertex operator in the TGT model, and using the result in Ref.\onlinecite{Hu2012a}, we can infer the following identity.
\be
 \mathcal{Z}_{DW}^{\mathbf{T}(X\times S^1)}(G,\omega)=\mathrm{GSD}_{TGT}^ {\mathbf{T}(\partial X\times S^1)}(G,\omega).
\ee

The correspondence deduced above justifies the claim that our TGT model may indeed be a Hamiltonian extension of the DW discrete gauge theory. And this can be generalized to higher dimensions without a problem.

Another remark is that the DW theory in $3+1$ dimensions is also a twisted version of the $BF$ topological field theory\cite{Horowitz1989,OOGURI1992} with only the $B\wedge F$ term and the gauge transformation twisted by a $4$-cocycle. In this $BF$ theory, apart from a usual $1$-form field $A$ whose curvature is $F$, there is also a $2$-form field $B$. As opposed to that of the field $A$, the flux of the field $B$ is obtained by integrating of $B$ over a two-dimensional surface. Recall that in the ground state spectrum of our TGT model, there are membranes that carry two flux indices. These observations imply that the TGT model can also be a Hamiltonian extension of the $BF$ theory that is twisted by \fcs.
Nevertheless, we shall leave more detailed discussion of this and the subtleties to future work.

\section{Discussions and outlook}\label{sec:disc}
Here, we shall remark on our main results and raise a few questions tied to these results, the answers to which deserve future exploration.

First, we constructed the TGT model, which is a lattice Hamiltonian model of three-dimensional topological phases. The model is defined by a $4$--cocycle $[\omega]\in H^4[G,U(1)]$ of a finite group $G$ on a simplicial triangulation of a closed, oriented $3$-manifold. Each edge of the triangulation is endowed with an element of $G$. This model describes a large class of topological phases in three spatial dimensions. The TGT model possess a set of topological observables, namely, GSD, the modular $\ttr$ and $\str$ matrices. The latter two observables give rise to fractional topological quantum numbers, e.g., topological spins, which together with GSD, characterize the topological properties of the TGT model.

Second, we found that these topological quantum numbers depend either directly on the defining $4$--cocycle of the model or indirectly via a doubly-twisted $2$--cocycle derived from the $4$--cocycle, which in turn classify the topological quantum numbers. Typically, two equivalent $4$--cocycles always define two equivalent TGT models that yield the same topological phase. 

Third, we also demonstrated that our TGT model is a Hamiltonian extension of the $3+1$-dimensional DW topological gauge theory. In fact, we have shown that the GSD of a TGT model defined by some $4$--cocycle on the boundary of a $4$--manifold coincides with the partition function of the DW theory in the $4$-manifold, whose topological action is determined by the same $4$--cocycle. This correspondence also implies a relation between our TGT model and certain type of $BF$ theory. The relation between $BF$ theories and topological phases is discussed has been discussed recently\cite{Kapustin2013,Gukov2013,Kapustin2013a,Kapustin2014}. Since more general $BF$ theories, especially those with a $B\wedge B$ term, may relate to gravity\cite{Freidel2012}, it would be very interesting to extend our model for such theories, which may shed light on the connection between gravity and condensed matter physics. 

We remark that, via the plausible duality mapping between symmetry-protected topological states (SPTs) and intrinsic topological orders\cite{Levin2012,Hung2012,Hung2012b,Mesaros2011}, one may associate the symmetry-protected topological terms of SPTs field theory to the topological terms of topological field theory for topological orders\cite{Wang2014d}. At least for (3+1)-d TGT described by Dijkgraaf-Witten theory and group cohomology, we shall be able to comprehend the topological terms via its dual SPTs. It is known in Ref.\onlinecite{Wang2014d} that for generic twisted Abelian gauge group $G=\prod_i \Z_{N_i}$, their topological terms have the forms, $A_i \wedge A_j \wedge dA_k $ or $A_i \wedge A_j \wedge A_k\wedge A_l$. These are the Type II,Type III and Type IV 4-cocycles examined in our Sec.\ref{sec:examples} and Ref.\onlinecite{Wang2014c}. Thus, the $B\wedge B$ term which is not among the topological terms of DW theory and group cohomology, would be beyond group cohomology.

In $2+1$ dimensions, the TQD model on the torus is also related to the rational conformal field theories (RCFT) that are the toric orbifolds by a finite group of a holomorphic CFT and are twisted by a nontrivial $3$--cocycle\cite{Hu2012a}. This correspondence identifies the ground states, the GSD, and the modular matrices of a TQD model, respectively, with the holomorphic characters, the number of primary fields, and also the modular matrices of the corresponding RCFT. Provided with the description of fractional quantum Hall effect by CFT, it is expected that the statistical and topological properties of the quasi-excitations and hence the topological phase of a TQD model can be investigated in terms of the modular matrices of the model. It is therefore important to examine whether and how  the TGT model may be related to three-dimensional conformal field theory.

In a two-dimensional topological phase, the quasi-excitations are in one-to-one correspondence with the ground states; hence, they share the same topological properties. In a three-dimensional topological phase, it is a question whether such a bijection still holds. As argued in Ref.\onlinecite{Wang2014c}, the bijection may cease to exist. Although it is clear that the three-dimensional excitations would share a few properties with the ground states, at this moment, we lack a systematic understanding of the excitations which cannot be dimensionally reduced from three to two dimension. Hence, we shall refrain from conclude affirmatively here and leave the answer for future work.  
\section*{Acknowledgements}
We thank Wojciech Kaminski, Huangjun Zhu, Ling-Yan Hung, Xiao-Gang Wen, Liang Kong, and Tian Lan for helpful discussions. YW appreciates his mentor, Xiao-Gang Wen, for his constant support and insightful conversations. YW is grateful to Yong-Shi Wu for his hospitality at Fudan University, where some parts of this work was done, and for his inspiring comments. This research is supported in part by Perimeter Institute for Theoretical Physics. Research at Perimeter Institute is supported by the Government of Canada through Industry Canada and by the Province of Ontario through the Ministry of Economic Development \& Innovation. This research is also supported by NSF Grant No.\;DMR-1005541, NSFC 11074140, NSFC 11274192, the BMO Financial Group and the John Templeton Foundation.

\begin{appendix}
\section{A brief introduction to cohomology groups $H^n[G,U(1)]$}\label{app:HnGU1}
In both Sections \ref{subsec:basic} and \ref{sec:DW}, we gave a physical account for the $4$-cocycles that define our models. Here, to make the mathematical content self-contained, we organize a few elementary but relevant definitions about the cohomology groups $H^n[G,U(1)]$ of finite groups $G$.

To define the $n$-th cohomology group of a finite group $G$, we need to learn a similar group structure, the $n$-th \textit{cochain group} $C^n[G,U(1)]$ of $G$. It is an Abelian group of $n$-\textit{cochains}, i.e., functions $c(g_1,\dots,g_n): G^{\times n}\to U(1)$, where $g_i\in G$ The group structure is fabricated by the product rule: $c(g_1,\dots,g_n)c'(g_1,\dots,g_n)=(cc')(g_1,\dots,g_n)$. 
The \textit{coboundary operator} $\delta$  maps $C^n$ to $C^{n+1}$, namely,
\begin{align*}
\delta &:C^n\to C^{n+1}\\
&: c(g_1,\dots,g_n)\mapsto \delta c(g_0,g_1\dots,g_n),
\end{align*}
where
\begin{align*}
&\delta c(g_0,g_1\dots,g_n)\\
=& \prod_{i=0}^{n+1}c(\dots,g_{i-2},g_{i-1}g_i,g_{i+1},\dots)^{(-1)^i}.
\end{align*}
At $i=0$, the series of variables starts at $g_0$, and at $i=n+1$, the series of variables ends at $g_{n-1}$. Equation (\ref{4CocycleCondition}) offers the example for $n=4$. 
One can easily check the nilpotency of $\delta$, i.e., $\delta^2c=1$, which leads to the following exact sequence:
\be\label{eq:exactSeq}
\cdots C^{n-1}\stackrel{\delta}{ \to} C^n\stackrel{\delta}{ \to} C^{n+1}\cdots.
\ee 
The images of the coboundary operator, namely, $\mathrm{im}(\delta:C^{n-1}\to C^n)$, form the $n$-th coboundary group, in which the $n$-cochains are dubbed $n$-\textit{coboundaries}. On the other hand, The kernel $\ker(\delta:C^n\to C^{n+1})$ is the group of $n$-\textit{\textbf{cocycles}}, which are the $n$-cochains satisfying the \textit{cocycle condition} $\delta c=1$. Equation. (\ref{4CocycleCondition}) is again the example for $n=4$. The definition of the $n$-th cohomology group then naturally follows from the exact sequence \eqref{eq:exactSeq}:
\[
H^n[G,U(1)]:=\frac{\ker(\delta:C^n\to C^{n+1})}{\mathrm{im}(\delta:C^{n-1}\to C^n)}.
\]
The group $H^n[G,U(1)]$ is Abelian; it consists of the equivalence classes of the $n$-cocyles that differ from one another by only an $n$-coboundary.
A trivial $n$-cocycle is equivalent to the unity and thus can be written as a $n$-coboundary. 

There also exists \emph{a slant product} that maps an $n$-cocycle $c$ to an $(n-1)$-cocycle $c_g$:
\begin{widetext}
\be\label{eq:slantProd}
c_g(g_1,g_2,\dots,g_{n-1})=c(g,g_1,g_2,\dots,g_{n-1})^{(-1)^{n-1}} \prod^{n-1}_{j=1} c(g_1,\dots,g_j,(g_1\cdots g_j)^{-1}g(g_1\cdots g_j),\dots,g_{n-1})^{(-1)^{n-1+j}}.
\ee
\end{widetext}
The twisted $3$-cocycles and the douly-twisted $2$-cocycles we have encountered in the main part of the paper are examples of this slant product.

\section{Algebra of the vertex and face operators}\label{app:algAvBf}
We prove that the vertex and face operators constituting the Hamiltonian (\ref{eq:Hamiltonian}) form a closed algebra as follows.
\begin{subequations}  \label{AlgebraAB}
\begin{alignat}{2}
& \mathrm{(i)}\quad && [B_{f'},B_f]=0,  \quad [B_f,A^g_v]=0,\label{eq:BfCommute}\\
& \mathrm{(ii)}\quad && [A^g_v,A^h_w]=0\text{ if } v\neq w, \label{eq:AvCommute}\\
& \mathrm{(iii)}\quad  && A^g_{v'}A^h_v = A^{g\cdot h}_v,\quad v'<v,v'v=h. \label{eq:AvProd}
\end{alignat}
\end{subequations}
The product rule \eqref{eq:AvProd} ensures that $A_v=\sum_gA^g_v/|G|$ is a projector. 
Note that for simplicity, because the vertex ordering are explicit in terms of the enumerations, we may drop the arrows on the edges in a graph without a prior notice.

\smallskip

\noindent(i).
The first commutator $[B_{f'},B_f]=0$ is a direct result of the  definition of $B_f$ in Eq. \eqref{eq:actionOfBf}. Whereas the second commutator should be proven in two cases. Consider an $A^g_v$ and a $B_f$ for $v\notin f$. Because $A_v$ varies only the group elements on the edges meeting at the vertex $v$, in this case, the zero-flux condition imposed by the $B_f$ on the $f$ cannot be affected by $A_v$, and vice versa. Thus, $[B_f,A_v^g]=0$ is obviously true when $v\notin f$. On the other hand,
if $v$ lies on the boundary of $f$, it suffices to  consider for example the two actions, 
$A_{v_4}^g B_{v_2v_3v_4}$ and $B_{v_2v_3v'_4}A_{v_4}^g$ with $v'_4v_4=g$, on the basis vector
\[
  \BLvert \fourTet{v_1}{v_2}{v_4}{v_3}{0}{}{0.5}{1} \Brangle.
\]
These two actions are equal:
\begin{align}
&A_{v_2}^g B_{v_2v_3v_4}\BLvert \fourTet{v_1}{v_2}{v_4}{v_3}{0}{}{0.5}{1}  \Brangle \\ =&\dots \frac{\dlt{v_2}{v_3}{v_4}}{[v_1v_2,v_2v_3,v_3v'_4,v'_4v_4]}\BLvert \fourTet{v_1}{v_2}{v'_4}{v_3}{0}{}{0.5}{1} \Brangle\nonumber\\
=&\dots \frac{\dlt{v_2}{v_3}{v'_4}}{[v_1v_2,v_2v_3,v_3v'_4,v'_4v_4]}\BLvert \fourTet{v_1}{v_2}{v'_4}{v_3}{0}{}{0.5}{1} \Brangle\nonumber\\
=&\dots \frac{B_{v_2v_3v'_4}}{[v_1v_2,v_2v_3,v_3v'_4,v'_4v_4]}\BLvert \fourTet{v_1}{v_2}{v'_4}{v_3}{0}{}{0.5}{1} \Brangle\nonumber\\
=&B_{v_2v_3v'_4} A_{v_4}^g\BLvert \fourTet{v_1}{v_2}{v_4}{v_3}{0}{}{0.5}{1} \Brangle,
\end{align}
where the dots $\dots$ omits all other irrelevant factors. 
The second equality above follows from the chain rule $v_2v'_4\cdot v'_4v_4=v_2v_4\Rightarrow v_4v_2=v_4v'_4\cdot v'_4v_2$. 
Therefore, we conclude that $[B_f,A_v^g]=0$ is true for all $f,v\in\Gamma$.

\smallskip

\noindent(ii).
The definition of $A_v$ makes $[A_v^g ,A_w^h]=0$ hold trivially if $v_1$ and $v_2$ do not bound any edge. The nontrivial case arises when $v$ and $w$ are the two ends of an edge. Since we consider closed graphs only, we focus on the basis graphs in which $vw$ is explicitly an interior edge. 
It is sufficient to check the action of $A_4^g A_2^h$ on the simplest relevant basis graph, 
where the edge $24$ is shared by three tetrahedra, $0124$, $0234$, and $1234$. We find:
\color{black}

\begin{align}
  \label{A4A2}
  &A_4^g A_2 ^h \BLvert \threeTet{0}{1}{2}{3}{4}{0.25} \Brangle
  \nonumber\\
  =&
  \frac{[02',2'2,23,34][01,12',2'2,23]}{[12',2'2,23,34][01,12',2'2,24]}  A_4^g \BLvert \threeTet{0}{1}{2'}{3}{4}{0.25} \Brangle \nonumber\\
  =&\frac{[02',2'2,23,34][01,12',2'2,23]}{[12',2'2,23,34][01,12',2'2,24]} [01,13,34',4'4]
  \nonumber\\
  &\x\frac{[01,12',2'4',4'4][02',2'3,34',4'4]}{[12',2'3,34',4'4]} \BLvert \threeTet{0}{1}{2'}{3}{4'}{0.25} \Brangle
\end{align}
with $[v'_2v_2]=h,[v'_4v_4]=g$. Only those \fcs relevant to the two vertices $2$ and $4$ are displayed.

Conversely, we apply $A^h_2A^g_4$
on the same basis graph and obtain
\begin{align}
  \label{A2A4}
  & A_2 ^h A_4^g\BLvert \threeTet{0}{1}{2}{3}{4}{0.25} \Brangle
  \nonumber\\
  =&
  \frac{[01,12,24',4'4][01,13,34',4'4]}{[12',2'3,34',4'4]}\nonumber\\
  &\x[02,23,34',4'4]  A_2^h \BLvert \threeTet{0}{1}{2}{3}{4'}{0.25} \Brangle \nonumber\\
  =&  \frac{[01,12,24',4'4][01,13,34',4'4]}{[12',2'3,34',4'4]}[02,23,34',4'4]
  \nonumber\\
  &\x\frac{[01,12',2'2,23][02',2'2,23,34]}{[01,12',2'2,24'][12',2'2,23,34']} \BLvert \threeTet{0}{1}{2'}{3}{4'}{0.25} \Brangle
\end{align} 
with $[v'_2v_2]=h,[v'_4v_4]=g$ as well.

Nevertheless, it is straightforward to show that the two amplitudes in Eqs. \eqref{A4A2} and \eqref{A2A4} are precisely the same by applying the chain rules \eqref{ChainRule} induced by each action of $A_v$ and the \fc condition \eqref{4CocycleCondition} three times, respectively with different group variables, as follows.
\begin{align*}
&\frac{[02,23,34',4'4][02',2'2,24',4'4][02',2'2,23,34']}{[2'2,23,34',4'4] [02',2'3,34',4'4][02',2'2,23,34]}=1,\\
&\frac{[2'2,23,34',4'4][12',2'3,34',4'4][12',2'2,23,34]}{[12,23,34',4'4] [12',2'2,24',4'4][12',2'2,23,34']}=1,\\
&\frac{[12',2'2,24',4'4][01,12,24',4'4][01,12',2'2,24]}{[02',2'2,24',4'4] [01,12',2'4',4'4][01,12',2'2,24']}=1.
\end{align*}

Notice that the chain rules \eqref{ChainRule}
guarantee that each group element indexed by the same pair
of enumerations is the same in the above evaluations.
Therefore we arrive at $A_4^g A_2^h =A_2^h A_4^g$. And clearly we can conclude that in general $[A_v^g ,A_w^h]=0, \forall v\neq w$. 

\begin{figure}[h!]
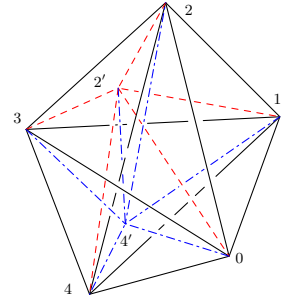

\centering
\twoAvCommute{0}{1}{2}{3}{4}{2'}{4'}{0.6}
\caption{(Color online) The $4$-complex encoding the actions $A^g_4A^h_2$ (absence of the edge $24'$) and $A^h_2A^g_4$ (absence of the edge $2'4$) in Eqs. \eqref{A4A2} and \eqref{A2A4}.}
\label{fig:A2A4commute}
\end{figure}

It would be illustrative to depict in Fig. \ref{fig:A2A4commute} the $4$-complex that encodes the two amplitudes in Eqs. \eqref{A4A2} and \eqref{A2A4} and also the three \fc\ conditions above. Here we remark that the \fc conditions one can use must not rely on any chain rule that does not exist for the three group elements along the boundary of a triangle. One can see that indeed our proof above does not assume the chain rule on any of the triangles in the basis graph in the first line of Eq. \eqref{A4A2}. 

Another subtlety in the proof above lies in determining the orientations of certain $4$-simplices and hence the signs of the associated \fcs. For instance, consider the \fc $[01,12',2'2,23]$, which is associated with the $4$-simplex $012'23$; however, since the combination $0123$ does not exist as an tetrahedron in the original basis graph, one cannot use Convention \ref{conv:4cocyleAv1Tet} and determine the orientation of $012'23$ by the orientation of $0123$ but has to resort to Convention \ref{conv:4cocycle4Tet} as if $012'23$ is merely a $3$-complex consisting of four tetrahedra except $0123$.

\smallskip
\noindent(iii). In the product rule $A^g_{v'} A^h_v = A^{g\cdot h}_v$, we assumed that $A_v^h$ acts first and turns the vertex $v$ to $v'<v$, such that $v'v=h$, and then $A^g_{v'} $ turns $v'$ to $v''<v'$, such that $v''v'=h$, whereas $A^{g\cdot h}_v$ replaces $v$ by $v''$ with $v''v=g\cdot h$. We consider the following generic basis graph consisting  of three tetrahedra, $0124$, $0234$, and $1234$, and  begin with the action of $A^g_{v'} A^h_v $ on the vertex $2$ therein.
\begin{align}
  &A^g_{2'} A^h_2\BLvert \threeTet{0}{1}{2}{3}{4}{0.25}\Brangle
  \nonumber\\
  =&\frac{[01,12',2'2,23][02',2'2,23,34]}{[01,12',2'2,24][12',2'2,23,34]}
  A^g_{v'}\BLvert \threeTet{0}{1}{2'}{3}{4}{0.25} \Brangle
  \nonumber\\
  =&\frac{[01,12',2'2,23][02',2'2,23,34]}{[01,12',2'2,24][12',2'2,23,34]}
    \label{eq:AgAh}\\
  &\x\frac{[01,12'',2''2',2'3][02'',2''2',2'3,34]}{[01,12'',2''2',2'4][12'',2''2',2  '3,34]} \BLvert \threeTet{0}{1}{2''}{3}{4}{0.25}\Brangle.\nonumber
\end{align}

\begin{figure}
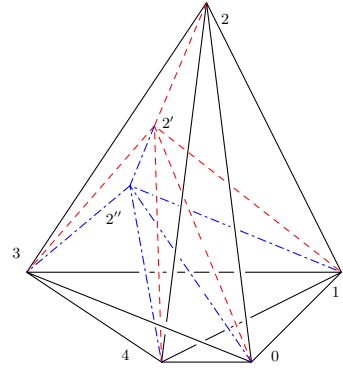

\centering
\AgAhISAgh{0}{1}{2}{3}{4}{2'}{2''}{0.6}
\caption{(Color online) The $4$-complex encoding the amplitudes in Eqs. \eqref{eq:AgAh} and \eqref{eq:AgAh=Agh} and the \fc conditions applied.}
\label{fig:AvProd}
\end{figure}

Now applying to the above amplitude made of eight \fcs the following four \fcs conditions in order,
\begin{align*}
&\frac{[02',2'2,23,34][02'',2''2',2'3,34][02'',2''2',2'2,23]}{[2''2',2'2,23,34] [02'',2''2,23,34][02'',2''2',2'2,24]}=1\\
&\frac{[2''2',2'2,23,34][12'',2''2,23,34][12'',2''2',2'2,24]}{[12',2'2,23,34] [12'',2''2',2'3,34][12'',2''2',2'2,23]}=1\\
&\frac{[02'',2''2',2'2,24][01,12'',2''2,24][01,12'',2''2',2'2]}{[12'',2''2',2'2,24] [01,12',2'2,24][01,12'',2''2',2'4]}=1\\
&\frac{[12'',2''2',2'2,23][01,12',2'2,23][01,12'',2''2',2'3]}{[02'',2''2',2'2,23] [01,12'',2''2,23][01,12'',2''2',2'2]}=1\blue{,}
\end{align*}
we obtain
\be\label{eq:AgAh=Agh}
\begin{aligned}
  &A^g_{2'} A^h_2\BLvert \threeTet{0}{1}{2}{3}{4}{0.25}\Brangle\\
  =&\frac{[01,12'',2''2,23][02'',2''2,23,34]}{[01,12'',2''2,24][12'',2''2,23,34]}
  \BLvert \threeTet{0}{1}{2''}{3}{4}{0.25} \Brangle\\
  =&A^{g\cdot h}_2\BLvert \threeTet{0}{1}{2''}{3}{4}{0.25}\Brangle,
\end{aligned}
\ee
where the last equality is immediate by acknowledging the chain rule $v''_2v_2=v''_2v'_2\cdot v'_2v_2=g\cdot h$.
It would be straightforward but just tedious to repeat the proof for vertices shared by any number of tetrahedra. And there is no need to do so. Therefore we conclude that
\[
  A^g_{v'} A^h_v = A^{g\cdot h}_v.
\]
Once again, for completeness and clarity, the ampltitudes in Eqs. \eqref{eq:AgAh} and \eqref{eq:AgAh=Agh}, and the four \fc conditions used above are implied in the $4$-complex in Fig. \ref{fig:AvProd}.

\section{Unitarity of mutations transformations}\label{app:Tunitarity}
In this appendix, we prove that the mutation transformations defined in Eqs. \eqref{eq:T1move} through \eqref{eq:T4move} are a unitary symmetry on the ground state Hilbert space of our model. Our proof consists of the following two two steps.

\subsection{Symmetry}
In the ground state space $\Hil^0_{\Gamma}\subset\Hil^{B_f=1}_{\Gamma}$, any triangle $v_iv_jv_k$ in $\Gamma$ must obey the following \textit{chain rule}:
 \begin{equation}
  \label{ChainRuleInGroundState}
   [v_i v_j]=[v_iv_j]\cdot[v_jv_k].
 \end{equation}
where we assume the three vertices are in the order $v_i<v_j<v_k$. Hence, we can forget about the operators $B_f$ in this subspace. The ground-state projector (\ref{eq:GSDprojector}) then reduces to $P^0_\Gamma=\prod_{v\in\Gamma}A_v$. Since we only concern whether the mutation transformations are a unitary symmetry of the ground states, we can restrict ourselves to the subspace $\Hil^{B_f=1}_{\Gamma}$ for all $\Gamma$. As such, to show that $T_iP^0_{\Gamma}=P^0_{\Gamma'}T_i$, where $\Gamma'=T_i(\Gamma)$, it suffices to show that\be\label{eq:TBcommute}
 T_i\prod_{v\in\Gamma}A_v=\prod_{v\in\Gamma'}A_v T_i
\ee
holds for all mutation transformation generators $T_i$, $i=1,2,3,4$, and any state in the $\Hil^{B_f=1}_{\Gamma}$ on any $\Gamma$. Note that a $T_i$ acts on at most four tetrahedra but preserves any other tetrahedron in the same graph, so $T_i$ certainly commutes with any $A_v$ at any vertex $v$ that does not lie in the tetrahedra the $T_i$ acts on. Thus, to show Eq. \eqref{eq:TBcommute}, we need only to focus on the part of $\Gamma$ within the scope of the action of $T_i$. Now we shall proceed to prove  Eq. \eqref{eq:TBcommute} for $T_2$. The proofs for $T_1$, $T_3$ and $T_4$ are similar and even simpler, and thus would not be detailed here.

It suffices to consider solely a basis $3$-complex that bounds a $4$-simplex in $\Hil^{B_f=1}_{\Gamma}$, namely $\BLvert\threeTet{0}{1}{2}{3}{4}{0.15}\Brangle$, where the vertex enumerations are in the obvious order. The $T_2$ operator will remove the edge $[24]$ by Definition \ref{eq:T2move}. For the sake of clearance, we will spilt the proof to two steps. First, prove $T_2$ commutes with $A_2A_4$; second, prove $T_2$ commutes with $A_0$, $A_1$ and $A_3$ respectively.

\begin{align}
  & A_2 A_4 T_2 \BLvert\threeTet{0}{1}{2}{3}{4}{0.25}\Brangle\nonumber\\
= & A_2 A_4 [01,12,23,34] \BLvert\twoTet{0}{1}{2}{3}{4}{0.25}\Brangle \nonumber\\
= & \frac{1}{|G|^2}\sum_{[22'],[44']\in~G} [01,12,23,34]\nonumber\\
 &[01,13,34,44']^{-1}[01,12,22',2'3] \BLvert\twoTet{0}{1}{2'}{3}{4'}{0.25}\Brangle
\end{align}

Note that we have two $4$-cocycle conditions respectively for the two 5-simplices $[012344']$ and $[0122'34'],$
\begin{align}
1=&[12,23,34,44'][02,23,34,44']^{-1}\nonumber\\
  &[01,13,34,44'][01,12,24,44']^{-1}\nonumber\\
  &[01,12,23,34'][01,12,23,34]^{-1}
\end{align}
and
\begin{align}
1=&[12,22',2'3,34'][02,22',2'3,34']^{-1}\nonumber\\
  &[01,12',2'3,34'][01,12,23,34']^{-1}\nonumber\\
  &[01,12,22',2'4'][01,12,22',2'3]^{-1}.
\end{align}

Applying the above $4$-cocycle conditions to the $4$-cocycles $[01,12,23,34]$, $[01,13,34,44']$, $[01,12,23,34']$, and $[01,12,22',2'3]$ leads to

\begin{align}
  & A_2 A_4 T_2 \BLvert\threeTet{0}{1}{2}{3}{4}{0.25}\Brangle\nonumber\\
= & \frac{1}{|G|^2}\sum_{[22'],[44']\in G} \frac{[01,12',2'3,34'][12,23,34,44']} {[02,23,34,44']}\nonumber\\
  & \frac{[12,22'2'3,34'][01,12,22'2'4']}{[01,12,24,44'][02,22',2'3,34']} \BLvert\twoTet{0}{1}{2'}{3}{4'}{0.25}\Brangle\nonumber\\
= & T_2 A_2 A_4 \BLvert\threeTet{0}{1}{2'}{3}{4'}{0.25}\Brangle
\end{align}

This concludes the first step: $A_2A_4T_2=T_2A_2A_4$.
Second, we show that $A_0T_2=T_2A_0$. The other two relations $A_1T_2=T_2A_1$, $A_3T_2=T_2A_3$ follow likewise.

\begin{align*}
  & A_0 T_2 \BLvert\threeTet{0}{1}{2}{3}{4}{0.25}\Brangle\\
= & A_0 [01,12,23,34] \BLvert\twoTet{0}{1}{2}{3}{4}{0.25}\Brangle\\
= & \frac{1}{|G|^2}\sum_{[00']\in~G} [01,12,23,34]\\
 &[00',0'1,12,23][00',0'1,13,34] \BLvert\twoTet{0'}{1}{2}{3}{4'}{0.25}\Brangle
\end{align*}

Using the $4$cocycle condition for the 5-simplex $[00'1234]$
\begin{align*}
1=& [0'1,12,23,34][01,12,23,34]^{-1}\nonumber\\
  & [00',0'2,23,34][00',0'1,13,34]^{-1}\nonumber\\
  & [00',0'1,12,24][00',0'1,12,23]^{-1}
\end{align*}
we obtain,
\begin{align*}
  & A_0 T_2 \BLvert\threeTet{0}{1}{2}{3}{4}{0.25}\Brangle\\
= & \frac{1}{|G|^2}\sum_{[00']\in~G} [0'1,12,23,34]\\
  & [00',0'2,23,34][00',0'1,12,24] \BLvert\twoTet{0'}{1}{2}{3}{4'}{0.25}\Brangle\\
= & T_2 A_0 \BLvert\threeTet{0}{1}{2}{3}{4}{0.25}\Brangle,
\end{align*}
which completes the second step, and therefore proves $A_0T_2=T_2A_0$. The proofs for $A_1T_2=T_2A_1$ and $A_3T_2=T_2A_3$ are the same. Combined them together, we arrive at the commutation relation $T_2\prod_{i=0}^{4}A_i=\prod_{i=0}^{4}A_i T_2$.
\subsection{Unitarity}
Now we demonstrate that all mutation transformations satisfy Eq. \eqref{unitary} and thus are unitary on the ground state Hilbert space $\Hil^0$. We need only to show this for the mutation transformation generators $T_1$, $T_2$, $T_3$, and $T_4$. 

We first show that $T_1T_2=1$ and $T_2T_1=1$ hold not only on the ground states but also over the entire subspace $\Hil^{B_f=1}$. Let us consider the action of $T_1T_2$ on a generic basis state as follows, where only the relevant part of the basis graph is displayed.
\begin{align*}
  & T_1T_2\BLvert\threeTet{0}{1}{2}{3}{4}{0.25}\Brangle\\
= & [01,12,23,34] T_1 \BLvert\twoTet{0}{1}{2}{3}{4}{0.25}\Brangle\\
= & \sum_{[24]\in G}[01,12,23,34][01,12,23,34]^{-1} \BLvert\threeTet{0}{1}{2}{3}{4}{0.25}\Brangle\\
= & \sum_{[24]\in G} \BLvert\threeTet{0}{1}{2}{3}{4}{0.25}\Brangle
=\BLvert\threeTet{0}{1}{2}{3}{4}{0.25}\Brangle.
\end{align*}
This is worth of more explanation. The first two equalities follow from the definitions of $T_2$ and $T_1$ respectively. The third equality is simple algebra because the group element $[01]$, $[12]$, $[23]$, $[34]$ and therefore $[01,12,23,34]$ do not change. The last equality owes to  the fact that $[24]$ is the only group element that satisfies the chain rule, as given in the initial state. That $T_2T_1=1$ follows likewise.

As a result, $T_2=T_1^{-1}$ ($T_1=T_2^{-1}$), such that we can define that $T_1^{\dag}=T_2$ ($T_2^{\dag}=T_1$) and unitarity is proved for $T_1$ and $T_2$.

Next, we prove that $T_4T_3=1$ on the entire subspace $\Hil^{B_f=1}$. 
We consider the action of $T_4T_3$ on a generic basis state as follows, in which only the relevant part of the graph is displayed.
\begin{align*}
& T_4 T_3 \BLvert\fourTet{1}{2}{4}{3}{0}{}{0.5}{0}\Brangle\\
=& \sum_{[01],[02],[03],[04]\in G}[01,12,23,34]T_4 \BLvert \fourTet{1}{2}{4}{3}{1}{0}{0.5}{0}\Brangle\\
=&\sum_{[01],[02],[03],[04]\in G} \frac{[01,12,23,34]}{ [01,12,23,34]} \BLvert\fourTet{1}{2}{4}{3}{0}{}{0.5}{0}\Brangle\\
=& \BLvert\fourTet{1}{2}{4}{3}{0}{}{0.5}{0}\Brangle,
\end{align*}
At this point, one may think that $T_4$ is the inverse of $T_3$ on $\Hil^{B_f=1}$. But this is not true because $T_3T_4\neq 1$ in general. Nevertheless, fortunately, as we now show, $T_3T_4=1$ on the ground states $\Hil^0$.

Since $T_3T_4P^0=T_3P^0T_4=P^0T_3T_4$ on $\Hil^0$, we have
\begin{align*}
& T_3T_4P^0(0,1,2,3,4) \BLvert\fourTet{1}{2}{4}{3}{1}{0}{0.5}{0}\Brangle\\
=& T_3 P^0(0,1,2,3,4) T_4 \BLvert\fourTet{1}{2}{4}{3}{1}{0}{0.5}{0}\Brangle\\
=& [01,12,23,34]^{-1} T_3 P^0(1,2,3,4) \BLvert\fourTet{1}{2}{4}{3}{0}{}{0.5}{0}\Brangle\\
=& [01,12,23,34]^{-1} P^0(1,2,3,4) T_3 \BLvert\fourTet{1}{2}{4}{3}{0}{}{0.5}{0}\Brangle\\
=& P^0(1,2,3,4) \sum_{[0'1],[0'2],[0'3],[0'4]\in G}[01,12,23,34]^{-1}\times\\
& [0'1,12,23,34]   \BLvert\fourTet{1}{2}{4}{3}{1}{0'}{0.5}{0}\Brangle\\
=& P^0(1,2,3,4)\frac{1}{|G|}\sum_{[00']\in G}[00',0'2,23,34]^{-1}\\
& \times [00',0'1,13,34][00',0'1,12,24]^{-1}\\
& \times [00',0'1,12,23]
\BLvert\fourTet{1}{2}{4}{3}{1}{0'}{0.5}{0}\Brangle\\
=& P^0(1,2,3,4)A_{0} \BLvert\fourTet{1}{2}{4}{3}{1}{0}{0.5}{0}\Brangle\\
=& P^0(0,1,2,3,4) \BLvert\fourTet{1}{2}{4}{3}{1}{0}{0.5}{0}\Brangle
\end{align*}
where $P^0(1,2,3,4)$ and $P^0(0,1,2,3,4)$ are projectors acting on the vertices of the corresponding basis graph. The fifth equality in the equation above is obtained by applying to the following $4$-cocycle condition for $[00'1234]$ in the fourth equality.
\begin{align*}
1=& [0'1,12,23,34][01,12,23,34]^{-1}\nonumber\\
  & [00',0'2,23,34][00',0'1,13,34]^{-1}\nonumber\\
  & [00',0'1,12,24][00',0'1,12,23]^{-1}
\end{align*}


The derivation above shows that $T_3T_4=1$ on the ground states, which together with $T_4T_3=1$ on $\Hil^{B_f=1}$, results in that $T_3=T_4^{\dag}$ and $T_4=T_3^{\dag}$ on $\Hil^0$. That is, $T_4$ and $T_3$ are unitary on the ground  states.

An amusing byproduct of our proof above is that in the subspace $\Hil^{B_f=1}$,
\be
T_3T_4=A_v,
\ee
where $v$ is the vertex annihilated by the action of $T_4$.
\section{Details of the ground-state projector}\label{app:Rewrite3torusAx}
Here we prove Eq. \eqref{eq:torusAxRewrite}.
We start with the amplitude in Eq. \eqref{eq:torusAx} and rewrite it compactly as 
\be\label{eq:AxTorusInIxabc}
\frac{I^x(a,b,c)I^x(b,c,a)I^x(c,a,b)}{I^x(a,c,b)I^x(c,b,a)I^x(b,a,c)},
\ee
where we define
\be\label{eq:Ixabc}
\begin{aligned}
  & I^x(a,b,c):= \frac{[b,c\bar x,x,a\bar x][x,a\bar x,xb\bar x,xc\bar x]}{[a,b,c\bar x,x][c\bar x,x,a\bar x,xb\bar x]},\\
& \quad\quad\quad\quad\quad\quad a=k,\quad b=g,\quad c=\bar gh.
\end{aligned}
\ee
So, clearly in the above, $a,b$, and $c$ commute with each other because $g,h$, and $k$ do. Using the normalization condition \eqref{NormalizationCondition} and the following \fc\ conditions
\begin{align}
  \label{dtdtdt}
  &\delta[a,b,c,\bar x,x]=1,
  \nonumber\\
  &\delta[c,\bar x, x,a\bar x,xb\bar x]=1,
  \nonumber\\
  &\delta[b,c,\bar x,x,a\bar x]=1,
  \nonumber\\
  &\delta[x,\bar x,xa\bar x,xb\bar x,xc\bar x]=1,
  \nonumber
\end{align}
the four \fcs\ in Eq. \eqref{eq:Ixabc} respectively become  
\begin{align}
  &[a,b,c\bar x,x]=\frac{[b,c,\bar x,x][a,bc,\bar x,x]}{[ab,c,\bar x,x][a,b,c,\bar x]}
  \nonumber\\
  &[c\bar x,x,a\bar x,xb\bar x]=\frac{[\bar x,x,a\bar x,xb\bar x][c,\bar x,x,ab\bar x]}{[c,\bar x,xa\bar x,xb\bar x][c,\bar x,x,a\bar x]}
  \nonumber\\
  &[b,c\bar x,x,a\bar x]^{-1}=\frac{[c,\bar x,x,a\bar x][b,c,\bar x,xa\bar x]}{[bc,\bar x,x,a\bar x][b,c,\bar x,x]}
  \nonumber\\
  &[x,a\bar x,xb\bar x,xc\bar x]^{-1}=\frac{[\bar x,xa\bar x,xb\bar x,xc\bar x][x,\bar x,xa\bar x,xbc\bar x]}{[x,\bar x,xab\bar x,xc\bar x][x,\bar x,xa\bar x,xb\bar x]}.
  \nonumber
\end{align}
And substituting in the above
\[
[\bar x,x,a\bar x,xb\bar x]=\frac{[c\bar x,x,\bar x,xab\bar x]}{[c,\bar x,xa\bar x,xb\bar x][c\bar x,x,\bar x,xa\bar x]},
\]
which is a result of the normalization condition \eqref{NormalizationCondition} and the \fc condition $\delta[c\bar x,x,\bar x,xa\bar x,xb\bar x]=1.$
Note that the above identities apply to the other five factors in Eq. \eqref{eq:Ixabc} under suitable permutations of $a,b$, and $c$. After some algebra, The expression \eqref{eq:AxTorusInIxabc} again consists of twenty-four \fcs and can be written as
\be\label{eq:Ax3torusTwisted3cocycles}
\frac{[\bar x,xc\bar x,xb\bar x]_a}{[b,c,\bar x]_a} \frac{[b,\bar x,xc\bar x]_a}{[c,\bar x,xb\bar x]_a} \frac{[c,b,\bar x]_a}{[\bar x,xb\bar x,xc\bar x]_a},
\ee
where
\be\label{eq:twisted3cocycleDef}
[x,y,z]_u\defeq \frac{[x,\bar xux,y,z][x,y,z,\overline{xyz}u(xyz)]}{[u,x,y,z][x,y, \overline{xy}u(xy),z]},
\ee 
which is in fact the slant product \eqref{eq:slantProd} at $n=4$. Let us call $[x,y,z]_u$ a normalized \textbf{twisted $3$-cocycle} by $u$ because it does not satisfy the usual $3$-cocycle condition but a twisted one, namely,
\be\label{eq:twisted3cocycleCond}
\begin{aligned}
&\widetilde\delta[w,x,y,z]_u=\frac{[x,y,z]_{\bar wuw}[w,xy,z]_u[w,x,y]_u}{[wx,y,z]_u[w,x,yz]_u}=1,\\
&[1,y,z]_u=[x,1,z]=[x,y,1]_u=1
\end{aligned}
\ee
$\forall w,x,y,z,u\in G$, which can be verified by the definition \eqref{eq:twisted3cocycleDef} and the following usual \fc conditions:
\begin{align*}
&\delta[w,x,y,z,\overline{wxyz}u(wxyz)]=1,\\
&\delta[w,x,y,\overline{wxy}u(wxy),z]=1,\\
&\delta[w,x,\overline{wx}u(wx),y,z]=1,\\
&\delta[w,\bar wuw,x,y,z]=1,\\
&\delta[a,w,x,y,z]=1. 
\end{align*}

Equation \eqref{eq:Ax3torusTwisted3cocycles} is yet not the end of the story; indeed, it can be casted in an even more compact form as follows.
\begin{align}
&\left( \frac{[b,c,\bar x]_a[c,\bar x,xb\bar x]_a}{[c,b,\bar x]_a}\right)^{-1} \frac{[b,\bar x,xc\bar x]_a[\bar x,xc\bar x,xb\bar x]_a}{[\bar x,xb\bar x,xc\bar x]_a}\nonumber  \\
=&\left( \frac{[c,\bar x]_{a,b}}{[\bar x,xc\bar x]_{a,b}}\right)^{-1} 
\label{eq:Ax3torusTwisted2cocycles},
\end{align}
where use is made of 
\be\label{eq:Dtwisted2cocycleDef}
[y,z]_{x,w}\defeq \frac{[w,y,z]_x[y,z,\overline{yz}w(yz)]_x}{[y,\bar y wy,z]_x }.
\ee
Note because $a,b$, and $c$ commute in $G$, the LHS of Eq. \eqref{eq:Ax3torusTwisted2cocycles} appear somewhat simpler than the definition \eqref{eq:Dtwisted2cocycleDef}. We dub $[y,z]_{x,w}$ a normalized \textbf{doubly-twisted} $2$-cochain by the pair $(x,w)$, where $x$ then $w$ induce the twisting. If $xw=wx$, $[y,z]_{x,w}$ turns out to be a doubly-twisted $2$-cocycle meeting the doubly-twisted $2$-cocycle condition:
\be\label{eq:Dtwisted2cocycleCond}
\widetilde\delta[x,y,z]_{w,u}=\frac{[y,z]_{\bar xwx ,\bar xux}[x,yz]_{w,u}}{[xy,z]_{w,u}[x,y]_{w,u}}\bigg\vert_{wu=uw}=1,
\ee
by definition \eqref{eq:Dtwisted2cocycleDef} and appropriate applications of the twisted $3$-cocycle condition \eqref{eq:twisted3cocycleCond}, namely, \begin{align*}
&\delta[x,y,z,\overline{xyz}u(xyz)]_w=1,\\
&\delta[x,y,\overline{xy}u(xy),z]_w=1,\\
&\delta[u,x,y,z]_w\big|_{uw=wu}=1, 
\end{align*}
where the first two identities hold for all $u,w\in G$. In fact, as far as the ground states are concerned, $xw=wx$ is always true (recall Eq. \eqref{eq:Ax3torusTwisted2cocycles} in which $ab=ba$ is understood). Besides, $[1,z]_{x,w}=[y,z]_{x,w}=1$. Clearly, if we restrict the arguments of a doubly-twisted $2$-cocycle $[y,z]_{x,w}$ to $Z_{x,w}=\{u\in G|ux=xu,uw=wu\}$, it would satisfy the usual $2$-cocycle condition.

Furthermore, we have,
\be\label{eq:DtwistedProperty1}
[y,z]_{x,w}=[z,y]_{w,x}^{-1},\quad \forall y,z\in Z_{x,w}.
\ee 
which can be easily verified by expanding the LHS in terms of the relevant \fcs and rearrange the \fcs to the form that gives rise to the RHS. Two more identities worth of note are
\begin{align}
&[c,d]_{a,b}=[c,d]_{b,a}^{-1},\quad \forall ab=ba, \label{eq:DtwistedIDabba}\\
&[b,c]_{a,b}=[c,b]_{a,b},\quad \forall a,b,c\in G,\label{eq:DtwistedIDonly3}
\end{align}   
which can be easily established by expansion in terms of twisted $3$-cocycles or $4$-cocycles.

One may wonder if Eq. \eqref{eq:Ax3torusTwisted2cocycles} is the only simplification of the expression \eqref{eq:AxTorusInIxabc}. The answer is ``No". In fact, it is not hard to see that the expression \eqref{eq:Ax3torusTwisted2cocycles} is invariant under all cyclic permutations of $a,b$, and $c$ but is sent to its inverse by any exchange of any two of $a,b$, and $c$. Therefore, if we follow our definition \eqref{eq:eta}, we can immediately pin down the following identities, indicating the equivalent simplifications of the expression \eqref{eq:AxTorusInIxabc}.
\be\label{eq:etaPermuteID}
\begin{aligned}
\frac{[c,\bar x]_{a,b}}{[\bar x,xc\bar x]_{a,b}}\defeq & \eta^{a,b}(c,x) =\eta^{b,c}(a,x) =\eta^{c,a}(b,x)\\
=&\eta^{b,a}(c,x)^{-1}=\eta^{c,b}(a,x)^{-1}=\eta^{a,c}(b,x)^{-1},
\end{aligned}
\ee   
where $a,b$, and $c$ commute with each other.
\section{Projective characters of the centralizer}\label{app:centralizer&projChi}
 We put forward and prove the following propositions. 

\begin{proposition}\label{prop:isoConjClass}
Consider any conjugacy class $C^A$ of a finite group G. For any $a,a'\in C^A$, the isomorphism between $Z_a$ and $Z_{a'}$ is a bijection between the conjugacy classes of $Z_a$ and those of $Z_{a'}$. 
\end{proposition}
\proof{Since $a,a'\in C^A$, we assume $a'=xa\bar x$ for some $x\in G$. This conjugation by $x$ is in fact an isomorphism: $Z_{a'}=xZ_a\bar x$. Now consider $b,b'\in C^B_{Z_a}$, a conjugacy class of $Z_a$, they are respectively mapped by the conjugaction to $xb\bar x$ and $xb'\bar x$. Let $b'=yb\bar y$ for some $y\in Z_a$, we can see that $xb'\bar x=(xy\bar x)xb\bar x(x\bar y \bar x)$. Note that $xy\bar x\in Z_{a'}$. Hence, $xb\bar x$ and $xb'\bar x$ belong to the same conjugacy class in $Z_{a'}$, and this class must be isomorphic to $C^B_{Z_a}$.}
\begin{proposition}\label{prop:isoZabFixed_a}
For any $b,b'\in C^B_{Z_a}$, some conjugacy class of $Z_a$, $Z_{a,b}\cong Z_{a,b'}$.
\end{proposition}
\proof{This is actually a trivial consequence of the fact that $Z_b\cong Z_{b'}$ for any $b,b'\in C^A_G$, with however, now $G$ is restricted to $Z_a$.}
\begin{proposition}\label{prop:isoZab}
For any $a,a'\in C^A_G$, where $a'=xa\bar x$ for some $x\in G$, any $b\in C^B_{Z_a}$, and $b'\in C^{B'}_{Z_{a'}}$ with $b'=xb\bar x$, $Z_{a,b}\cong Z_{a',b'}$.
\end{proposition}
\proof{For any $c\in Z_{a,b}$, clearly $xc\bar x$ commute with both $a'$ and $b'$ and is thus in $Z_{a',b'}$, which establishes the proposition. Note that we have $C^B_{Z_a}\cong C^{B'}_{Z_{a'}}$ by Proposition \ref{prop:isoConjClass}.}

Bearing Proposition \ref{prop:isoZab} in mind, given the representation $\widetilde{\rho}^{k,g}_{\mu}$ of $Z_{k,g}$ for a fixed $k\in C^A$ and $g\in C^B_{Z_{k}}$,   we construct $\widetilde{\rho}^{xk\bar x,xg\bar x}_{\mu}$ as follows. For all $x\in G$, the elements $xh'\bar x$ runs over all elements in $xZ_{k,g}\bar x$, while $h$ runs over all elements in $Z_{k,g}$. We can then define a projective representation $\widetilde{\rho}^{xk\bar x,xg\bar x}_{\mu}$ of $xZ_{k,g}\bar x$ from a given representation $\widetilde{\rho}^{k,g}_{\mu}$ of $Z_{k,g}$, by
\begin{align}
  \label{eq:rhoxgx}
  \widetilde{\rho}^{xk\bar x,xg\bar x}_\mu(xh'\bar x)
  =\eta^{k,g}(h',x)^{-1}\widetilde{\rho}^{k,g}_{\mu}(h')
\end{align}

We now verify that $\widetilde{\rho}^{xk\bar x,xg\bar x}_\mu$ is truly a $\beta_{xk\bar x,xg\bar x}$-representation by checking the the multiplication rule for all $h'_1,h'_2\in Z_{k,g}$
\begin{align}
  \label{eq:betaxgxrepresentation}
  &\widetilde{\rho}^{xk\bar x,xg\bar x}_\mu(xh'_1\bar x)\widetilde{\rho}^{xk\bar x,xg\bar x}_\mu(xh'_2\bar x)
  \nonumber\\
  =&\eta^{k,g}(h'_1,x)^{-1}\eta^{k,g}(h'_2,x)^{-1}
  \widetilde{\rho}^{k,g}_{\mu}(h'_1)\widetilde{\rho}^{k,g}_{\mu}(h'_2)
  \nonumber\\
  =&\eta^{k,g}(h'_1,x)^{-1}\eta^{k,g}(h'_2,x)^{-1}\beta_{k,g}(h'_1,h'_2)\widetilde{\rho}^{k,g}_{\mu}(h'_1h'_2)
  \nonumber\\
  =&\frac{\eta^{k,g}(h'_1h'_2,x)}{\eta^{k,g}(h'_1,x)\eta^{k,g}(h'_2,x)}\beta_{k,g}(h'_1,h'_2) \widetilde{\rho}^{xk\bar x,xg\bar x}(xh'_1h'_2\bar x)\nonumber\\
  =&\beta_{xk\bar x,xg\bar x}(xh'_1\bar x,xh'_2\bar x)\widetilde{\rho}^{xk\bar x,xg\bar x}(xh'_1h'_2\bar x),
\end{align}
where the last equality holds because of the following relation:
\be\label{eq:3etaRel}
\frac{\eta^{k,g}(h'_1h'_2,x)}{\eta^{k,g}(h'_1,x)\eta^{k,g}(h'_2,x)}=   \frac{\beta_{xk\bar x,xg\bar x}(xh'_1\bar x,xh'_2\bar x)}{\beta_{k,g}(h'_1,h'_2)},
\ee
which can be checked by applying the following doubly-twisted 2--cocycle conditions \eqref{eq:Dtwisted2cocycleCond}
to the LHS of Eq. \eqref{eq:3etaRel}
\begin{align*}
\widetilde\delta\beta_{k,g}(h'_1,h'_2,\bar x)=1,\\
\widetilde\delta\beta_{k,g}(h'_1,\bar x,xh'_2\bar x)=1,\\
\widetilde\delta\beta_{k,g}(\bar x,xh'_1\bar x,xh'_2\bar x)=1.
\end{align*}
Note that the second and third identities in the above are the usual $2$-cocycle condition because $h'_1,h'_2\in Z_{k.g}$. 

An immediate consequence of the above isomorphism is the relation between the projective characters,
\be
\label{eq:CharacterInConjugacyClassesAppendix}
\begin{aligned}
\widetilde{\chi}^{xk\bar x,xg\bar x}_\mu(xh'\bar x) &= \eta^{k,g}(h',x)^{-1}  \widetilde{\chi}^{k,g}_{\mu}(h'),\\
\overline{\widetilde{\chi}^{xk\bar x,xg\bar x}_\mu}(xh'\bar x) &= \eta^{k,g}(h',x)^{} \overline{\widetilde{\chi}^{k,g}_{\mu}}(h'),
\end{aligned}
\ee
for all $x\in G$, which is the very Eq. (\ref{eq:CharacterInConjugacyClasses}). We remark that in general, $\overline{\widetilde\chi^{k,g}_{\mu}}(h')\neq \widetilde\chi^{k,g}_{\mu}(\overline{h'})$ for such projective representations. Rather, we have 
\be\label{eq:chiChiBar}
 \widetilde\chi^{k,g}_{\mu}(\overline{h'}) = \overline{\widetilde\chi^{k,g}_{\mu}}(h') \beta^{k,g}(h,\overline{h'}),
\ee 
where $\beta^{k,g}(h,\overline{h'})\equiv \beta^{k,g}(\overline{h'},h'),\, \forall h'\in Z_{k,g}$ due to the the $2$-cocycle condition \eqref{eq:DtwistedUsual2cocycleCond}. Equation \eqref{eq:chiChiBar} is a direct consequence of the definition \eqref{eq:betarepresentation} of the $\beta$-representation.   

Equation \eqref{eq:CharacterInConjugacyClassesAppendix} leads to the following proposition. 
\begin{proposition}\label{prop:chiIs0}
If $h'\in Z_{k,g}$ is not $\beta_{k,g}$-regular, $\widetilde{\chi}^{k,g}_{\mu}(h')=0$.
\end{proposition}
\proof{
The proof is straightforward. If $h'\in Z_{k,g}$ is not $\beta_{k,g}$-regular, there must exist $l\in Z_{k,g,h'}$, such that $\beta_{k,g}(h',l)\neq\beta_{k,g}(l,h')$. Let $x=\bar l$ in Eq. \eqref{eq:CharacterInConjugacyClassesAppendix}, we have
\[
\widetilde{\chi}^{k',g'}_{\mu}(h')= \frac{\beta_g(l,h')} {\beta_g(h',l)}  
\widetilde{\chi}^{k',g'}_{\mu}(h')\Longrightarrow\widetilde{\chi}^{k',g'}_{\mu}(h')=0.
\]
}
In view of Proposition \ref{prop:isoZabFixed_a}, we can consider simultaneous conjugation of the $g$ and $h'$ in certain $\widetilde{\chi}^{k,g}_{\mu}(h')$. Since $g\in Z_{k}$ for any fixed $k$ must hold under such a conjugation, we simply can restrict Eq. \eqref{eq:CharacterInConjugacyClassesAppendix} to this case and obtain
\be\label{eq:rhoygy}
\begin{aligned}
\widetilde{\chi}^{k,yg\bar y}_\mu(yh'\bar y) &= \eta^{k,g}(h',y)^{-1}  \widetilde{\chi}^{k,g}_{\mu}(h'),\\
\overline{\widetilde{\chi}^{k,yg\bar y}_\mu}(yh'\bar y) &= \eta^{k,g}(h',y)^{} \overline{\widetilde{\chi}^{k,g}_{\mu}}(h'),
\end{aligned}
\ee
for all $y\in Z_{k}$.

We now establish the following theorem.
\begin{theorem}\label{theo:betaRegClass}
A group element $c\in Z_{a,b}$ is $\beta_{a,b}$-regular if and only if the entire conjugacy class $[c]\subset Z_{a,b}$ is $\beta_{a,b}$-regular.
\end{theorem}
\proof{Let us proof this for an arbitrary $c'\in [c]$, then the statement for the entire conjugacy class follows immediately. Assume $c'=yc\bar y$ for some $y\in Z_{a,b}$. Since $c$ is $\beta_{a,b}$-regular, we have
\be\label{eq:cISbetaReg}
\frac{\beta_{a,b}(c,d)}{\beta_{a,b}(d,c)}=\eta^{a,b}(c,\bar d)=1,\quad\forall d\in Z_{a,b,c}.
\ee
Our goal is to show that 
\be\label{eq:c'ISbetaReg}
\frac{\beta_{a,b}(yc\bar y,d)}{\beta_{a,b}(d,yc\bar y)}=\eta^{a,b}(yc\bar y,\bar d)=1,\quad\forall d\in Z_{a,b,c}.
\ee
To this end, let us plug in Eq. \eqref{eq:3etaRel} the relations $h_1=y$, $h_2=c\bar y$, and $x=\bar d$, then we obtain
\be\label{eq:c'ISbetaRegUseEta1}
\frac{\eta^{a,b}(yc\bar y,\bar d)}{\eta^{a,b}(y,\bar d)\eta^{a,b}(c\bar y,\bar d)} =\frac{\beta_{a,b}(y,c\bar y)}{\beta_{a,b}(y,c\bar y)}=1.
\ee
In the above and in the subsequent derivation, keep in mind that $d$ commutes with $a,b,c$, and $y$, also that $y$ commutes with $a,b$, and $d$. Let us again plug in Eq. \eqref{eq:3etaRel} but with $h'_1=c$, $h'_2=\bar y$, and $x=\bar d$, and get
\be\label{eq:c'ISbetaRegUseEta2}
\frac{\eta^{a,b}(c\bar y,\bar d)}{\eta^{a,b}(c,\bar d)\eta^{a,b}(\bar y,\bar d)} =\frac{\beta_{a,b}(c,\bar y)}{\beta_{a,b}(c,\bar y)}=1.
\ee
In view of Eqs. \eqref{eq:c'ISbetaRegUseEta2} and \eqref{eq:cISbetaReg}, Eq. \eqref{eq:c'ISbetaRegUseEta1} becomes
\be\label{eq:c'ISbetaRegUseEta3}
\frac{\eta^{a,b}(yc\bar y,\bar d)}{\eta^{a,b}(y,\bar d)\eta^{a,b}(\bar y,\bar d)} =1.
\ee
As such, proving Eq. \eqref{eq:c'ISbetaReg} amounts to proving that the denominator on the LHS of the equation above is unity. We observe that
\be
\frac{1}{\eta^{a,b}(y,\bar d)\eta^{a,b}(\bar y,\bar d)}=\eta^{a,b}(d,\bar y)\eta^{a,b}(d,y),
\ee
simply by the definition of $\eta$ in Eq. \eqref{eq:eta} and that $yd=dy$. Now as pointed out below Eq. \eqref{eq:rhoEta}, $\eta^{a,b}(d,\cdot)$ is a $1$d representation of $Z_{a,b,d}$, where the `$\cdot$' is a wild card for any element in $y\in Z_{a,b,d}$. Hence, we must have 
\[
\eta^{a,b}(d,\bar y)\eta^{a,b}(d,y)=\eta^{a,b}(d,y\bar y)=\eta^{a,b}(d,1)=1,
\] 
where the last equality holds because $\beta_{a,b}$ is normalized. Therefore by Eq. \eqref{eq:c'ISbetaRegUseEta3}, we finally have $\eta^{a,b}(yc\bar y,\bar d)=1$, which is the very Eq. \eqref{eq:c'ISbetaReg}.

Since $c'$ is arbitrary, our steps above applies to the entire class $[c]$. The necessary side of the theorem is obvious, concluding the theorem.}

The following property of the representations $\widetilde\rho^{k,g}_\mu(h')$ is also key to deriving the topological quantum numbers of the model.
\be\label{eq:rhogg}
\widetilde\rho^{k,g}_\mu(g)=\frac{\widetilde\chi^{k^A,g^B}_\mu(g^B)}{\dim_\mu} \mathds{1},
\ee
where $g^B$ is any representative of $C^B_{Z_k}$. The proof contains two steps. First, we have $\forall h'\in Z_{k,g}$,
\begin{align*}
\widetilde\rho^{k,g}_\mu(g)\widetilde\rho^{k,g}_\mu(h')=&\beta_{k,g}(g,h') \widetilde\rho^{k,g}_\mu(gh')\\
\xeq{Eq. \eqref{eq:DtwistedIDonly3}}&\beta_{k,g}(h',g) \widetilde\rho^{k,g}_\mu(h'g)\\
=&\rho^{k,g}_\mu(h')\widetilde\rho^{k,g}_\mu(g).
\end{align*} 
Then by Schur's Lemma, we infer that
\[
\widetilde\rho^{k,g}_\mu(g)\propto\mathds{1}.
\]
Second, Eqs. \eqref{eq:etah=g'Is1} and \eqref{eq:rhoygy} indicate that
\[
\widetilde\rho^{xk\bar x,xg\bar x}_\mu(xg\bar x)=\widetilde\rho^{k,g}_\mu(g), \forall x\in G.
\]  
Therefore, for $g\in C^B_{Z_k}$, $\widetilde\rho^{k,g}_\mu(g)$ is a constant of the conjugacy class pair $(C^A,C^B)$ and is a multiple of the identity matrix, leading to Eq. \eqref{eq:rhogg}.
\section{Modular transformations}\label{app:modularTrans}
\begin{figure}[h!]
\centering
\includegraphics[scale=2]{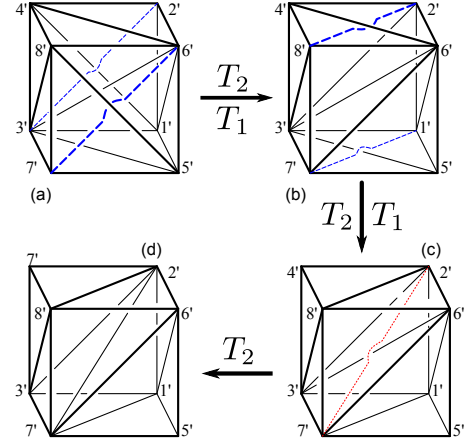}
\caption{(Color online) Mutation moves used in the final step of the $\str$ transformation. Certain edges are bent to emphasize that these moves should be viewed  in $4$D. (a) already encodes a $T_1$ move by the blue (dashed) lines $2'3'$ and $6'7'$. The two $3$-complexes $1'2'3'4'6'$ and $3'5'6'7'8'$ are in fact identified, so the two parallel lines $2'3'$ and $6'7'$ refers to the same mutation move. The situation is similar in (b). The other five moves are clear. (d) fits in the final state in Eq. \eqref{eq:SxRaw}.}
\label{fig:StransPachner}
\end{figure}

In this appendix, we detail our construction of the $SL(3,\Z)$ generators as the modular matrix operators $\str^x$ and $\ttr^x$.
To better visualize the $\str$ transformation, we shear Fig. \ref{fig:3torusA} to the left a bit, which does not change the triangulation at all, as follows.

\begin{align}
 & \str^x:  \BLvert\torusBeforeS{1}{5}{2}{6}{3}{7}{4}{8}{1.75}\Brangle
  \nonumber\\
\mapsto &[8'2,23,34,48]^{-1}[8'2,23,37,78][8'2,26,67,78]^{-1}\x \nonumber \\
&\frac{[6'1,12,23,37][6'8',8'2,26',67][6'1,15,56,67]}{[6'8',8'2,23,37][6'1,12,26,67]}\x \nonumber \\
&\frac{[4'6',6'1,12,26]}{[4'6',6'8',8'2,26][4'6',6'1,15,56]}\x \nonumber \\
&[2'4',4'6',6'1,15][7'8',8'2,23,34] \x \nonumber \\
&\frac{[5'6',6'8',8'2,23]}{[5'7',7'8',8'2,23][5'6',6'1,12,23]}\x \nonumber \\
&\frac{[3'5',5'7',7'8',8'2][3'4',4'6',6'8',8'2]}{[3'5',5'6',6'8',8'2][3'4',4'6',6'1,12]}\x \nonumber \\
&\frac{[3'5',5'6',6'1,12][1'3',3'4',4'6',6'1]}{[1'3',3'5',5'6',6'1][1'2',2'4',4'6',6'1]}\x \nonumber\\
  &\BLvert\torusBeforeS{1'}{2'}{3'}{4'}{5'}{6'}{7'}{8'}{1.75} \Brangle\nonumber\\
  \mapsto
 &[8'2,23,34,48]^{-1}[8'2,23,37,78][8'2,26,67,78]^{-1}\x \nonumber \\
&\frac{[6'1,12,23,37][6'8',8'2,26',67][6'1,15,56,67]}{[6'8',8'2,23,37][6'1,12,26,67]}\x \nonumber \\
&\frac{[4'6',6'1,12,26]}{[4'6',6'8',8'2,26][4'6',6'1,15,56]}\x \nonumber \\
&[2'4',4'6',6'1,15] [7'8',8'2,23,34] \x \nonumber \\
&\frac{[5'6',6'8',8'2,23]}{[5'7',7'8',8'2,23][5'6',6'1,12,23]}\x \nonumber \\
&\frac{[3'5',5'7',7'8',8'2][3'4',4'6',6'8',8'2]}{[3'5',5'6',6'8',8'2][3'4',4'6',6'1,12]}\x \nonumber \\
&\frac{[3'5',5'6',6'1,12][1'3',3'4',4'6',6'1]}{[1'3',3'5',5'6',6'1][1'2',2'4',4'6',6'1]}\x \nonumber\\
  &\frac{[1'2',2'3',3'4',4'6']^2[1'2',2'3',3'6',6'7']}{[2'3',3'4',4'6',6'8']^2[2'3',3'6',6'7',7'8']} \nonumber\\
&\BLvert \torusAfterS{1'}{2'}{3'}{4'}{5'}{6'}{7'}{8'}{1.75}\Brangle,\label{eq:SxRaw}
\end{align}
where $x=1'1=2'5=3'2=4'6=5'3=6'7=7'4=8'8$, and $1'<2'<3'<4'<5'<6'<7'<8'<1<2<3<4<5<6<7<8$. The factors after the first `$\mapsto$' is obtained by making the transformations $8\rightarrow 8'$, $7\rightarrow 6'$, $6\rightarrow 4'$, $5\rightarrow 2'$, $4\rightarrow 7'$, $3\rightarrow 5'$, $2\rightarrow 3'$, and $1\rightarrow 1'$ in order, giving rise to the rows of factors respectively. Each such transformation can be thought as realized by a vertex operator $A^x$ acting on the vertex being transformed, yielding the corresponding \fcs. Nevertheless, the state after the vertex transformations is yet not the right state after the transformation, to obtain which a few mutation transformations (defined in Eqs. \eqref{eq:T1move} and \eqref{eq:T2move}) are required. Figure \ref{fig:StransPachner} illustrates the procedure of applying the necessary mutation transformations.

Six mutation moves in order, namely $T_1(1'2'3'4'6')$, $T_2(1'2'3'4'6')$, $T_1(2'3'4'6'8')$, $T_2(2'3'4'6'8')$, $T_1(2'3'6'7'8')$ and $T_2(1'2'3'6'7'),$ contribute the factor of four \fcs in the last row of Eq. \eqref{eq:SxRaw}. Worth of note is that the two blue (dash-dot) lines in Fig. \ref{fig:StransPachner}(a) seem indicating two different $T_1$ moves; however, since the two $3$-complexes $1'2'3'4'6'$ and $3'5'6'7'8'$ are in fact identified, the $T_1$ move about $1'2'3'4'6'$ and that about $3'5'6'7'8'$ are done simultaneously and must be thought as a single $T_1$ move. Hence, one can choose either $1'2'3'4'6'$ or $3'5'6'7'8'$ as the one being acted on by the $T_1$. And we choose the former in Eq. \eqref{eq:SxRaw}. Similarly, for the $T_2$ move taking Fig. \ref{fig:StransPachner}(a) to Fig. \ref{fig:StransPachner}(b), we choose the $3$-complex $2'3'4'6'8'$. Moreover, since $T_1(1'2'3'4'6')$ and $T_2(1'2'3'4'6')$ yield the same amplitude, so do $T_1(2'3'4'6'8')$ and $T_2(2'3'4'6'8')$, we obtain the square in the  last row of Eq. \eqref{eq:SxRaw}.

Plugged in with the group elements, Eq. \eqref{eq:SxRaw} becomes
\begin{align}
&\str^x\ket{k,g,h}\\
= &[x\overline{hk},\bar gh,g,k]^{-1}[x\overline{hk},\bar gh,k,g][x\overline{hk}.k,\bar gh,g]^{-1}\x \nonumber \\
&\frac{[x\overline{hk},g,\bar gh,k][xg\bar x,x\overline{hk},k,\bar gh][x\overline{hk},k,g,\bar gh]}{[xg\bar x,x\overline{hk},\bar gh,k][x\overline{hk},g,k,\bar gh]}\x \nonumber \\
&\frac{[x\bar gh\bar x,x\overline{hk},g,k]}{[x\bar gh\bar x,xg\bar x,x\overline{hk},k][x\bar gh\bar x,x\overline{hk},k,g]}\x \nonumber \\
&[xg\bar x,x\bar gh\bar x,x\overline{hk},k] [xk\bar x,x\overline{hk},\bar gh,g] \x \nonumber \\
&\frac{[xk\bar x,xg\bar x,x\overline{hk},\bar gh]}{[xg\bar x,xk\bar x,x\overline{hk},\bar gh][xk\bar x,x\overline{hk},g,\bar gh]}\x \nonumber \\
&\frac{[x\bar gh\bar x,xg\bar x,xk\bar x,x\overline{hk}][xk\bar x,x\bar gh\bar x,xg\bar x,x\overline{hk}]}{[x\bar gh\bar x,xk\bar x,xg\bar x,x\overline{hk}][xk\bar x,x\bar gh\bar x,x\overline{hk},g]}\x \nonumber \\
&\frac{[x\bar gh\bar x,xk\bar x,x\overline{hk},g][xg\bar x,xk\bar x,x\bar gh\bar x,x\overline{hk}]}{[xg\bar x,x\bar gh\bar x,xk\bar x,x\overline{hk}][xk\bar x,xg\bar x,x\bar gh\bar x,x\overline{hk}]}\x \nonumber\\
  &\frac{[xk\bar x,x\bar kg\bar x,xk\bar x,x\bar gh\bar x]^2[xk\bar x,x\bar kg\bar x,x\bar ghk\bar x,x\bar kg\bar x]}{[x\bar kg\bar x,xk\bar x,x\bar gh\bar x,xg\bar x]^2[x\bar kg\bar x,x\bar ghk\bar x,x\bar kg\bar x,xk\bar x]}\x \nonumber\\
& \ket{xh\bar x,xk\bar x,xg\bar x},\label{eq:SxRawGroupElement}
\end{align}
where we used the natural basis in Fig. \ref{fig:3torusA} for notational simplicity, and we can freely switch to the physical basis in Fig. \ref{fig:3torusB} at anytime. The $\str^x$ transforms the basis vector $\ket{k,g,h}$ to $\ket{xh\bar x,xk\bar x,xg\bar x}$ because $g=12\mapsto 1'2'=xk\bar x$, $h=13\mapsto 1'3'=xg\bar x$, and $k=15\mapsto 1'5'=xh\bar x$. As to the physical basis, we readily see that 
\be\label{eq:physBasisStrans}
\str^x:\ket{k,g,h'}\mapsto\ket{xgh'\bar x,xk\bar x,x\bar kg\bar x}.
\ee
After some manipulations of the \fcs\ in Eq. \eqref{eq:SxRawGroupElement}, one can turn the equation  into a compact form:
\be\label{eq:SxReduced}
\begin{aligned}
&\str^x\ket{k,g,h}\\
=&\eta^{xk\bar x,xg\bar x}(x\bar gh\bar x,hk\bar x)\\
&\x\frac{\beta_{xh\bar x,xk\bar x}(x\bar kg\bar x,xk\bar x)}{\beta_{xk\bar x,xg\bar x}(x\bar gh\bar x  ,xg\bar x)}\ket{xh\bar x,xk\bar x,xg\bar x}.
\end{aligned}
\ee
For completeness and the sake of the interested reader, the two doubly-twisted $2$-cocycles in the expression above are obtained by applying to the $4$-cocycle factors \[
\frac{[xk\bar x,x\bar kg\bar x,xk\bar x,x\bar gh\bar x]^2[xk\bar x,x\bar kg\bar x,x\bar ghk\bar x,x\bar kg\bar x]}{[x\bar kg\bar x,xk\bar x,x\bar gh\bar x,xg\bar x]^2[x\bar kg\bar x,x\bar ghk\bar x,x\bar kg\bar x,xk\bar x]}
\]
in Eq. \eqref{eq:SxRawGroupElement} these three $4$-cocycle conditions:
\begin{align*}
&\delta[xg\bar x,x\bar gh\bar x,xk\bar x,x\bar kg\bar x,xk\bar x]=1,\\
&\delta[xk\bar x,x\bar kg\bar x,xk\bar x,x\bar gh\bar x,xg\bar x]=1,\\
&\delta[xk\bar x,x\bar kg\bar x,x\bar ghk\bar x,x\bar kg\bar x,xk\bar x]=1.
\end{align*} 

\begin{figure}[h!]
\centering
\includegraphics[scale=1.75]{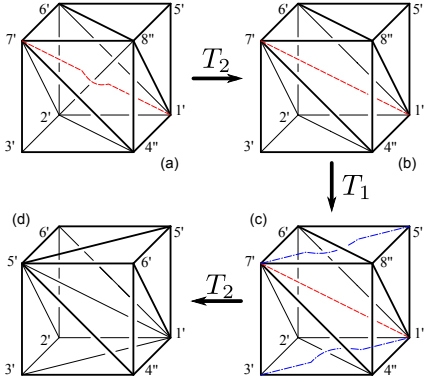}
\caption{(Color online) Mutation moves used in the final step of the $\ttr$ transformation. Certain edges are bent to emphasize that these moves should be viewed  in $4$D. (a) encodes a $T_1$ move already, about the $4$-simplex $1'2'4''8''7'$. There is only one $T_1$ move from (b) to (c) because the two four simplices $1'5'6'8''7'$ and $1'2'4''3'7'$ are the same simplex.}
\label{fig:TtransPachner}
\end{figure}

Likewise, we can work out the $\ttr^x$ as follows.
\begin{align}
 & \ttr^x :\BLvert\torusCubeNB{4}{3}{2}{1}{8}{7}{6}{5}{1.75}\Brangle
  \nonumber\\
\mapsto &[7'2,23,34,48]^{-1}[7'2,23,37,78][7'2,26,67,78]^{-1}\x \nonumber \\
&\frac{[8''7',7'2,26,67][8''1,12,23,37][8''1,15,56,67]}{[8''7',7'2,23,37][8''1,12,26,67]}\x \nonumber \\
&\frac{[6'8'',8''1,12,26]}{[6'8'',8''1,15,56][6'8'',8''7',7'2,26]}\x \nonumber \\
&[5'6',6'8'',8''1,15] \x \nonumber \\
&[3'7',7'2,23,34] \x \nonumber \\
&\frac{[4''8'',8''7',7'2,23]}{[4''3',3'7',7'2,23][4''8'',8''1,12,23]}\x \nonumber \\
&\frac{[2'4'',4''3',3'7',7'2][2'6',6'8'',8''7',7'2]}{[2'4'',4''8'',8''7',7'2][2'6',6'8'',8''1,12]}\x \nonumber \\
&\frac{[2'4'',4''8'',8''1,12][1'2',2'6',6'8'',8''1]}{[1'2',2'4'',4''8'',8''1][1'5',5'6',6'8'',8''1]}\x \nonumber \\
 &\BLvert\torusCubeNlabel{3'}{4''}{2'}{1'}{7'}{8''}{6'}{5'}{1.75}\Brangle
  \nonumber\\
\mapsto &[7'2,23,34,48]^{-1}[7'2,23,37,78][7'2,26,67,78]^{-1}\x \nonumber\\
&\frac{[8''7',7'2,26,67][8''1,12,23,37][8''1,15,56,67]}{[8''7',7'2,23,37][8''1,12,26,67]}\x \nonumber \\
&\frac{[6'8'',8''1,12,26]}{[6'8'',8''1,15,56][6'8'',8''7',7'2,26]}\x \nonumber \\
&[5'6',6'8'',8''1,15] \x \nonumber \\
&[3'7',7'2,23,34] \x \nonumber \\
&\frac{[4''8'',8''7',7'2,23]}{[4''3',3'7',7'2,23][4''8'',8''1,12,23]}\x \nonumber \\
&\frac{[2'4'',4''3',3'7',7'2][2'6',6'8'',8''7',7'2]}{[2'4'',4''8'',8''7',7'2][2'6',6'8'',8''1,12]}\x \nonumber \\
&\frac{[2'4'',4''8'',8''1,12][1'2',2'6',6'8'',8''1]}{[1'2',2'4'',4''8'',8''1][1'5',5'6',6'8'',8''1]}\x \nonumber \\
 &\frac{[1'2',2'4'',4''8'',8''7']}{[1'2',2'6',6'8'',8''7']} \x\nonumber \\
&[1'5',5'6',6'8'',8''7']^2 \BLvert \torusAfterT{3'}{4''}{1'}{2'}{7'}{8''}{5'}{6'}{1.75}\Brangle,\label{eq:TxRaw}
\end{align}
where $x=1'1=2'2=4''3=3'4=5'5=6'6=8''7=7'8$, and $1'<2'<4''<3'<5'<6'<8''<7'<1<2<3<4<5<6<7<8$. The factors after the first `$\mapsto$' is obtained by making the transformations $8\rightarrow 7'$, $7\rightarrow 8''$, $6\rightarrow 6'$, $5\rightarrow 5'$, $4\rightarrow 3'$, $3\rightarrow 4''$, $2\rightarrow 2'$, and $1\rightarrow 1'$ in order. This produces the rows of factors respectively. The state after the vertex transformations is yet not the right state after the $\ttr$ transformation, to obtain which a few mutation transformations (Eqs. \eqref{eq:T1move} and \eqref{eq:T2move}) are needed. Figure \ref{fig:TtransPachner} illustrates the procedure of applying the necessary mutation transformations. Four mutation moves $T_1$, $T_2$, $T_1$, and $T_2$ in order contribute the last four \fcs in Eq. \eqref{eq:TxRaw}. Like the case of the $\str$ transformation, the two blue (dash-dot) lines in Fig. \ref{fig:TtransPachner}(c) also seem indicating two different $T_1$ moves; however, since the two $3$-complexes $1'5'6'8''7'$ and $1'2'4''3'7'$ are identified in the triangulation, the $T_1$ move about $1'5'6'8''7'$ and that about $1'2'4''3'7'$ are done simultaneously and must be thought as a single $T_1$ move. Hence, one can choose either $1'2'4''3'5'$ or $3'5'6'8''7'$ as the one being acted on by the $T_1$. We choose the former in Eq. \eqref{eq:TxRaw}. The same logic applies to the $T_2$ taking Fig. \ref{fig:TtransPachner}(c) to (d). Note also that here, the $T_1$ and $T_2$ on $1'5'6'8''7'$ produce the same \fc $[1'5',5'6',6'8'',8''7']$. Had we chosen the $4$-simplex $1'2'4''3'7'$, we would obtain $[1'2',2'4'',4''3',3'7']^{-1}$ for the $T_1$ and the same for the $T_2$, which should not change the result of $\ttr^x$. Thus,
\be\label{eq:3torus4cocycleIDdue2T1T2}
[1'5',5'6',6'8'',8''7']=[1'2',2'4'',4''3',3'7']^{-1}
\ee 

In terms of the group elements explicitly, the action of $\ttr^x$ on the natural basis reads
\begin{align}
&\ttr^x\ket{k,g,h}\nonumber\\
=
&[x\overline{hk},\bar gh,g,k]^{-1}[x\overline{hk},\bar gh,k,g][x\overline{hk},k,\bar gh,g]^{-1}\x \nonumber\\
&\frac{[xg\bar x,x\overline{hk},k,\bar gh][x\overline{hk},g,\bar gh,k][x\overline{hk},k,g,\bar gh]}{[xg\bar x,x\overline{hk},\bar gh,k][x\overline{hk},g,k,\bar gh]}\x \nonumber \\
&\frac{[x\bar gh\bar x,x \overline{hk},g,k]}{[x\bar gh\bar x,x\overline{hk},k,g][x\bar gh\bar x,xg\bar x,x\overline{hk},k]}\x \nonumber \\
&[xg\bar x,x\bar gh\bar x,x\overline{hk},k] [xk\bar x,x\overline{hk},\bar gh,g] \x \nonumber \\
&\frac{[xk\bar x,xg\bar x,x\overline{hk},\bar gh]}{[xg\bar x,xk\bar x,x\overline{hk},\bar gh][xk\bar x,x\overline{hk},g,\bar gh]}\x \nonumber \\
&\frac{[x\bar gh\bar x,xg\bar x,xk\bar x,x\overline{hk}][xk\bar x,x\bar gh\bar x,xg\bar x,x\overline{hk}]}{[x\bar gh\bar x,xk\bar x,xg\bar x,x\overline{hk}][xk\bar x,x\bar gh\bar x,x\overline{hk},g]}\x \nonumber \\
&\frac{[x\bar gh\bar x,xk\bar x,x\overline{hk},g][xg\bar x,xk\bar x,x\bar gh\bar x,x\overline{hk}]}{[xg\bar x,x\bar gh\bar x,xk\bar x,x\overline{hk}][xk\bar x,xg\bar x,x\bar gh\bar x,x\overline{hk}]}\x \nonumber \\
 &\frac{[xg\bar x,x\bar gh\bar x,xk\bar x,xg\bar x]}{[xg\bar x,xk\bar x,x\bar gh\bar x,xg\bar x]}\frac{[xk\bar x,xg\bar x,x\bar gh\bar x,xg\bar x]}{[xg\bar x,x\bar gh\bar x,xg\bar x,xk\bar x]}\x \nonumber\\  
&\ket{xk\bar x,xg\bar x,xgh\bar x},\label{eq:TxRawGroupElement}
\end{align} 
where Eq. \eqref{eq:3torus4cocycleIDdue2T1T2} is used to obtain the last fraction of two \fcs. The $\ttr^x$ transforms the basis vector $\ket{k,g,h}$ to $\ket{xk\bar x,xg\bar x,xgh\bar x}$ because $h=13\mapsto 1'3'=xhh\bar x$, $g=12\mapsto 1'2'=xhg\bar x$, and $k=15\mapsto 1'5'=xk\bar x$. As to the physical basis, we readily see that 
\be\label{eq:physBasisTtrans}
\ttr^x:\ket{k,g,h'}\mapsto\ket{xkx,xgx,xgh'\bar x}.
\ee
The RHS of Eq. \eqref{eq:TxRawGroupElement} is already in such a form that by rearranging the \fcs, one reduces its complexity:
\be\label{eq:TxNatBasis}
\begin{aligned}
&\ttr^x\ket{k,g,h}\\
=&\frac{[x\bar gh\bar x,x\overline{hk}]_{xk\bar x,xg\bar x}}{[x\overline{hk},\bar gh]_{xk\bar x,xg\bar x}}\\ 
&\x[x\bar gh\bar x,xg\bar x]^{-1}_{xk\bar x,xg\bar x}\ket{xhx,xg\bar x,xgh\bar x}\\
=&\eta^{xk\bar x,xg\bar x}(x\bar gh\bar x,hk\bar x)\\ 
&\x\beta_{xk\bar x,xg\bar x}(x\bar gh\bar x,xg\bar x)^{-1}\ket{xhx,xg\bar x,xgh\bar x},
\end{aligned}
\ee
where definitions \eqref{eq:betaDef} and \eqref{eq:eta} are assumed.

\section{Solutions for the $\str$ and $\ttr$ matrices}\label{app:solModularTrans}
We derive Eq. \eqref{eq:TonAmu} as follows.
We first write down the action of $\ttr\ket{A,B,\mu}$ explicitly.
\begin{align*}
&\ttr\ket{A,B,\mu}\\
=&\ttr^1\ket{A,B,\mu}\\
=&\frac{1}{\sqrt{|G|}}\sum_{\substack{k\in C^A,g \in C^B_{Z_k}\\h'\in Z_{k,g}}} \eta^{k,g}(h',h'gk)\beta_{k,g}(h',g)^{-1}\\
&\x\widetilde\chi^{k,g}_\mu(h')   \ket{k,g,gh'}\\
  =&\frac{1}{|G|}\sum_{\substack{k\in C^A,g \in C^B_{Z_k}\\h'\in Z_{k,g}}} \sum_{\nu=1}^{r(Z^{A,B},\beta_{k^A,g^B})}\frac{\eta^{k,g}(h',h'gk)}{\beta_{k,g}(h',g)}\\
&\x \widetilde\chi^{k,g}_\mu(h')\overline{\widetilde\chi^{k,g}_\nu}(gh') \ket{A,B,\nu},
\end{align*}
where use is made of the inverse transformation \eqref{eq:ghToAmu}. Since $k$ and $g$ each must belong to  only one conjugacy class, in the second equality above, we can just keep the sum over $C^A$ and $C^B_{Z_k}$. Owing to Proposition \ref{prop:chiIs0}, $\chi^{k,g}_\mu(h')=0$ if $h'$ is not $\beta_{k,g}$-regular. Thus, in the sum above, one should keep only the terms with $h'$ being $\beta_{k,g}$-regular, for which $\beta_{k,g}(h',a) =\beta_{k,g}(a,h'),\forall a\in Z_{k,g,h'}$. Clearly, $h'gk\in Z_{k,g,h'}$; hence, $\eta^{k,g}(h',h'gk)=1$ holds. We then have
\begin{align*}
  &\ttr\ket{A,B,\mu}\\
  =&\frac{1}{|G|}\sum_{\substack{k\in C^A,g \in C^B_{Z_k}\\h'\in Z_{k,g}}} \sum_{\nu=1}^{r(Z^{A,B},\beta_{k^A,g^B})}\beta_{k,g}(h',g)^{-1}\\
 & \x \widetilde\chi^{k,g}_\mu(h')\overline{\widetilde\chi^{k,g}_\nu}(gh') \ket{A,B,\nu}.
\end{align*}

Now by definition \eqref{eq:betarepresentation} of $\beta_{k,g}$-representation and Eq. \eqref{eq:rhogg}, we have
\begin{align*}
\widetilde{\chi}^{k,g}_{\mu}(h')\beta_{k,g}(h',g)^{-1}
  & =\mathrm{tr}\left[\widetilde{\rho}^{k,g}_{\mu}(h')\beta_{k,g}(h',g)^{-1}\right]\\
  &=\mathrm{tr}[\widetilde{\rho}^{k,g}_{\mu}(h'g)\widetilde{\rho}^{k,g}_{\mu}(g)^{-1}]\\
  &=\mathrm{tr}[\widetilde{\rho}^{k,g}_{\mu}(h'g)\frac{\overline{\widetilde{\chi}^{k^A,g^B}_{\mu}}(g^B)}{\dim_{\mu}} \mathds{1}]\\
&= \frac{\overline{\widetilde{\chi}^{k^A,g^B}_{\mu}}(g^B)}{\dim_{\mu}}\widetilde{\chi}^{k,g}_{\mu}(h'g).
\end{align*}

Hence, by the orthogonality condition in Eq. \eqref{eq:characterrelation}, and noting that $\sum_{h'}=\sum_{gh'}$, we finally obtain
\be
\begin{aligned}
&\ttr\ket{A,B,\mu}\\
=&\frac{1}{|G|}\sum_{\substack{k\in C^A,g \in C^B_{Z_k}\\h'\in Z_{k,g}}} \sum_{\nu=1}^{r(Z^{A,B})}\frac{\overline{\widetilde{\chi}^{k^A,g^B}_{\mu}}(g^B)}{\dim_{\mu}} \\
&\x \widetilde{\chi}^{k,g}_{\mu}(h'g)\overline{\widetilde\chi^{k,g}_\nu}(gh') \ket{A,B,\nu}\\
=&\frac{\overline{\widetilde{\chi}^{k^A,g^B}_{\mu}}(g^B)}{\dim_{\mu}}\sum_{k\in C^A}\frac{|Z_k|}{|G|} \sum_{g\in C^B_{Z_k}}\frac{|Z_{k,g}|}{|Z_k|}\\
&\x \frac{1}{|Z_{k,g}|}\sum_{h'\in Z_{k,g}}\widetilde{\chi}^{k,g}_{\mu}(h'g)\overline{\widetilde\chi^{k,g}_\nu}(gh') \ket{A,B,\nu}\\
=&\frac{\overline{\widetilde{\chi}^{k^A,g^B}_{\mu}}(g^B)}{\dim_{\mu}} \ket{A,B,\mu},
\end{aligned}
\ee
confirming Eq. \eqref{eq:TonAmu}.

Next, we derive the $\str$ matrix elements in Eq. (\ref{eq:smatrix}). We act the $\str$ operator on a generic eigenvector $\ket{A',B',\nu}$ of the $\ttr$ operator. By Eq. \eqref{eq:SxPhysBasis} with $x$ set to $1$, we have
\begin{align*}
&\str\ket{A',B',\nu}\\
=&\str^1\ket{A',B',\nu}\\
  =&\frac{1}{\sqrt{|G|}}\sum_{\substack{w\in C^{A'},u\in C^{B'}_{Z_{w}}\\v'\in Z_{w,u}}} \widetilde{\chi}^{w,u}_{\nu}(v')\eta^{w,u}(v',v'uw)\\
&\x\frac{\beta_{v'u,w}(\bar wu,w)}{\beta_{w,u}(v',u)}\ket{v'u,w,\bar wu}\\
=&\frac{1}{\sqrt{|G|}}\sum_{\substack{w\in C^{A'},u\in C^{B'}_{Z_w}\\v'\in Z_{w,u}}} \widetilde{\chi}^{w,u}_{\nu}(v')\frac{\beta_{v'u,w}(\bar wu,w)}{\beta_{w,u}(v'  ,u)}\ket{v'u,w,\bar wu},
\end{align*}
where the fact that $\eta^{w,u}(v',v'uw)=1$ for $\beta_{w,u}$-regular characters $\widetilde{\chi}^{w,u}_{\nu}(v')$ that is otherwise zero is taken into account to obtain the third quality. We then have
\be\label{eq:SmatrixRaw}
\begin{aligned}
&\bra{A,B,\mu}\str\ket{A',B',\nu}\\
=&\frac{1}{|G|}\sum_{\substack{k\in C^A,g\in C^B_{Z_k}\\h'\in Z_{k,g}}}  \sum_{\substack{w\in C^{A'},u\in C^{B'}_{Z_w}\\v'\in Z_{w,u}}}\frac{\widetilde{\chi}^{w,u}_{\nu}(v')}{\beta_{w,u}(v',u)}\\
&\x\beta_{v'u,w}(\bar wu,w)\overline{\widetilde\chi^{k,g}_\mu}(h')       \braket{k,g,h'}{v'u,w,\bar wu}\\
=&\frac{1}{|G|}\sum_{\substack{k\in C^A,g\in C^B_{Z_k}\\h'\in Z_{k,g}}}  \sum_{\substack{w\in C^{A'},u\in C^{B'}_{Z_w}\\v'\in Z_{w,u}}}
\frac{\widetilde{\chi}^{w,u}_{\nu}(v')}{\beta_{w,u}(v',u)}\\
&\x\beta_{k,g}(h',g)\overline{\widetilde\chi^{k,g}_\mu}(h')\delta_{k,v'u} \delta_{g,w}\delta_{h',\bar wu},
\end{aligned}
\ee
where the second equality relies on the orthogonality condition $\braket{k,g,h'}{v'u,w,\bar wu}=\delta_{k,v'u} \delta_{g,w}\delta_{h',\bar wu}$. Aided by Eq. \eqref{eq:rhogg} and the following relations, 
\begin{align*}
&\beta_{k,g}(h',g)\widetilde\rho^{k,g}_\mu(h')^\dag=\widetilde\rho^{k,g}_\mu(h'g)^\dag \widetilde\rho^{k,g}_\mu(g)\\
\Leftrightarrow\quad & \beta_{k,g}(h',g)\overline{\widetilde\chi^{k,g}_\mu}(h')= \overline{\widetilde\chi^{k,g}_\mu}(h'g)\frac{\widetilde\chi^{k^A,g^B}_\mu(g^B)}{\dim_\mu}
\end{align*} 
and
\begin{align*}
&\beta_{w,u}(v',u)^{-1}\widetilde{\rho}^{w,u}_{\nu}(v')=\widetilde{\rho}^{w,u}_{\nu}(v'u) \widetilde{\rho}^{w,u}_{\nu}(u)^\dag\\
\Leftrightarrow\quad & \beta_{w,u}(v',u)^{-1}\widetilde{\chi}^{w,u}_{\nu}(v')= \widetilde{\chi}^{w,u}_{\nu}(v'u)\frac{\overline{\widetilde\chi^{w^{A'},u^{B'}}_\nu}(u^{B'})}{\dim_\nu},
\end{align*}
which are a result of Definition \eqref{eq:betarepresentation}, Eq \eqref{eq:SmatrixRaw} becomes
\begin{align*}
&\bra{A,B,\mu}\str\ket{A',B',\nu}\\
=&\frac{1}{|G|}\frac{\widetilde\chi^{k^A,g^B}_\mu(g^B)}{\dim_\mu}
\frac{\overline{\widetilde\chi^{w^{A'},u^{B'}}_\nu}(u^{B'})}{\dim_\nu}
\sum_{\substack{k\in C^A,g\in C^B_{Z_k}\\h'\in Z_{k,g}}}  \sum_{\substack{w\in C^{A'},u\in C^{B'}_{Z_w}\\v'\in Z_{w,u}}}\\
&\x \widetilde{\chi}^{w,u}_{\nu}(v'u)\overline{\widetilde\chi^{k,g}_\mu}(h'g)\delta_{k,v'u} \delta_{g,w}\delta_{h',\bar wu}.
\end{align*}
Finally, by absorbing the $\delta$-functions in the above equation and noticing that $\sum_{h'}=\sum_{h'g}$, we obtain 
\begin{align*}
&\bra{A,B,\mu}\str\ket{A',B',\nu}\\
=&\frac{\widetilde\chi^{k^A,g^B}_\mu(g^B)\overline{\widetilde\chi^{g^{A'},{h'}^{B'}}_\nu}({h'}^{B'})} {|G|\dim_\mu\dim_\nu}
\sum_{\substack{k\in C^A\cap Z_{g,h'},\\ g\in C^B_{Z_k}\cap C^{A'}\\h'\in Z_{k,g}\cap C^{B'}_{Z_g}}} \widetilde{\chi}^{g,h'}_{\nu}(k)\overline{\widetilde\chi^{k,g}_\mu}(h'),
\end{align*}
which is precisely Eq. (\ref{eq:smatrix}).
\end{appendix}

\bibliographystyle{apsrev}
\bibliography{StringNet}

\end{document}